\documentclass[a4paper,fleqn,usenatbib]{mnras}

\usepackage[T1]{fontenc}
\usepackage{ae,aecompl}

\usepackage{graphicx}	
\usepackage{amsmath}	
\usepackage{amssymb}	

\usepackage[utf8x]{inputenc}
\usepackage{amsmath}
\usepackage{graphicx}
\usepackage{multirow}
\usepackage{caption}
\usepackage{subfig}
\usepackage{array}
\usepackage{pdfpages}

\DeclareMathAlphabet{\mathbfsf}{\encodingdefault}{\sfdefault}{bx}{n}
\newcommand{\tens}[1]{\mathbfsf{#1}}
\newcommand{\lf}{\hbox{\it lens}fit}
\def \farcs{\hbox{$.\!\!^{\prime\prime}$}}

\title[KiDS shear calibration]{Calibration of weak-lensing shear in the Kilo-Degree Survey}

\author[Fenech Conti, Herbonnet, Hoekstra, Merten, Miller, Viola]{
I. Fenech Conti$^{1,2}$\thanks{E-mail: ianfc89@gmail.com},
R. Herbonnet$^{3}$,
H. Hoekstra$^{3}$,
J. Merten$^{4}$,
L. Miller$^{4}$,
M. Viola$^{3}$\thanks{Email: viola@strw.leidenuniv.nl}
\\
$^{1}$Department of Physics, University of Malta, Malta.\\
$^{2}$Institute of Space Sciences and Astronomy, University of Malta.\\
$^{3}$Leiden Observatory, Leiden University, PO Box 9513, 2300 RA, Leiden, The Netherlands.\\
$^{4}$Department of Physics, University of Oxford, Keble Road, Oxford OX1 3RH, U.K.
}

\date{Accepted XXX. Received YYY; in original form ZZZ}

\pubyear{2016}

\newcommand{\imsim}{SCHOo{\it l}}

\begin{document}
\label{firstpage}
\pagerange{\pageref{firstpage}--\pageref{lastpage}}
\maketitle

\begin{abstract}
We describe and test the pipeline used to measure the weak lensing shear signal from the Kilo Degree Survey (KiDS). It includes a novel method of `self-calibration' that partially corrects for the effect of noise bias. We also discuss the `weight bias' that may arise in optimally-weighted measurements, and present a scheme to mitigate that bias. To study the residual biases arising from both galaxy selection and shear measurement, and to derive an empirical correction to reduce the shear biases  to $\lesssim 1\%$, we create a suite of simulated images whose properties are close to those of the KiDS survey observations.
We find that the use of  `self-calibration' reduces the additive and multiplicative shear biases significantly,
although further correction via a calibration scheme is required, which also corrects for a dependence of the bias
on galaxy properties.  We find that the calibration relation itself is biased by the use of noisy, measured
galaxy properties, which may limit the final accuracy that can be achieved. We assess the accuracy of the 
calibration in the tomographic bins used for the KiDS cosmic shear analysis, testing in particular the effect
of possible variations in the uncertain distributions of galaxy size, magnitude and ellipticity, and conclude that 
the calibration procedure is accurate at the level of multiplicative bias $\lesssim 1\%$ required for the KiDS cosmic shear analysis.
\end{abstract}




\section{Introduction \label{introduction}}

The matter distribution in the Universe changes the geometry of spacetime, thus altering the paths of light rays. As this mimics the effects of a lens, with the gravitational potential taking the role of the index of refraction, this phenomenon is referred to as gravitational lensing. If the deflector is massive and the light rays pass sufficiently close, multiple  images of the same source may be observed. More typically the source position only appears shifted by an unknown amount. The variation in the deflection across the image results, however, in a stretching (shear) and changes the observed size (magnification). This regime is commonly referred to as weak gravitational lensing \citep[see e.g.][for an extensive introduction]{bartelmann01a}.

The original source properties are unknown, and thus the measurement of  a single galaxy does not provide meaningful information. However, sources that are close on the sky have experienced similar deflections and consequently their observed orientations are correlated. The changes in the shapes of the observed galaxies are small, typically at the level of a few percent, much smaller than their intrinsic shapes. Hence, the weak lensing signal can only be determined statistically by averaging the shapes of many sources, under the assumption that there are no intrinsic correlations \citep[but see e.g.,][for a review on intrinsic alignments]{joachimi15a}.

The ellipticity correlations can be related directly to the statistics of matter density fluctuations
\citep[e.g.][]{blandford91a, miralda-escude91a, kaiser92a} and can thus be used to infer the cosmological model. This application, commonly known as cosmic shear, is one of the most powerful ways to study the nature of dark energy and constrain modified gravity theories \citep[see][for a recent review]{kilbinger15a}. Since the first detections in 2000 \citep{bacon00a,kaiser00a,vanwaerbeke00a,wittman00a} the precision of the measurements has improved dramatically thanks to deep imaging surveys of ever larger areas
\citep[e.g.][]{hoekstra06a, fu08a}. Moreover, observations in multiple pass-bands allowed for the determination of photometric redshifts, which are essential to improve constraints on cosmological parameters 
\citep{schrabback10a, heymans13a, jee15a}. The measurement of cosmic shear is also a major science driver for a number of ongoing large imaging surveys, such as the Kilo Degree Survey  \citep[KiDS;][]{dejong15a,kuijken15a}, the Dark Energy Survey \citep[DES;][]{becker15a,jarvis15a} and the Hyper-Suprime Cam Survey \footnote{http://www.naoj.org/Projects/HSC/surveyplan.html}. 

The increase in precision afforded by these surveys needs to be matched by a corresponding improvement in the accuracy with which galaxy shapes can be measured. The main complications are (i) that the true galaxy image is convolved with a point spread function (PSF) due to atmospheric effects and telescope optics; (ii) the resulting image is pixelised by the detector; (iii) the images contain noise from various sources. Each effect introduces systematic changes in the galaxy shapes, or affects our ability to correct for it. Although shape measurement algorithms differ in their sensitivity to some of the systematics, because of differences in their implementation or the assumptions that are made, they are all affected by noise in the data. 

Fortunately, it is well understood how the galaxy surface brightness is transformed into an image, and this process can be emulated. Creating mock images of telescope observations can thus be used to understand the impact of systematic effects and their propagation throughout the shear measurements. Moreover, by comparing the output shears to the input values the biases can be quantified. The biases themselves
are classified in additive and multiplicative bias. The former arises from an incomplete correction for the convolution by the (typically) anisotropic PSF, or by residual errors in the PSF model itself. The data themselves can be used to examine the presence of additive biases \citep[see e.g.][]{heymans12a}.
Multiplicative bias, a change in the amplitude of the lensing signal, can only be reliably studied using simulated data. The Shear TEsting Programme \citep[STEP;][]{STEP1,massey07a} represented the first community-wide effort to benchmark the performance of various weak lensing pipelines using simulated images. Although simplistic in many regards, the simulated data included some of the complexity of real data, such as blending of objects. To examine the differences between algorithms more systematically, the Gravitational LEnsing Accuracy Testing \citep[GREAT;][]{GREAT08, GREAT10, mandelbaum15a} challenges focused on more idealised scenarios. 

When applying an algorithm to actual data, evaluating the performance on realistic mock data is essential
\citep{miller13a,hoekstra15a}. An essential step in this process is to ensure that the simulations are sufficiently realistic, such that the inferred bias is robust given the uncertainties of the input parameters. One approach is to match the observed properties of the simulated images to those of the real data by modifying the input distributions in case differences are found \citep[e.g.][]{bruderer15a}. Alternatively, the simulated output can be used to account for differences with the actual data by parameterising the bias as a function of observed galaxy properties. In \citet{kuijken15a} and \citet{jarvis15a} the shear biases for KiDS DR1/2 and DES, respectively, were corrected using a function of size and signal-to-noise ratio (hereafter SNR). Another option we explore is to re-weight the catalogue entries such that they match the observations. 

In this paper we focus on \lf{} \citep{miller13a}, a likelihood based algorithm, which fits observed galaxy profiles with an elliptical surface brightness model that is convolved with a model of the PSF.  This algorithm has been used to measure the lensing signal from CFHTLenS \citep{heymans13a} and RCSLens \citep{hildebrandt16a}, as well as the initial release of KiDS \citep{kuijken15a}. Like any other method, the \lf{} measurements are biased if the SNR is low \citep[this is commonly referred to as noise bias; e.g.][]{melchior12a,refregier12a,miller13a}. In the latest of these challenges, GREAT3 \citep{mandelbaum15a} an improved version of \lf{} was introduced and tested: a new self-calibrating algorithm was added to alleviate the effect of noise bias. This improvement reduced the biases from tens of percents to a percent level. In this paper we expand on this formalism and apply the algorithm to simulated images that are designed to mimic KiDS data. 

The third public data release of KiDS \citep[KiDS-450 hereafter;][]{KiDS450} comprises 360.3 square degrees of unmasked area with an effective number density of 8.3 galaxies per square arcminute. \cite{KiDS450} calculate that the required level of bias in shape measurements that can be tolerated given the precision afforded by KiDS-450 implies that 
the multiplicative bias needs to be determined to better than $\sim 1\%$. In spite of the fact that the performance of the self-calibrating version of \lf{} is close to this requirement, a final adjustment is nonetheless required to reduce the bias further. Although this is only a small correction in absolute terms when compared to the improvement by self-calibration itself, we note that the actual implementation can be rather complex .

To reduce the biases in the shear determination for KiDS-450 to the required level of accuracy, we present \imsim \ for KiDS, the Simulations Code for Heuristic Optimization of \lf{} for the Kilo Degree Survey, which was used to obtain a shear bias calibration for the latest KiDS-450 lensing catalogues obtained with a new version of \lf{}. \imsim \ was designed to carry out the following: i) testing of the newest version of the \lf{} algorithm; ii) deriving bias calibration functions for the KiDS-450 data; iii) evaluating the robustness of the final calibration functions to the input of the calibration data. The main modifications to \lf{} are presented in \S\ref{sec:selfcal}. The image simulations are described in detail in \S\ref{galsims}. These are used to quantify and account for the residual bias in the self-calibrating \lf{} algorithms in 
\S\ref{sec:shearCalib}. In \S\ref{sec:resampling} we examine how differences between the simulated and observed data can be accounted for using a resampling of the the simulated measurements.  In \S\ref{sec:sensitivity} we examine the robustness of the results.

\section{The shear measurement method}
\label{sec:selfcal}

\subsection{{\em lens}fit}

The shear measurement method used in the analysis of KiDS data is \lf{} 
\citep{miller07a,kitching08a,miller13a},
which has also been used to measure the lensing signal from CFHTLenS \citep{heymans13a}, RCSLenS \citep{hildebrandt16a} and the initial release of KiDS \citep{kuijken15a}.  
It is a likelihood based algorithm that fits observed galaxy profiles with a surface brightness model that is convolved with a model of the PSF. 
The PSF model is obtained from a fit to the pixel values of stars, normalised in flux, with a polynomial
variation across individual CCD images and across the full field of each individual exposure.
Galaxies are modelled as an exponential disk plus a bulge (S\'{e}rsic index $n=4$) component.
There are seven free parameters (flux, size, ellipticity, position and bulge-to-total flux ratio).
To reduce the model complexity, the ratio of disk and bulge
scale lengths is a fixed parameter and the ellipticities of the disk and bulge are set equal.
The likelihood for each galaxy, as a function of these parameters, is obtained from a joint fit to
each individual exposure, taking into account the local camera distortion.
The measured ellipticity parameters are deduced from the  likelihood-weighted mean parameter value, 
marginalised over the other parameters, adopting priors for their distribution.
To determine the lensing signal, the ellipticities of the galaxy models are combined with a weight, which takes care of the uncertainty in the ellipticity measurement, to form an estimate of the shear from the weighted average. The complexity of the galaxy model has been designed to be sufficient to
capture the dominant variation in galaxy surface brightness distributions visible in ground-based data,
without unduly overfitting a model that is too complex to noisy data (SNR$\gtrsim 10$).
In principle, we may be concerned that differences between the \lf{} model and actual surface
brightness distributions may introduce model bias \citep[e.g.][]{kacprzak14a, zuntz13a},
however \citet{miller13a} have argued that the possible model
bias should be sub-dominant in ground-based data analyses, an argument that is supported by the 
performance of \lf{} on simulated realistic galaxies in the {\sc great3} challenge
\citep{mandelbaum15a}.  

We investigate the possible amplitude of such model bias in Appendix\,\ref{sec:modelbias_sims} and 
conclude that indeed the effect is expected to be small in the KiDS-450 analysis.

For the latest analysis of KiDS-450 data \citep{KiDS450} we use an updated version of \lf{}, which is based largely on the methods adopted for CFHTLenS as described by \citet{miller13a}, but with some modifications and improvements to the algorithms. The most prominent changes are the self-calibration for
noise bias and the procedure to calibrate for weight bias, which are described in more detail below
in \S\ref{sec:selfcal_detail} and \S\ref{sec:weight_bias}, respectively. 
Moreover, the handling of neighbouring objects, and the sampling of the likelihood surface were improved.

In surveys at the depth of CFHTLenS or KiDS, it is essential to deal with  contamination by closely neighbouring galaxies (or stars). The \lf{} algorithm fits only individual galaxies, so contaminating stars or galaxies in the same postage stamp as the target galaxy are masked out during the fitting process.
The masks are generated from an image segmentation and deblending algorithm, similar to that employed in \textsc{SExtractor} \citep{bertin96a}. However, the CFHTLenS version rejected target galaxies that were too close to its neighbours. For KiDS, a revised deblending algorithm was adopted that resulted in fewer rejections and thus a 
higher density of measured galaxies. The distance to the nearest neighbour was recorded in the catalogue output so that any bias as a function of neighbour distance could be identified and potentially
rectified by selecting on that measure. The sampling of the likelihood surface was improved in both speed and accuracy, by first identifying the location of the maximum likelihood and only then applying the adaptive sampling strategy described by \citet{miller13a}. More accurate marginalisation over the galaxy size parameter was also implemented.

In the following analysis, the identical version of {\em lens}fit, with the same data handling setup, was used for
the simulations as for the KiDS-450 data analysis of \citet{KiDS450}.

\subsection{Self Calibration of Noise Bias}
\label{sec:selfcal_detail}

In common with other shear measurement methods, {\em lens}fit measurements of
galaxy ellipticity are biased by the presence of pixel noise: even if the pixel
noise is Gaussian or Poissonian in nature, the non-linear transformation to
ellipticity causes a skewness of the likelihood and a bias in any single-point
estimate of individual galaxy ellipticity that propagates into a bias on
measured shear values in a survey
\citep{refregier12a, melchior12a, miller13a}. The bias is a complex function of
of SNR, size, ellipticity and surface brightness distribution
of the galaxies, but also depends on the point spread function (PSF) morphology.
Given that we only have noisy estimates of galaxy properties, it is difficult to
predict the bias with sufficient accuracy, and to date published shear surveys have
used empirical methods to calibrate the bias, typically by creating simulations that
match the properties of the survey, measuring the bias in the simulation as a function
of observed (noisy) galaxy properties and applying a calibration relation derived from
those measurements to the survey data
\citep{miller13a, kuijken15a, jarvis15a, hoekstra15a}.

In the current analysis we first apply an approximate correction for noise bias that is
derived from the measurements themselves, which we refer to as self-calibration.  The method
was first used for the ``MaltaOx'' submission in the {\sc great3} challenge \citep{mandelbaum15a}.
When a galaxy is measured, a nominal model is obtained for that galaxy, whose parameters are
obtained from a mean likelihood estimate. The idea of self-calibration is to create a
simulated test galaxy with those parameters, remeasure the test galaxy 
using the same measurement pipeline, and measure
the difference between the remeasured ellipticity and the known test model ellipticity. It is
assumed that the measured difference is an estimate of the true bias in ellipticity for that
galaxy, which may be subtracted from the data measurement. The estimate of a galaxy's size
is also simultaneously corrected with the ellipticity.
Ideally, when the test galaxy
is remeasured, we would like to add multiple realisations of pixel noise and marginalise
over the pixel noise: however such a procedure is computationally expensive, so in the
current self-calibration algorithm we adopt an approximate method in which the noise-free
test galaxy model is measured, but the likelihood is calculated as if noise were present.
Mathematically we may represent the log likelihood of a measurement, $\log\mathcal{L}$ as
\begin{eqnarray}
\log\mathcal{L}(p) &=& -\frac{1}{2} (\vec{D}-\vec{M}(p))^T \tens{C}^{-1} (\vec{D}-\vec{M}(p)) \nonumber \\
&=& (\vec{M_0}+\vec{N}-\vec{M}(p))^T \tens{C}^{-1} (\vec{M_0}+\vec{N}-\vec{M}(p)) \nonumber \\
&=& (\vec{M_0}-\vec{M}(p))^T \tens{C}^{-1} (\vec{M_0}-\vec{M}(p)) \nonumber \\
& & + 2 (\vec{M_0}-\vec{M}(p))^T \tens{C}^{-1} \vec{N} \nonumber \\
& & + \vec{N}^T \tens{C}^{-1} \vec{N} \label{eqn:likel-selfcal}
\end{eqnarray}
where we express the data as a vector $\vec{D}$, the model obtained with parameters $p$ as
$\vec{M}(p)$ and the pixel noise covariance matrix as $\tens{C}$, and where we decompose
the data into a true model $\vec{M_0}$ and a noise vector $\vec{N}$. Our self-calibration 
procedure corresponds to generating a test galaxy whose model $\vec{M_0}$ is described
by the parameters measured from the data for that galaxy  and where we only calculate the leading term
in the likelihood, equation\,\ref{eqn:likel-selfcal}, for this test galaxy, ignoring terms involving $\vec{N}$,
when estimating the bias. In the case where the noise is uncorrelated with the galaxy, corresponding to the background-limited case of a faint galaxy, the noise-model cross-term would disappear if we were to marginalise $\log\mathcal{L}$ over the noise, the final term would be a constant, and the leading term would provide a good estimate of the expected distribution. Unfortunately, when estimating the ellipticity, we are interested in the likelihood $\mathcal{L}$ and not its logarithm, $\log\mathcal{L}$, and so ignoring the 
noise-model cross-term may lead to an error in the derived bias. However, we also make the approximation that the values of the model parameters measured from the data are close to the true galaxy parameters, which at low SNR may not be true. Hence our procedure can only be approximate. 

However, self-calibration has the advantage that, unlike calibration from an external simulation, it does not rely on an assumed distribution of galaxy parameter values: the input model parameter values are taken from those measured on each individual galaxy in the data analysis.  The method appears particularly useful in removing PSF-dependent additive bias,
which is otherwise hard to mitigate using external simulations, which typically do not reproduce the PSF for 
each observed galaxy.

In making the self-calibration likelihood measurements, we are careful to ensure that the galaxy ellipticity and size parameters are sampled at the same values as in the data measurement for each galaxy, so that sampling variations do not cause an additional source of noise in the self-calibration. This procedure also makes self-calibration computationally fast, as the step of identifying which samples to use is not repeated.

The GREAT3 results \citep{mandelbaum15a} showed that the self-calibration correction does, on average, reduce the shear bias to the percent level and that the amplitude of the residual bias is almost independent of the morphology of the simulated galaxies. Importantly, the reduction in noise bias improves both the multiplicative and additive biases, and the self-calibration procedure therefore has been applied to the survey data measurements presented in \cite{KiDS450}. The residual bias, however, is still correlated with galaxy properties such as SNR and size. As the distributions of those properties are redshift- and magnitude-dependent, the residual bias may be large enough to lead to a significant bias in tomographic shear analyses. We therefore seek to empirically calibrate the residual bias using conventional methods, employing realistic image simulations as described in \S\ref{galsims}.

\subsection{Weight bias correction}
\label{sec:weight_bias}

In our standard analysis, we apply a weight to each galaxy that takes account of both the shape noise variance and the ellipticity measurement noise variance, following \citet{miller13a}. The ellipticity noise variance is measured from the ellipticity likelihood surface for each galaxy, after marginalisation over other parameters, with a correction for the finite support imposed by requiring ellipticity to be less than unity. This contrasts with approaches such as that of 
\citet{jarvis15a}, where an average correction as a function of galaxy parameters, such as 
flux signal-to-noise ratio, is derived and applied.

Our scheme should result in optimal SNR in the final shear measurements,
but any bias in the weights would introduce a shear bias.  Inspection of the distribution of
weight values shows that indeed there are two sources of weight bias that arise.  First, the
measurement variance is a systematic function of the ellipticity of the galaxy, with a tendency
for galaxies to have smaller measurement variance, and hence higher weight, at intermediate values
of ellipticity, compared with either low or high ellipticity, for galaxies of comparable
isophotal area and SNR. This results in a tendency to overestimate shear at intermediate and low
values of SNR, to an extent that is sensitive to the distribution of galaxy ellipticities.

A second bias that arises is correlated with the PSF anisotropy.  Galaxies of a given total flux
that are aligned with the PSF tend to have a higher SNR than galaxies that are cross-aligned
with the PSF, and also tend to have a smaller measurement variance. This orientation bias
has the same origin as that discussed by \citet{kaiser00b} and \citet{bernstein02a} and results in a net
anisotropy in the overall distribution of weights which, if uncorrected, would result in a
net shear bias.

In the KiDS-450 analysis, we adopt an empirical correction for these effects by 
determining the mean measurement variance for the full sample of galaxies as a function of their
2D ellipticity, $e_1$, $e_2$, and as a function of their SNR and isophotal area. 
From that mean variance, a correction is derived that may be applied to the weights to 
ensure that, on average, the distribution of weights is neither a strong function of
ellipticity nor of position angle. The anisotropic bias depends on the size and ellipticity
of the PSF, so to accommodate variations in the PSF across the survey, galaxies from the 
entire completed survey are binned according to their PSF properties, and the weights
correction is derived in each PSF bin \citep{KiDS450}.  In the simulations,
we apply the equivalent weight bias correction to each of 13 sets of PSFs that are simulated (see \S\ref{sec:sim_setup}).

\section{Image simulations}
\label{galsims}

\subsection{The simulation of galaxies}
The performance of shape measurement algorithms can only be evaluated using simulated images. 
To this end, a number of community-wide efforts have been undertaken to benchmark methods.
The self-calibrating version of \lf{} performed well on simulated images from GREAT3 \citep{mandelbaum15a}, the latest of these challenges, with an average shear bias of about a percent. Whilst useful to test new algorithms and to better understand common sources of bias in shape measurements, these general image simulations cannot be used to evaluate the actual performance. First of all, they ignore the effects neighbouring objects can have on the shape measurement, which was shown to be important by \cite{hoekstra15a}. Moreover, to calibrate the performance with high accuracy, the simulations should match the real data in terms of  survey depth, number of exposures, noise level, telescope PSF and pixelisation.

To quantify and calibrate the shear biases of the self-calibrating version of \lf{} for the new KiDS-450 dataset we created the \imsim \ for KiDS pipeline, Simulations Code for Heuristic Optimization of \lf{} for the Kilo Degree Survey. We use it to generate a suite of image simulations that mimic the $r$-band KiDS observations that were used in \cite{KiDS450} to measure the cosmic shear signal. As discussed below, we match the dither pattern, instrument footprint, average noise level, seeing and PSF properties. The simulated images are created using \textsc{GalSim} \citep{rowe15a}, a widely used galaxy simulation software tool developed for GREAT3. Note that we do not aim to test the PSF modelling \citep[this was presented in][]{kuijken15a}.

\subsection{Simulation volume}
\label{sec:imsim:lfcat}

The precision with which biases are measured can be improved by creating and analysing more simulated images. However, it is a waste of computational resources if the biases are already known sufficiently well compared to the statistical uncertainties of the cosmic shear signal. Moreover, as a result of simplifications in the simulated data, residual biases may remain. It is therefore useful to establish the level of accuracy that is required, given the KiDS-450 data set, and use these results to determine the simulation volume that is needed. \cite{KiDS450} showed that the \lf{} shear multiplicative bias has to be known with an accuracy of at least 1\% for the error bars on cosmological parameters not to increase by more than 10\% (see their Appendix A3).
\cite{KiDS450} do not set requirements on the knowledge of the additive bias from the simulations. In fact the residual additive bias is measured from the data themselves \citep{heymans12a} as there are a number of steps in the data acquisition, processing and analysis which are not simulated and might contribute to amplitude of the additive bias (e.g. cosmic rays, asteroids, binary stars, imperfect PSF modelling, non-linear response of CCD...). The observed level of residual bias may be used to determine the maximum scale where the cosmic shear signal is robust, in contrast to multiplicative shear bias, which affects all angular scales.

In our simulations we apply a shear with a modulus $|g|=0.04$ to all galaxies. This is a compromise
between the small shears we aim to recover reliably, whilst minimising the number of simulated images.
For a fiducial intrinsic dispersion of ellipticities $\sigma_\epsilon=0.25$, the minimum required number of galaxies to reach a precision of 0.01 on the multiplicative bias is then $N_{\rm gal}=(\sigma_{\epsilon}/(0.01|g|))^2 \approx 3.9\times 10^5$. This number should be considered the bare minimum, because in practice we wish to explore the amplitude of the bias as a function of galaxy and PSF properties. 

The dominant source of uncertainty is the intrinsic dispersion of ellipticities. This source of noise can, however, be reduced in simulations using a shape noise cancellation scheme  \citep{massey07a}. This results in a significant reduction in the number of simulated galaxies, without affecting the precision with which the biases can be determined. Previous studies have done so by introducing a copy of each galaxy, rotated in position angle by $90^\circ$ before applying a shear and convolution by the PSF, such that the mean of the intrinsic ellipticity $\epsilon^s$ satisfies $\langle\epsilon^s\rangle = 0$ \citep[e.g.][]{massey07a, hoekstra15a}.
Although this reduces the shape noise caused by galaxies, such a scheme does not guarantee that the mean of the {\it observed} ellipticity values $\langle \epsilon \rangle = g$. That condition is only satisfied by a population of galaxies that are uniformly distributed around circles of $\epsilon^s$.  Fortunately, even a small number of rotated copies of each galaxy suffices to meet this criterion to adequate accuracy.

In this work we create four copies of each galaxy, separated in intrinsic position angle by $45^\circ$. 
If we write the first copy as having intrinsic ellipticity $\epsilon^s$, we may write the complex intrinsic ellipticity of each copy as $\epsilon^s_n = {\mathrm i}^n\epsilon^s$ for each rotation, $n=0 \ldots 3$.
The relation between the sheared ellipticity $\epsilon_n$, the reduced shear $g$ and 
$\epsilon^{s}_n$, for each rotation, is 

\begin{equation}
\epsilon_n = \frac{\epsilon^{s}_n + g}{1+ g^* \epsilon^{s}_n} = \frac{{\mathrm i}^n\epsilon^{s} + g}{1+ g^* {\mathrm i}^n\epsilon^{s}} \, ,
\label{shear_eps}
\end{equation}

\noindent where the asterisk denotes the complex conjugate.  A shear estimate $\tilde{g} = \langle \epsilon_n \rangle$ then reduces to

\begin{equation}
 \tilde{g} = \frac{g - {g^*}^3  \left(\epsilon^{s} \right)^4}{1-\left( g^* \epsilon^{s} \right)^4} \, .
\end{equation}

\noindent For the same fiducial values, $\lvert \epsilon^s \rvert \simeq 0.25$ and $\vert g \rvert= 0.04$,
this expression differs from $g$ with a relative error of order
$\Delta g / g \simeq \lvert g \rvert^2 \lvert\epsilon^{s}\rvert^4 \simeq 6 \times 10^{-6}$, 
compared with
$\Delta g / g \simeq \lvert\epsilon^{s}\rvert^2 \simeq 0.06$ 
for the shape noise reduction achieved using only pairs of galaxies \citep{massey07a}.
The four-rotation method has significantly higher
accuracy relative to the two-rotation method at the highest values of $\epsilon^{s}$.
 
Using a larger number of rotated galaxies reduces the shear measurement error further,  to $\Delta g /g \sim 10^{-13}$ for 8 duplicated galaxies. However, for a given simulation volume, this reduces the diversity in other galaxy properties.
Moreover, pixel noise in the simulated images reduces the effectiveness of shape noise cancellation for
galaxies with low SNR, which are the most numerous. Furthermore, not all rotated galaxy copies may be detected, thus breaking the assumed symmetry in the analytical estimate. The weighted dispersion of the mean input ellipticities of the set of four catalogues is 0.084, a factor about $3$ reduction compared to the case without shape noise cancellation. This corresponds to a decrease of a factor about $9$ in the number of simulated galaxies required to achieve a fixed uncertainty in shear bias measurement. 

\begin{figure}
 \centering
  \includegraphics[width=0.5\textwidth]{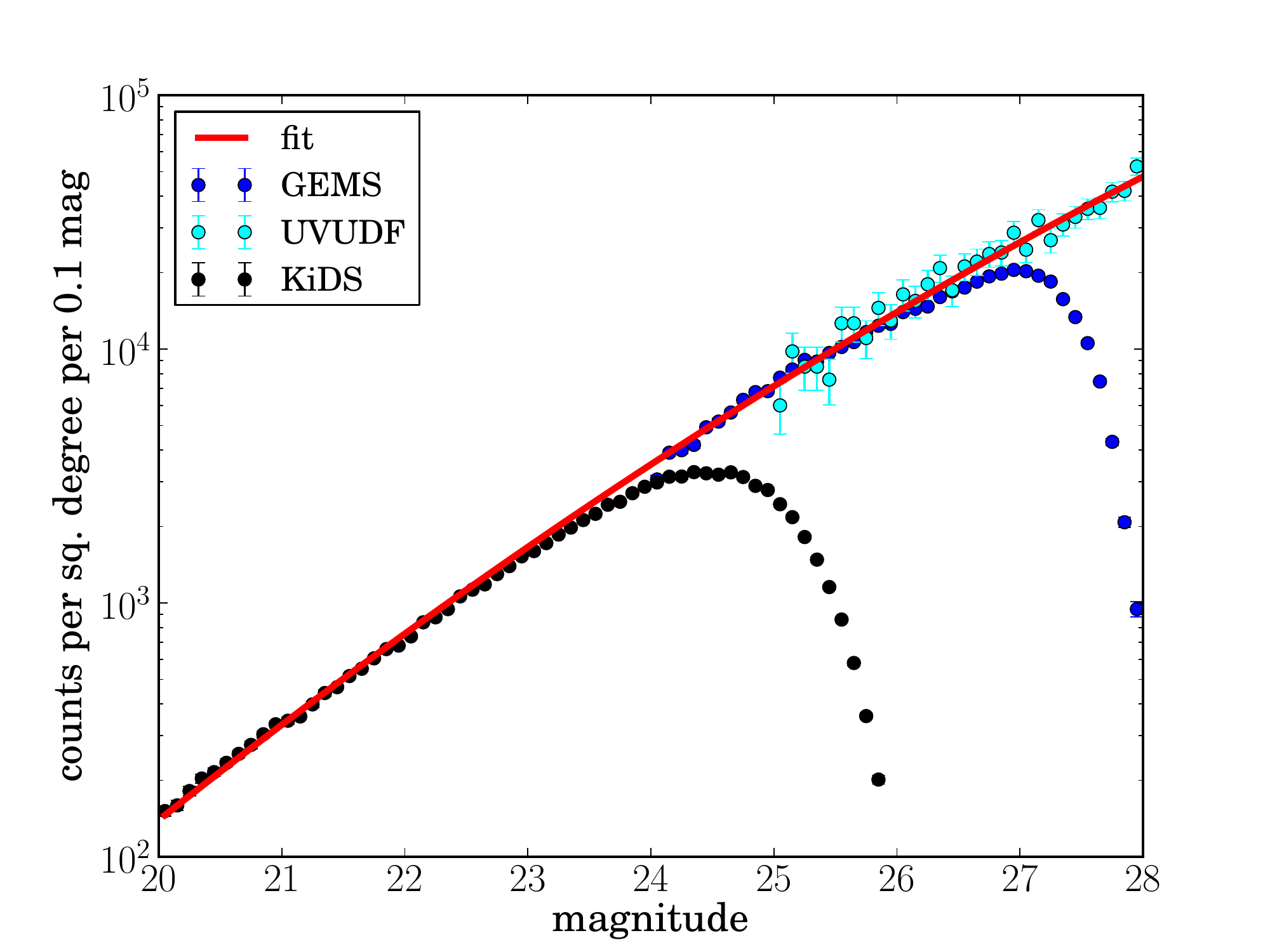}
   \caption{$r$-band magnitude histograms of KiDS-450 data (black), GEMS survey data (blue) and UVUDF survey (cyan), with uncertainties given by the Poisson errors of each point. The red line is the best fit through KiDS-450 $20<m_r<23$ points, GEMS $25<m_r<26$ points and UVUDF $26<m_r<29$ datapoints and is used as the input magnitude distribution of the simulations.}
 \label{fig:magdistr}
\end{figure}

\subsection{Input object catalogue} 
\label{sec:input_cat}

To measure meaningful shear biases from the simulated data it is essential that the properties of the simulated objects are sufficiently realistic. For instance, neighbouring galaxies affect shape measurements \citep{dawson14a}, and therefore the correct number density of galaxies needs to be determined. Moreover, \citet{hoekstra15a} highlighted the importance of simulating galaxies well beyond the detection limit of the survey in order to derive a robust shear calibration. Galaxies just below the detection limit can still blend with brighter galaxies,  directly affecting the measurement of the object ellipticity, whereas even fainter galaxies affect the background and noise determination by acting as a source of correlated noise. Hence we include in our simulations galaxies as faint as 28th magnitude, which should be adequate given the depth of KiDS.

We place the objects at random positions, and thus ignore the additional complication from clustering. The fraction of blended objects in the simulations might therefore be low compared to the true Universe. Alternatively, galaxies could be positioned in the simulations according to their positions in observations \citep[e.g.][]{miller13a,jarvis15a}. This would naturally include realistic clustering, but cannot be used for the galaxies below the detection limit, and thus unusable for our deep magnitude distribution. However, we examined the impact of varying number density and found the changes in bias to be negligible for the KiDS-450 analysis (see \S\ref{sec:shearCalib:lf} for details).

\begin{figure*}
 \centering
  \includegraphics[width=1.0\textwidth]{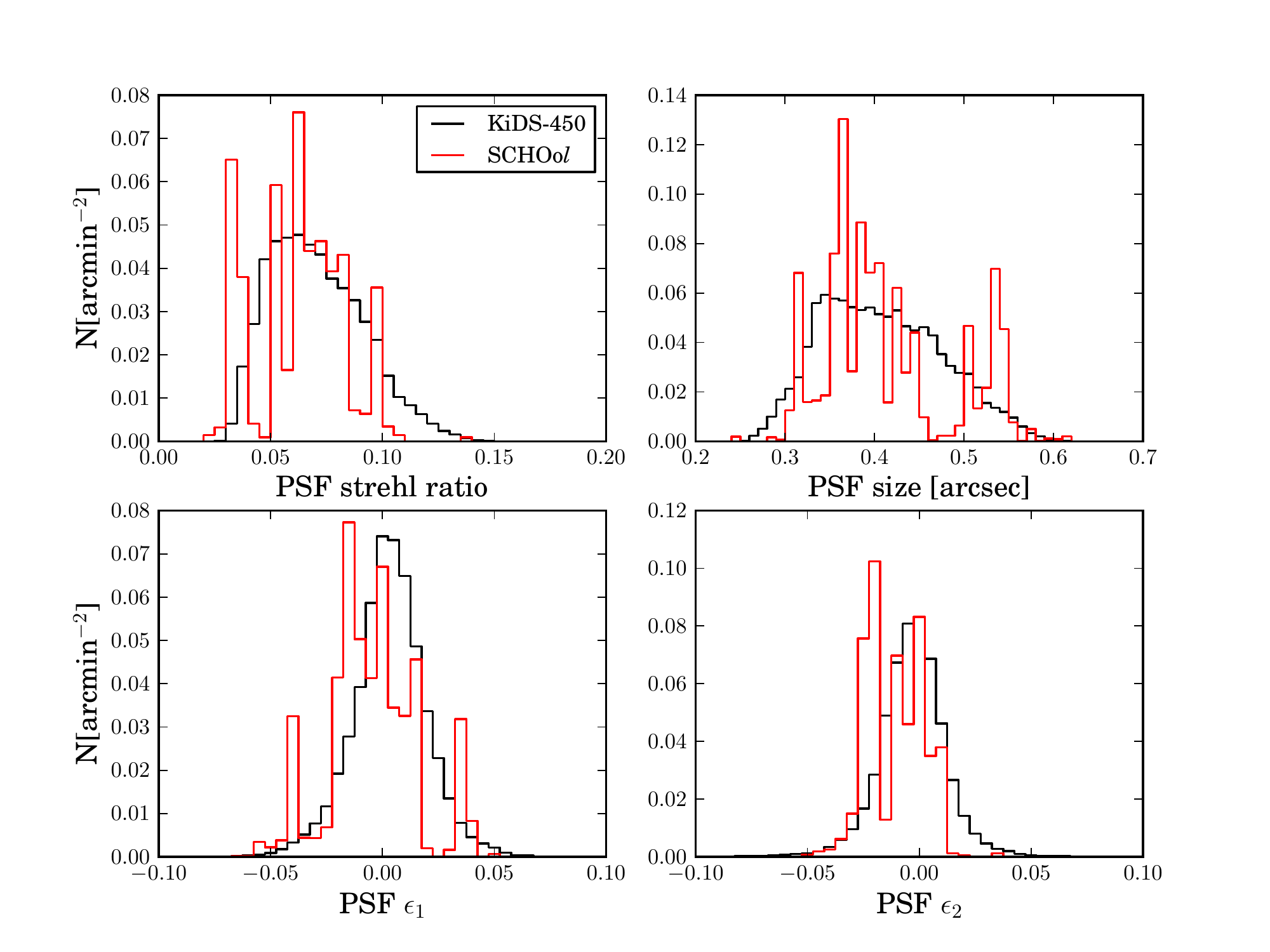}
   \caption{Distributions of PSF parameters in the simulations (red) and KiDS-450 (black) measured by \lf{} using a 2.5 pixel weighting function. Shown are the distributions of measured pseudo-Strehl ratio, size and the two components of the ellipticity. The constant PSFs (for individual exposures) in the \imsim \ images give rise to very peaky distributions, but overall the range in properties in the data are matched by the image simulations.}
 \label{fig:sim_psfdistr}
\end{figure*}

To create a realistic magnitude distribution that extends to 28th magnitude, we augment the measured KiDS-450 galaxy counts with measurements from deeper {\it Hubble Space Telescope} (HST) images.
We use the HST/ACS $F606W$ counts from GEMS \citep{rix04a} and UVUDF \citep{rafelski15a}, because
this filter resembles the KiDS $r$ filter fairly well.  We remove objects classified as stars from all three data sets, and exclude masked areas in the KiDS-450 data. Fig.~\ref{fig:magdistr} shows the magnitude distributions of a subsample of KiDS-450 data (black), GEMS data (blue) and UVUDF data (cyan). The error bars show the Poisson errors of the data points.

We fit a second order polynomial to the logarithm of the number counts, using KiDS-450 data between $20<m_r<23$, GEMS data between $25<m_r<26$ and UVUDF data between $26<m_r<29$. The resulting magnitude distribution for the simulated galaxies is given by: 

\begin{equation}
 \mathrm{log}N(m_r) = -8.85 +  0.71 m_r - 0.008 m_r^2,
 \label{eq:magprior}
\end{equation}
where $N(m_r)$ is the number of objects with $r$-band magnitude $m_r$ per square degree. The fit is mostly constrained by the KiDS data, with the ancillary data driving the flattening of the curve at faint magnitudes. Magnitudes are converted to counts to be used by \textsc{GalSim} using a magnitude zeropoint of 24.79, the median magnitude zeropoint in the KiDS-450 data.

Creating images of large numbers of faint galaxies with $m \geq 25$ by \textsc{GalSim} would be rather time consuming. However, we are not interested in their individual properties, because they are too faint to enter the sample used for the lensing analysis. Instead we only need to ensure that their impact on shape measurements is captured, for which it is sufficient that their number densities and sizes are realistic. 
To improve the speed of the pipeline, we therefore create postage stamps for a representative sample of these faint galaxies, and use these to populate the simulations by randomly drawing from this sample, whilst ensuring that the magnitude distribution in equation\,\ref{eq:magprior} is obeyed. These faint galaxies also have lensing shear applied.

Realistic galaxy morphologies, in particular the distribution of surface brightness profiles, and consequently sizes and ellipticities, are another essential ingredient for image simulations. The intrinsic ellipticity distribution for galaxies is the same as in the CFHTLenS image simulations and the functional form is taken from Appendix~B2 in \cite{miller13a}. It corresponds, as is the case for the size distribution, to the prior used by \lf{} to measure galaxy shapes.
We model the galaxies as the linear combination of a de Vaucouleur profile for the bulge and an exponential profile for the disk. The bulge flux to total flux ratio, $B/T$, is randomly sampled from a truncated Gaussian distribution between 0 and 1 with its maximum at 0 and a width of 0.1, the same as was used for the CFHTLenS simulations presented in \citet{miller13a}. Ten percent of all galaxies are set to be bulge-only galaxies with $B/T=1$, and the rest have a disk with random values for the bulge fraction. 

The sizes of the galaxies are defined in terms of the scale length of the exponential disk along the major axis, and are randomly drawn from the distribution
\begin{equation}
 P(r) \propto r \hspace{1pt} \mathrm{exp}(-(r/A)^{4/3}),
 \label{eq:sizeprior}
\end{equation}
where $A$ is related to the median of the distribution, $r_{\rm med}$, by $A = r_{\rm med}/1.13$\footnote{There was an error in Appendix B1 of \citet{miller13a}: the factor 1.13 shown here was also used for the CFHTLenS analysis, instead of the incorrectly reported value of 0.833.} and where the relationship between $r_{\rm med}$ and magnitude is given
by $r_{\rm med} = \mathrm{exp}(-1.31-0.27 (m_r-23))$. 
This distribution is the same as given by \citet{miller13a} but with the $r_{\rm med}$ relation shifted to be appropriate for observations in the KiDS $r$ filter \citep[see][]{kuijken15a}. The distribution corresponds also to the \lf{} prior used in the analysis of the KiDS observations. For the bulge-plus-disk galaxies simulated here, the halflight radius of the
bulge component is set equal to the exponential scale length of the disk component \citep[see][for details]{miller13a}.
Galaxies are simulated using 
\textsc{GalSim}, which defines the size as $r_{ab} = \sqrt{ab}$, where $a$ and $b$ are the semi-major and semi-minor axis of the object, respectively, so the sizes sampled from equation\,\ref{eq:sizeprior} were converted to $r_{ab}$ prior to simulation.

\begin{table*}
 \caption{Overview and specifications of all simulated images created with the \imsim \ pipeline}
 \label{tbl:imsims}
 \begin{tabular}{lc}
  Total simulated area & 416 square degrees \\
  \hline
  Tile & 5 exposures of $\sim$1 square degree dithered by 25 arcsec, 85 arcsec \\  
  Exposure & 32 chips of $\sim$ 2000x4000 pixels with 70 pixel wide chip gaps in between\\
  Applied shears   &  (0.0,0.04) (0.0283,0.0283) (0.04,0.0) (0.0283,-0.0283) \\
   & (0.0,-0.04) (-0.0283,-0.0283) (-0.04,0.0) (-0.0283,0.0283) \\
   & The same shear is applied to all galaxies in a tile \\
  Applied PSF & 13 sets; each set contains 5 different PSF models of KiDS-450 observations \\
   & Each PSF model is applied to all galaxies in an exposure  \\
  Shape noise reduction & Each tile is copied with galaxies rotated by 45, 90 and 135 degrees  \\
  \hline
 \end{tabular}
\end{table*}

We also include stars in the simulations, as they might contaminate the galaxy sample and blend with real galaxies \citep[see][for a discussion of the effect of stars on shear measurements]{hoekstra15a}. The
simulated stars are perfect representations of the PSF in the simulated exposure and we do not include realistic CCD features around bright stars, such as bleeding, stellar spikes or ghosts, as these effects are masked in the real data. The stellar $r$-band magnitude distribution is derived using the Besan\c{c}on model\footnote{model.obs-besancon.fr} \citep{robin03a,czekaj14a} for a right ascension $\alpha=175^\circ$ and a declination $\delta=0^\circ$, corresponding to one of the pointings in the KiDS-450 footprint. We note that the star density in that pointing is higher than average. This is not a concern for the bias calibration, as discussed in \S\ref{sec:shearCalib:lf}. We do not include very bright ($m_r<20$) stars, because they would be masked in real observations and we exclude stars fainter than $m_r>25$.

\subsection{Simulation setup}
\label{sec:sim_setup}

As described in detail in \cite{dejong15a} and \cite{kuijken15a}, \lf{} measures galaxy shapes
using the five $r-$band exposures that make up a tile covering roughly one square degree of the sky.
The KiDS-450 data are analysed tile-by-tile, i.e. data from the overlap of tiles is ignored. It is thus sufficient to simulate individual tiles.  Each VST/OmegaCam exposure is seen by a grid of 8 $\times$ 4 CCD chips, where each chip consists of  $2040\times 4080$ pixels that subtend 0\farcs214. There are gaps of around 70 pixels between the chips and to fill the gaps the exposures are dithered. To capture the resulting variation in depth due to this dither pattern we simulate individual tiles of data, using the same dither pattern described in \citet{dejong15a}, which we incorporate by adding artificial astrometry.  We also add a small random shift in pointing between the exposures, so that the same galaxy is mapped on a slightly different location in the pixel grid for each exposure. This extra shift is accounted for when stacking the exposures. Gaussian background noise is added to the simulated exposures, where the 
root mean square of the noise background $\sigma_{\rm bg}=17.03$ was determined as the median value from a sub-sample of 100 KiDS-450 tiles. When exposures are stacked, the noise level varies with position in the simulated tile as in the real data, owing to the chip gaps.

The simulated images for each exposure are created using \textsc{GalSim} \citep{rowe15a} which renders the surface brightness profiles of stars and sheared galaxies using
the input catalogues detailed in \S\ref{sec:input_cat}. The five exposures for each tile are created using the same input catalogue. The 32 individual chips in each of the five exposures are coadded using \textsc{SWARP}\footnote{Note that we do not use the resampling option of \textsc{SWARP} to reduce the processing time. This might introduce some incorrect sub-pixel matching of the pixels in the coadded image, but does not affect the \lf{} measurements,
which are made by jointly fitting to the original individual exposures.} \citep{bertin10a}.  Finally we run \textsc{SExtractor} \citep{bertin96a} to detect objects in the coadded image. We use the same version of the software and configuration file as is used in the analysis of the KiDS-450 data \citep{dejong15a} to ensure homogeneity. Only the magnitude zeropoint is set to the value of 24.79 
which was used to create the simulations. 

Eight shear values are sampled isotropically from a circle of radius $ \lvert g\rvert\ =  0.04 $ and using evenly spaced position angles (see Table\,\ref{tbl:imsims} for the exact values).  We apply the same shear to each simulated galaxy in the five exposures in a simulated tile, using the \textsc{GalSim} \texttt{Shear} function which preserves galaxy area, but vary the shear between tiles.  The sheared galaxies are convolved with an elliptical Moffat PSF, whose parameters are representative of the ones measured in KiDS-DR1/2
\citep{dejong15a}. To obtain the PSF parameters, we ran \textsc{PSFEx} \citep{bertin13a} on KiDS-DR1/2 data. As the VST seeing conditions change over time, so that different exposures have different PSFs, we mimic this temporal variation of the PSF in the \imsim \ simulations. To this end we selected a series of PSF parameters corresponding to 5 subsequently observed dithered exposures of KiDS data. This gave us a set of Moffat parameters for the PSF in each of the 5 exposures of a tile. All galaxies in a simulated exposure were convolved with the same Moffat profile. All galaxies in the first simulated exposure thus have the PSF in the first exposure of the observed KiDS tile.  The second simulated exposure has galaxies convolved with the observed PSF in the second exposure of the KiDS tile. And so on for all five exposures of the simulated tile. This ensures that the PSFs in the simulations are the same as in the KiDS observations. We used the PSF parameters from 13 KiDS tiles, so that we have in total 65 different PSFs in the simulations. This number of PSFs gave us enough statistical power to reach the required precision. The 13 tiles were chosen so that the distributions of PSF parameters in the simulations would match the distribution of the full KiDS data. The distributions of simulated PSF properties measured by \lf{} on the \imsim \ images are shown in the red histograms in Fig.\,\ref{fig:sim_psfdistr}. We define the PSF size in terms of the weighted quadrupole moments $P_{ij}$ of the surface brightness of the PSF:

\begin{equation}
r^2_{\mathrm{PSF}}:=\sqrt{P_{20} P_{02} - P_{11}^2},
\end{equation}
where we measure the moments employing a Gaussian weighting function with a size of 2.5 pixels.
The bottom panels show the two components of the weighted $\epsilon$ ellipticity. Comparison with the distributions measured in the  KiDS-450 data (shown in black) shows that the simulations sample the range in PSF properties. The median full width to half maximum (FWHM) of 0\farcs64 in our sample is very similar to the value of  0\farcs65 from the full KiDS sample. However, the lack of spatial variation in the simulations produces very spiky distributions. This also leads to an over-representation of large and elliptical PSFs in the simulations.

In total we have simulated 416 deg$^2$ of KiDS observations, slightly more than the unmasked area of the KiDS-450 dataset. However, the use of shape noise reduction ensures that we have ample statistical power in the calibration, because the simulated data are equivalent to an area of $\sim 3750$ deg$^2$
without the shape noise cancellation. A summary of the set of simulations created with the \imsim \ pipeline
is provided in Table\,\ref{tbl:imsims}.

\begin{figure*}
 \centering
  \includegraphics[width=1.0\textwidth]{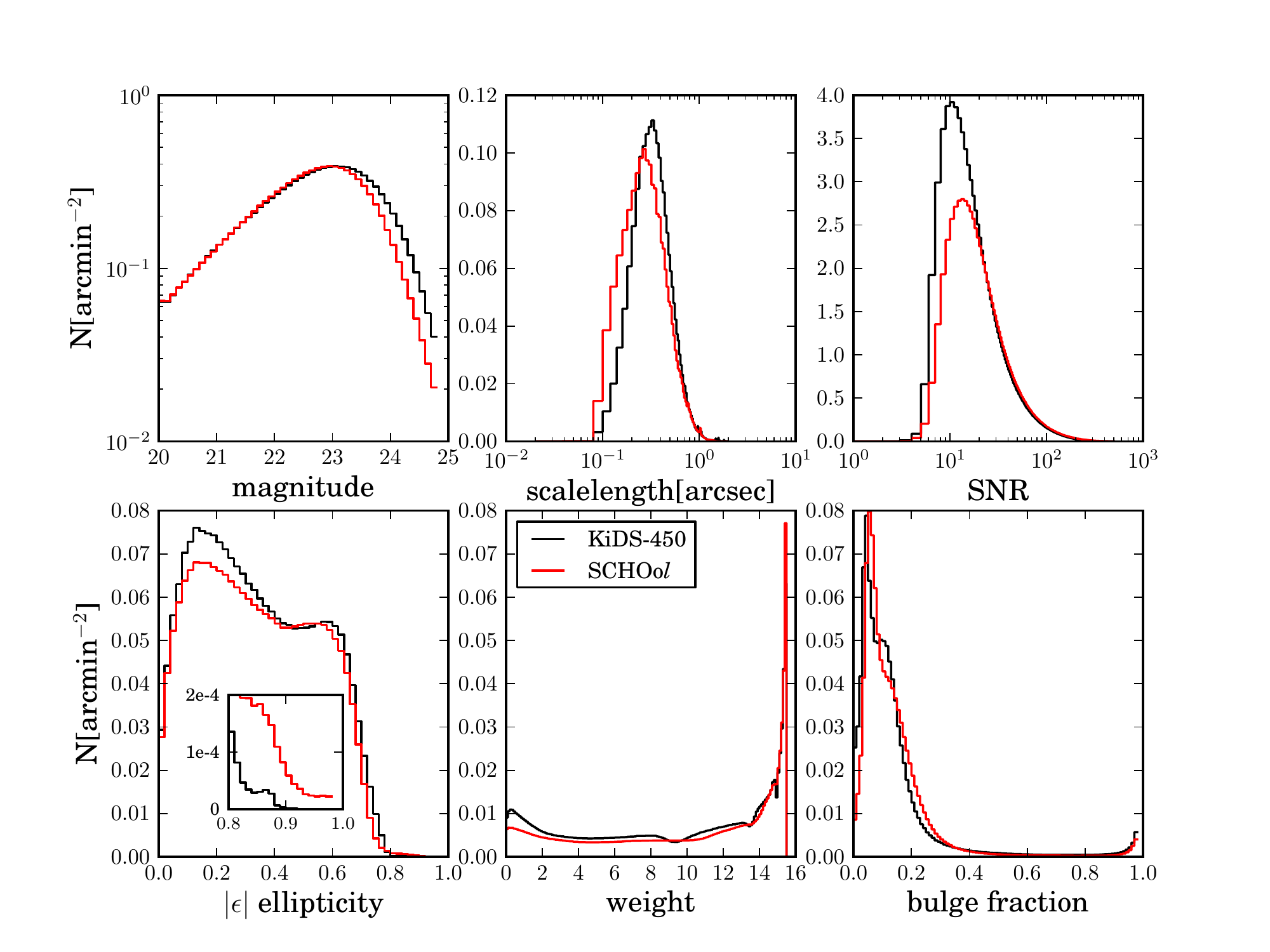}
   \caption{Comparison of KiDS-450 data (black) and \imsim \ simulations (red) for weighted normalised distributions of galaxy properties. From left to right, top to bottom: magnitude, size, SNR, modulus of the ellipticity $| \epsilon |$, \lf{} weights, bulge fraction. The inset shows a zoom in of the ellipticity distributions for $\epsilon>0.8$.} 
 \label{fig:distr_sim_data}
\end{figure*}

\subsection{Comparison to data}
\label{sec:compsimdata}

Although our input catalogue is based on realistic prior distributions, it is important to verify whether the simulated data are a good representation of the observations. Differences with the actual KiDS-450 measurements may occur because of simplifying assumptions or errors in the prior distributions. For instance, in the simulations the PSF is constant over one square degree and the noise level does not vary. Therefore, the resulting \lf{} measurements are not identical to those in KiDS-450 data and the average shear biases inferred from the simulations may differ from the actual shear biases in the data. Rather than adjusting the input catalogue such that the agreement with the data is improved \citep{bruderer15a}, we instead aim to model the biases as a function of observed properties  (see \S\ref{sec:shearCalib}). This approach does not require perfect simulations, but does require that the simulations capture the variation in galaxy properties seen in the data. To examine whether this is indeed the case, we compare the measured galaxy properties in the simulations to those in the KiDS-450 data.

We run \lf{} on the entire volume of the simulations, using the \textsc{SExtractor} detection catalogue as input. For each detected object \lf{} returns a measurements of the ellipticities, weights as well as measurements of the galaxy properties such as SNR and size. A measurement of the observed magnitude is provided by \textsc{SExtractor}.  In order for the comparison with the data to be meaningful the same cuts have to be applied to both datasets. In both cases we consider only measurements of galaxy shapes for objects fainter than $m_r=20$. Moreover, to study selection biases (see \S\ref{sec:selectionbias}) we create a catalogue that contains for each detected object its input properties and those measured by \textsc{SExtractor} and \lf{}. This is done using a kD-tree based matching routine which combines each \lf{} output catalogue with the input catalogue used to create the galaxy images. 

For each object in a given \lf{} catalogue we find its five nearest neighbours in the input catalogue, according to the L2-norm spatial separation. We discard all candidates with a separation larger than three pixels and select from the remainder  the one with the smallest difference in measured magnitude and input magnitude as the final match. This last step introduces a sensible metric to discard by-chance close-neighbour pairs of physically different objects. 
This matching process removes spurious detections from the catalogue. This is not a problem for the bias characterisation, as \lf{} would have assigned a vanishing weight to such spurious detections.

After the matching we apply a series of cuts to the data, starting with the removal of all objects with a vanishing \lf{} weight to reduce the size of the analysis catalogues. This does not have any effect on the recovered shear since this is calculated as a weighted average of the measured ellipticities. This initial selection automatically removes the following:

\begin{enumerate}
\item Objects identified as point sources ({\tt fitclass}  = 1)
\item Objects that are unmeasurable, usually because they are too faint ({\tt fitclass}  = -3)
\item Objects whose marginalised centroid from the model fit is further from the \textsc{SExtractor} input centroid that the positional error tolerance set to 4 pixels ({\tt fitclass} = -7). 
\item Objects where insufficient data is found, for example an object at the edge of an image or defect ({\tt fitclass}  = -1)   
\end{enumerate}

Additionally, in order to match the cuts applied to the KiDS-450 data (see Appendix D in \cite{KiDS450}), we also remove: 
\begin{enumerate}
\setcounter{enumi}{4}
\item Objects with a reduced $\chi^2>1.4$ for their respective \lf{} model, meaning that they are poorly fit by a bulge plus disk galaxy model  ({\tt fitclass} = -4).    
\item Objects whose \lf{} segmentation maps contain more than one catalogue object ({\tt fitclass} = -10).\footnote{
In order to remove contamination from nearby objects, \lf{} builds a dilated segmentation map that is used to mask out a target galaxy's neighbours. It was found that a small fraction of targets had two input catalogue target galaxies within a single segmented region associated with the target, owing to differing deblending criteria being applied in the 
\textsc{SExtractor} catalogue generation stage from the \lf{} image analysis. When measured, this leads to two catalogue objects being measured using the same set of pixels, and thus the inclusion of two correlated, high ellipticity values in the output. As these accounted for a very small fraction of the catalogue, these instances were flagged in the output and excluded from subsequent analysis.}
\item Objects that are flagged as potentially blended, defined to have a neighbouring object with significant light extending within a contamination radius $>4.25$ pixels of the  \textsc{SExtractor} centroid.  
\item Objects that have a measured size smaller than 0.5 pixels.
\end{enumerate}

After these cuts, considering all image rotations, shear and PSF realisations, we obtain a sample of $\sim 16$ million galaxies which are used in the analysis. Fig.\,\ref{fig:distr_sim_data} shows the resulting weighted distributions of magnitude, scale length, modulus of ellipticity, bulge fraction, 
SNR and weight measured from KiDS-450 data (black) and the \imsim \  simulations (red). 

The distributions of the \lf{} measurement weight and bulge fraction are in good agreement with the data, although the measured bulge fractions are extremely noisy, and are eliminated from the shear measurement by a marginalisation step. However, the agreement in the simulated and observed distributions gives some reassurance that the simple parametric galaxy profiles are an adequate representation of the KiDS-450 data.
The simulated galaxy counts are in good agreement with the observations for bright galaxies, but the magnitude and SNR distributions suggest that the simulations lack faint, low SNR objects. The paucity in the simulated catalogues might be attributed partly to the fixed noise level or the spatially constant PSF in the simulations, which is not fully representative of KiDS-450 observations, but also partly to a difference in intrinsic size distributions of the data and simulations, which may also be seen in Fig.\,\ref{fig:distr_sim_data}.

The shear measurement bias that we seek to calibrate depends primarily on galaxy size and SNR \citep[e.g.][]{miller13a}, and differences in the distributions of these quantities between the data and the simulations mean that we cannot simply measure the total bias from the simulations and apply the result to the data. Furthermore, this consideration applies to the bias for any sub-selection  of the data, such as the analysis of shear in tomographic bins of \citet{KiDS450}. Even if the data and simulations were a perfect match in Fig.\,\ref{fig:distr_sim_data}, any
dependence of bias on galaxy properties would mean that a `global' bias for the simulations might not be
appropriate to the galaxy selection in tomographic bins.
Thus, in this paper we derive a shear calibration that includes a dependence on size and SNR, but also investigate the sensitivity of the final shear calibration to modifications of the assumed distributions, in \S\ref{sec:sensitivity_mag} and \S\ref{sec:sensitivity_size}.

The ellipticity distributions also differ, both at low and high ellipticity. Both the simulations and the KiDS-450 data contain very elliptical galaxies galaxies, as is clear from the inset in the lower left panel of
Fig.\,\ref{fig:distr_sim_data}, which shows the high ellipticity tail of the distribution. In the simulations these high ellipticities are caused by noise or blending with neighbours, as there are no galaxies with an intrinsic ellipticity $\epsilon>0.804$. However, in the data this is not necessarily the case. Differences in the ellipticity distribution may lead to an incorrect estimate of the shear bias and this is especially worrying for highly elliptical objects \citep{melchior12a,viola14a}. 
In \S\ref{sec:sensitivity} we investigate the (origin of the) discrepancy and also quantify the resulting uncertainty in shear bias that arises from the differences between the data and the simulations.

As noted above, the observed differences suggest that the simulations cannot be used directly to infer the shear biases, and in the remainder of this paper we explore calibration strategies that use observed properties to estimate the bias  for a given selection of galaxies \citep{miller13a,hoekstra15a}. For this to work, it is important that the simulations at least cover the multi-dimensional space of relevant parameters. Moreover, differences in selection effects should be minimal. Before we explore these issues in more detail, we first examine the distributions of the two most relevant parameters, namely the SNR and the ratio of the PSF size and the galaxy size \citep[e.g.][]{massey13a}. The latter parameter, which we define as,
\begin{equation}
 {\cal R} := \frac{r^2_{\textrm{PSF}}}{\left(r_{ab}^{2}+r^2_{\textrm{PSF}}\right)},
 \label{CAL::EQU:resolution}
\end{equation}
quantifies how the shape is affected by the convolution by the PSF. 
For the analysis, we adopt the $r_{ab}$ size definition, because it has
significantly lower correlation with the measured ellipticity in noisy data (cf. \S\ref{sec:calibselectionbias}).

\begin{figure}
\includegraphics[width=9cm, angle=0]{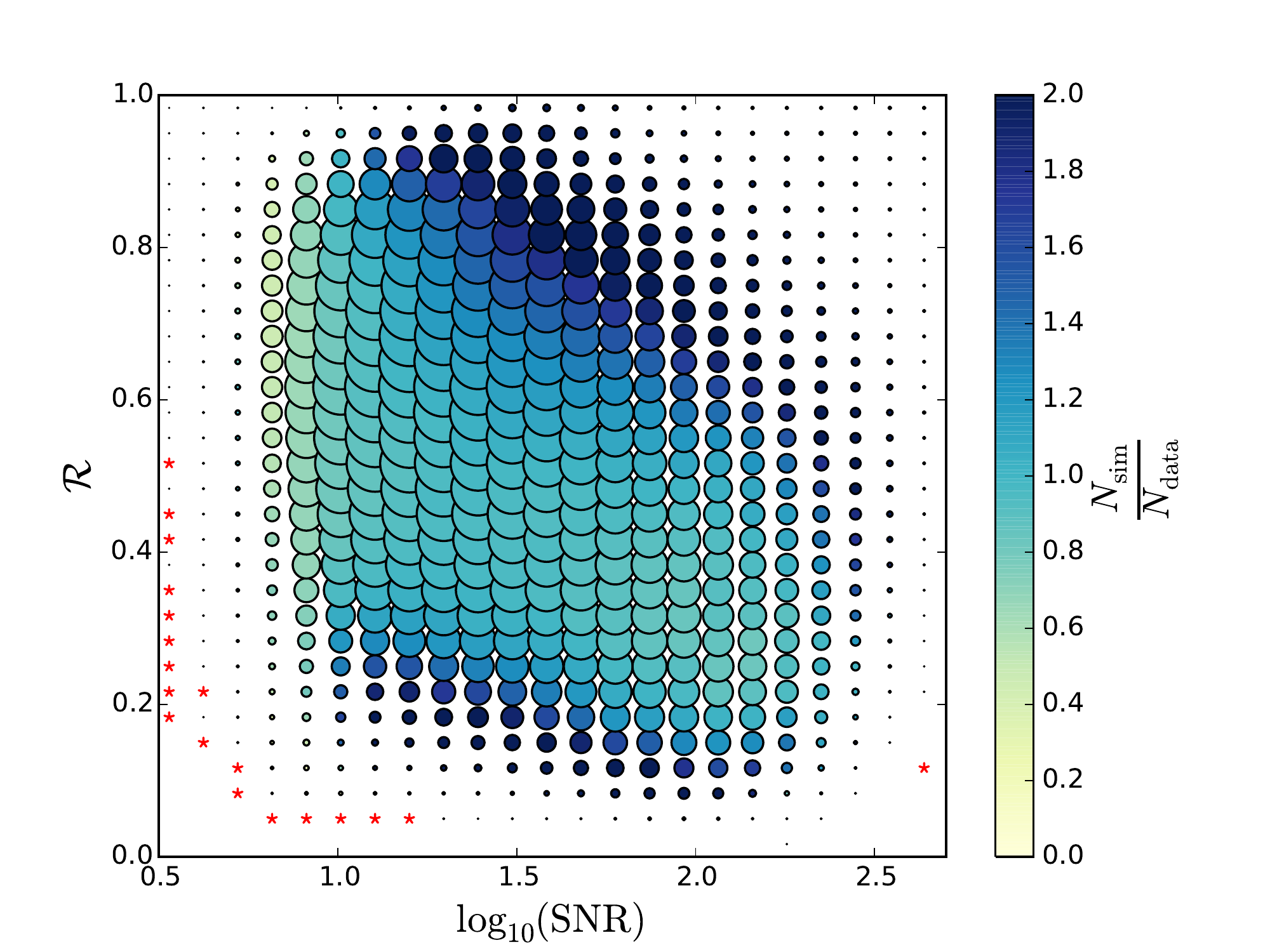}
\caption{Ratio between the number of galaxies in the simulation and the data on a SNR and resolution grid defined using the real galaxies. The size of each data point is proportional to the total \lf{} weight in each grid cell. The red stars indicate the grid points with a ratio of 0.}\label{2Dsurface_res}
\end{figure}

Fig.\,\ref{2Dsurface_res}, shows the ratio between the number of simulated and real galaxies on a 
grid in SNR and ${\cal R}$ defined using the KiDS-450 data. The size of each data point is proportional to the sum of the \lf{} weight in each grid cell. The red stars indicate the region where the ratio is 0; i.e. the simulations do not contain objects with that SNR and resolution. The simulations are lacking very large objects (low ${\cal R}$) and with low SNR. Those objects contribute only 0.001 \% of the total weight and hence the fact that they are not present in the simulations can be safely ignored.

\section{KiDS Calibration Method}\label{sec:shearCalib}

\subsection{The evaluation of shear bias}

As our image simulations are a good, but not perfect representation of the KiDS-450 data, and as in our data
analyses \citep[e.g.][]{KiDS450} we select sub-samples of galaxies with differing distributions of intrinsic properties, it would be incorrect to simply compute the average multiplicative and additive bias from the simulations and use the result as a scalar calibration of the KiDS-450 shear measurements.  
This is because previous analyses 
\citep[e.g.][]{miller13a,hoekstra15a}, and analytical arguments \citep[e.g.][]{massey13a} have demonstrated
that  the shear bias depends on galaxy and PSF properties. In particular, we expect the bias to be a function of the galaxy SNR and size, and to depend on the PSF size and ellipticity.  Estimating those functional dependencies is crucial in order to derive a shear calibration that may be robustly applied to the data. 

A practical procedure for estimating the bias and its dependences from the simulations is to bin
the simulated data, and compute the  multiplicative and additive shear bias in each bin. To do so, we use the \lf{} measurements of the galaxy ellipticities $\epsilon_j$ in combination with the re-calibrated weights $w_j$ (see \S\ref{sec:weight_bias}) to compute the two components of the measured shear $g_j$:
\begin{equation}
g^{\textrm{meas}}_j=\frac{\sum_i w_i \epsilon_{ij}}{\sum_i w_i} \, .
\label{eq:elshear}
\end{equation}
Following \citet{STEP1} we quantify the shear bias in terms of a multiplicative term $m$ and an additive term $c$:
\begin{equation}
 g^{\textrm{meas}}_{j}=\left(1+m_{j}\right)g^{\textrm{true}}_{j}+c_{j} \, ,
 \label{EQU::CAL::BIAS}
\end{equation}
where we consider the biases for each of the ellipticity components separately. In our analysis below, 
we designate $m,c$ values for components evaluated in the original `sky' co-ordinate frame by $m_{1,2},c_{1,2}$. 
When investigating PSF-dependent anisotropy, we also investigate biases 
on components where the ellipticity and shear values have been
first rotated to a co-ordinate frame that is aligned with the orientation of the major axis of each galaxy's PSF
\citep[c.f.][]{mandelbaum15a}. We designate the latter linear bias components as 
$m_{\vert\vert},c_{\vert\vert},m_{\times},c_{\times}$ for the components parallel to and at $45^{\circ}$ to the 
PSF orientation, respectively.

Several calibration binning schemes may be considered, such as fixed linear or logarithmic bin sizes, or a scheme that equalises the number of objects in each bin. In the following, we choose a binning scheme that equalises the total \lf{} weight in each bin and assign the median as the centre of each bin for each respective data sample. The multiplicative and additive biases for both shear components are then obtained by a linear regression with intersection of all measured average ellipticity values $\left<\epsilon\right>_{j}$ against the true input reduced shear values $g^{\textrm{true}}_{j}$. 

We use two different methods to assign errors to the respective biases in $m$ and $c$ in each bin. In the first method, the uncertainties are estimated from the scatter of the measurements around the best fit line.
The other method is to bootstrap resample the sets of galaxies that share the same input shear values. The number of bootstrap realisations is chosen to be large enough for the resulting errors to stabilise. We find this to be the case after the creation of 20 bootstrap realisations.  

\begin{figure*}
  \centering
  \begin{tabular}{cc}
  \includegraphics[width=0.48\textwidth]{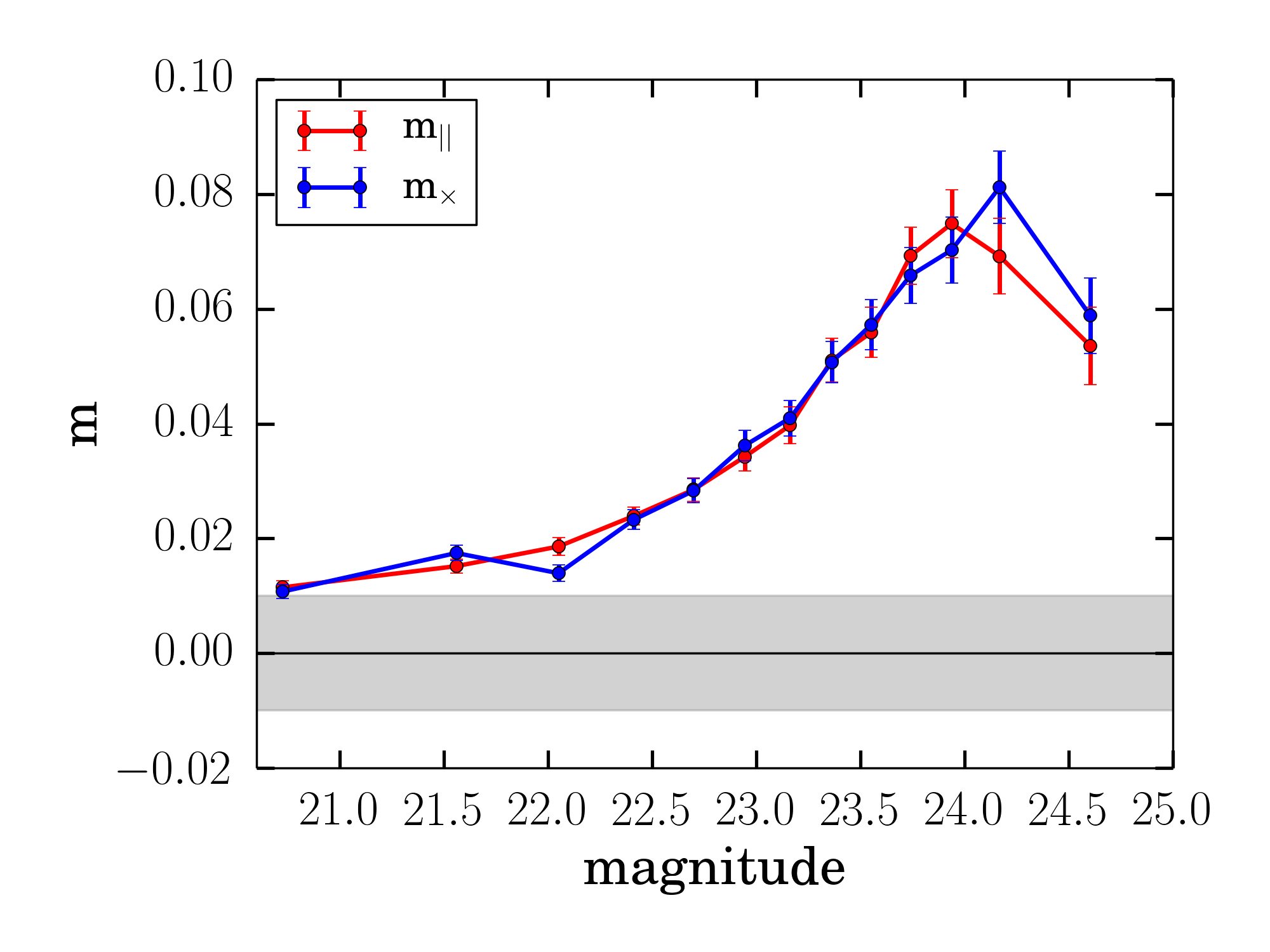} &
  \includegraphics[width=0.48\textwidth]{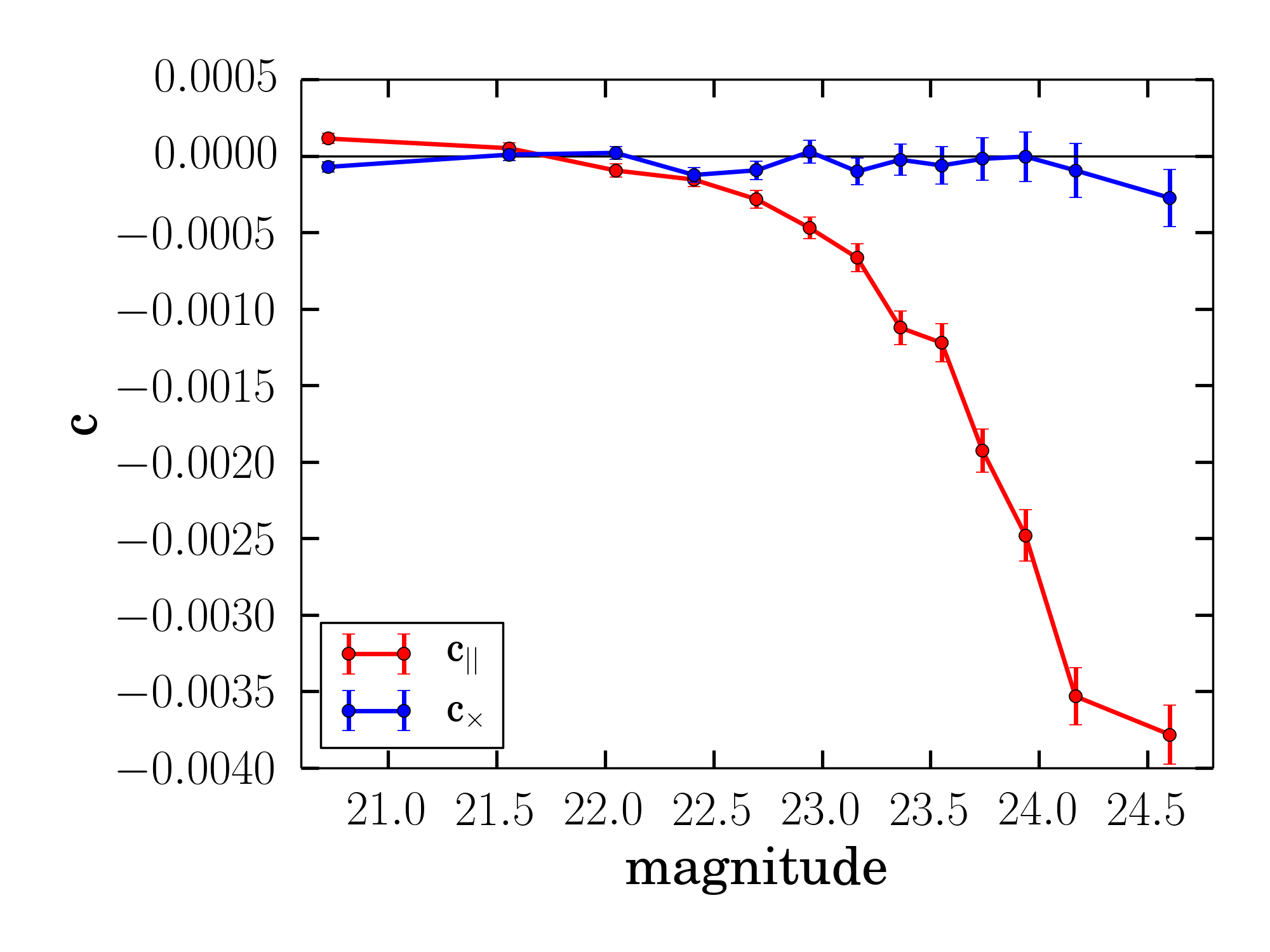} \\
  \end{tabular}
  \caption{
Multiplicative (left panel) and additive (right panel) selection bias, $m$ and $c$, for the components aligned ($m_{\vert\vert},c_{\vert\vert}$) or 
cross-aligned ($m_{\times},c_{\times}$) with the PSF major axis orientation, as a function of galaxy magnitude, as discussed in \S\ref{sec:selectionbias}. The grey band in the left panel indicates the requirement on the knowledge of the multiplicative bias set by \citet{KiDS450} in the context of a cosmic shear analysis.}
\label{fig:selectionbias}
\end{figure*}

\subsection{Selection bias}
\label{sec:selectionbias}

Bias in the measurement of the shear arises from the combined processes of galaxy detection or selection (`selection bias') and the shear measurement itself (`model bias' and `noise bias'). In this section, we inspect the individual selection bias contributions. Selection biases may occur if the intrinsic ellipticity distribution of galaxies is anisotropic \citep{kaiser00b, bernstein02a, hirata03a}, which may happen if galaxies are preferentially detected when they are aligned with the shear or the PSF, or if an anisotropic weighting function is employed in the measurement. Multiplicative shear bias
may also arise if the distribution of ellipticities that are selected is systematically biased with respect to the 
underlying distribution.
Such anisotropic or multiplicative selection effects may arise at two stages of the process. First, galaxies and stars are detected on stacked images using \textsc{SExtractor}. In principle, the dependence of the SNR on galaxy size, ellipticity, orientation and PSF properties may result in biases at this detection stage. Second, the \lf{} shear measurement process may not be able to measure useful ellipticity values for some galaxies, leading to an additional contribution to selection bias.

We investigate these biases by inserting the `true' sheared ellipticity value of each simulated galaxy
into our shear measurement framework, characterising a linear relation between shear estimates formed from these quantities and the true shear.  In this approach, there is no contribution to the bias estimate, or to
its measurement uncertainty, from noise bias. The only potential source of bias is sampling noise, but
in our simulations ellipticity shape noise has largely been `cancelled' (see \S\ref{sec:imsim:lfcat}), apart from the
effect of galaxies that are not detected.
In this test, we find a small bias, $m_{\vert\vert} \simeq m_{\textrm{x}} \simeq -0.005 \pm 0.001$, $c_{\vert\vert} \simeq 0.0002 \pm 0.00004$, $c_{\textrm{x}} \simeq 0.00005 \pm 0.00004$, as a result of the \textsc{SExtractor} stage.
However, if we measure the shear bias after the \lf{} stage by selecting those galaxies that are both detected by
\textsc{SExtractor} and with
shear measurement weight greater than zero, we do find a significant multiplicative bias, of $4.4$\,percent when averaged across the sample, with little difference between biases whether the true shear values are unweighted or weighted by the \lf{} weight, for those galaxies with non-zero weight. As shown in Fig.\,\ref{fig:selectionbias} the bias is strongly magnitude-dependent, with a maximum bias around 8\,percent.  By rotating galaxy ellipticity and shear values to the coordinate frame aligned with the PSF major axis (the PSF orientation varies in our simulations), we may also look for additive selection bias that  is correlated with the PSF: Fig.\,\ref{fig:selectionbias} also demonstrates the existence of such an additive selection bias, with a significant aligned $c$ term (there is no significant bias detected in the cross-aligned $c$ term).

The bias is caused by the inability to measure small galaxies: if an object has a \lf{} star-galaxy discrimination classification that favours the object being a star over a galaxy \citep[see][]{miller13a}, it is classified as a star and given zero weight in the subsequent analysis. This step introduces a significant selection bias, because galaxies are more easily measured and distinguished from stars if they are more elliptical: thus galaxies whose intrinsic ellipticity is aligned with its shear value are more likely to be selected as measurable galaxies, than those whose intrinsic ellipticity and shear values are cross-aligned.  This results in a significant bias in the average intrinsic ellipticity of the measured galaxies, and thus a significant shear bias.

This measurement selection bias should arise in both the data and the simulations, and thus our calibration derived from
the simulations should remove the effect from the data.  We note however that the selection bias is not small relative to
our target accuracy (grey band in Fig.\,\ref{fig:selectionbias}), and is comparable to the noise bias that has received more attention in the literature. 
We expect the selection bias to have some sensitivity to the distributions of size and ellipticity and thus not to be
precisely reproduced in our fiducial simulations: as previously
mentioned, in \S\ref{sec:resampling} we resample the simulations to match the observed distributions in the KiDS
tomographic bins, and in \S\ref{sec:sensitivity_size} 
we further test the effect of modifying the size distribution. 
We also consider the possible contribution of object selection bias to the PSF
leakage in \S\ref{sec:shearCalib:cbias}.

\begin{figure*}
  \centering
  \begin{tabular}{cc}
  \includegraphics[width=0.48\textwidth]{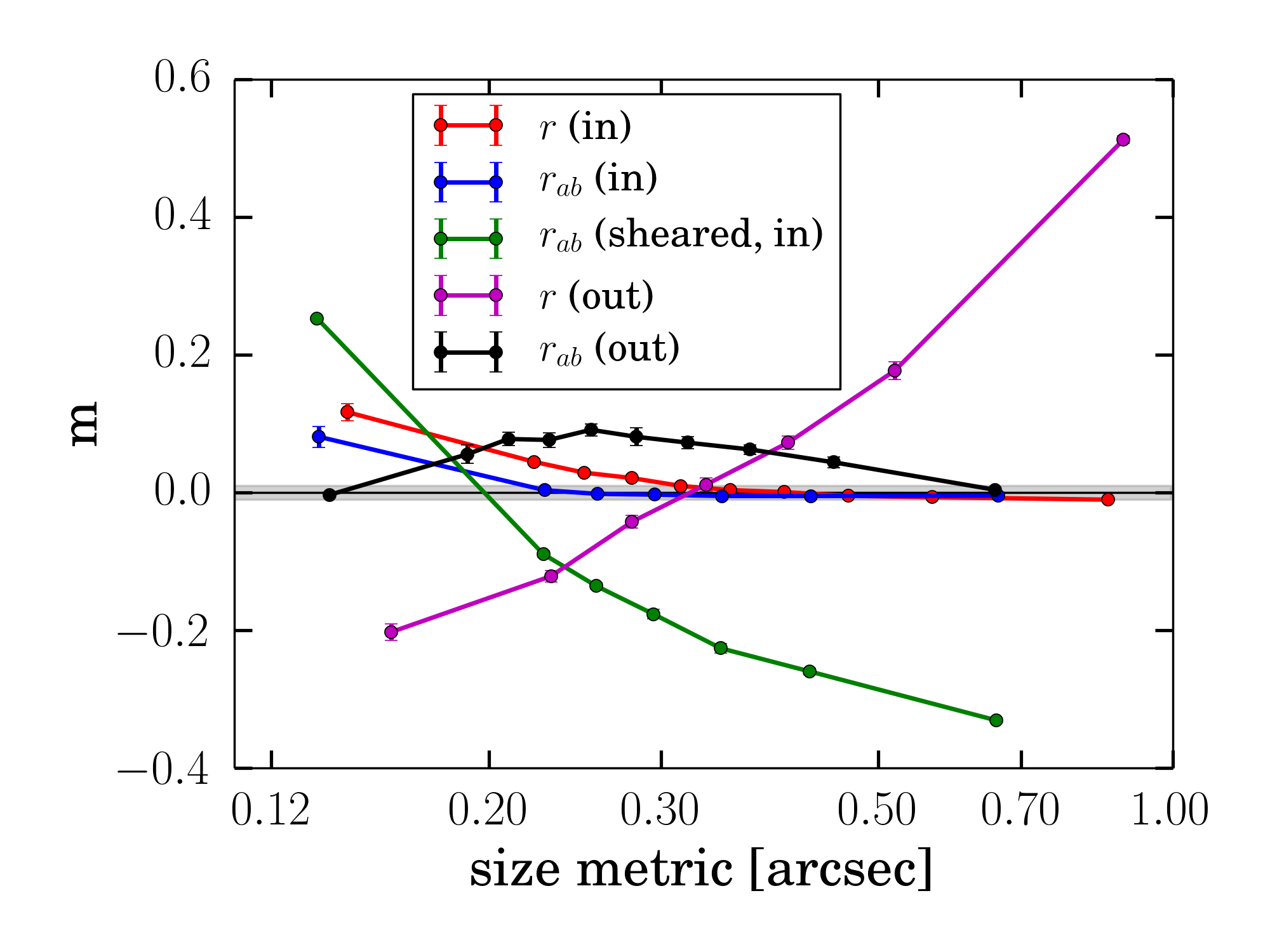} & 
  \includegraphics[width=0.48\textwidth]{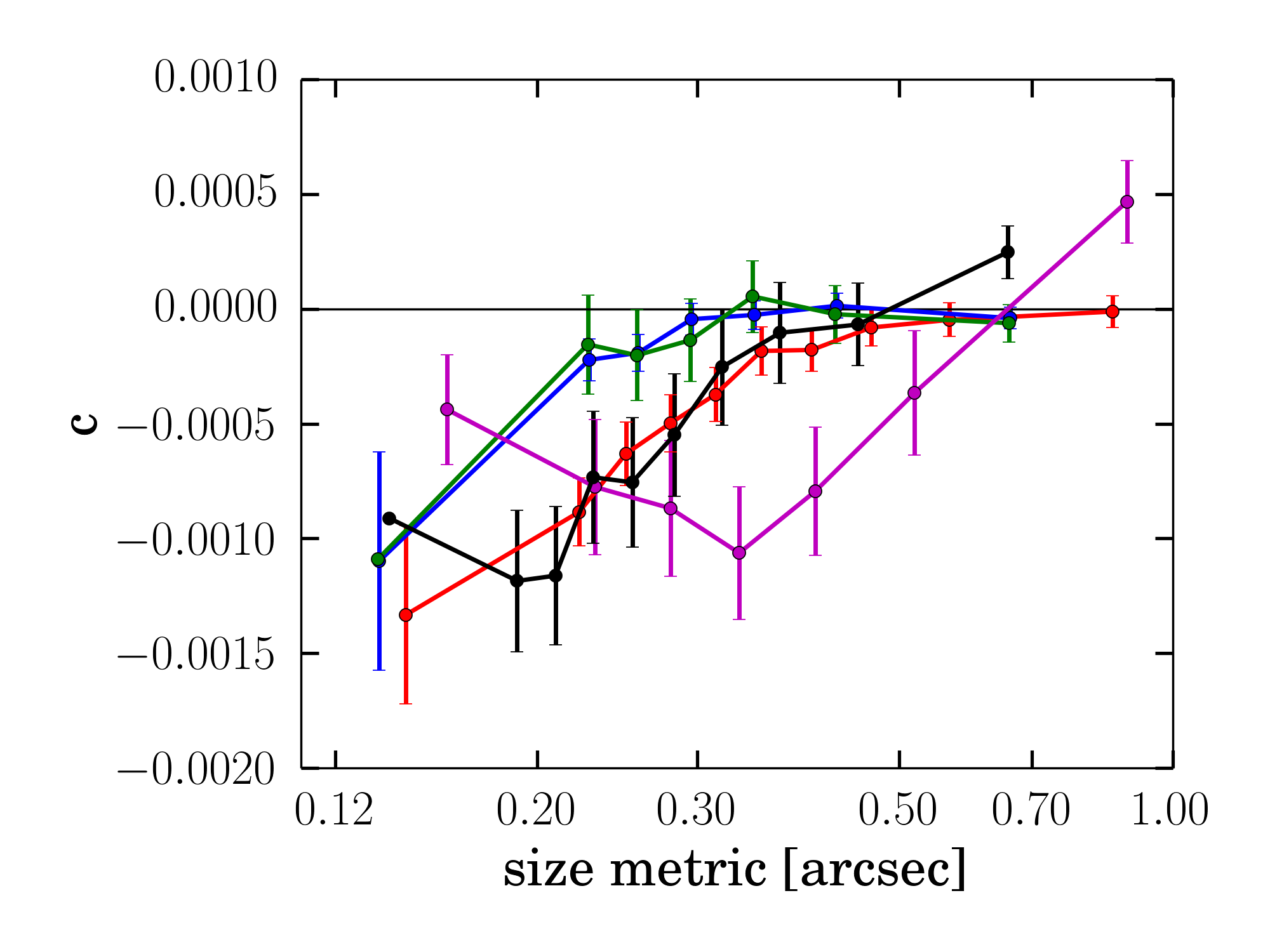} \\
  \end{tabular}
  \caption{
The apparent multiplicative (left panel) and additive (right panel) calibration selection bias, $m$ and $c$, deduced from the analysis of true, noise-free, sheared galaxy ellipticity values,
as a function of galaxy size. 
Relations are shown for five definitions of galaxy size: 
(red) size $r$ measured from true input major axis values;
(magenta) size $r$ measured from noisy output major axis values;
(blue) $r_{ab}$ size, measured from true input, unsheared major and minor axis values;
(green) $r_{ab}$ size, measured from true input, sheared major and minor values;
(black) $r_{ab}$ size, measured from noisy output major and minor values.  
The additive bias $c$ is shown for the component aligned with the PSF major axis.
See \S\ref{sec:calibselectionbias}.
}
\label{fig:sizeplot}
\end{figure*}

\subsection{Calibration selection bias}
\label{sec:calibselectionbias}

In a conventional approach to shear calibration,
the objective is to establish a shear calibration relation, whose parameters are observed quantities, which may be applied to the survey data. Ideally, to ensure that unbiased measurements of the cosmology are obtained, after shear calibration has been applied, we should aim for a lack of residual dependence on true, {\em intrinsic} galaxy properties (such as size or flux) in the simulations, even though the  calibration relation must be derived from observed quantities. The absence of such dependencies would imply that the results are not sensitive to changes in the input distributions.

\begin{figure*}
 \begin{tabular}{cc}
  \includegraphics[width=.48\textwidth]{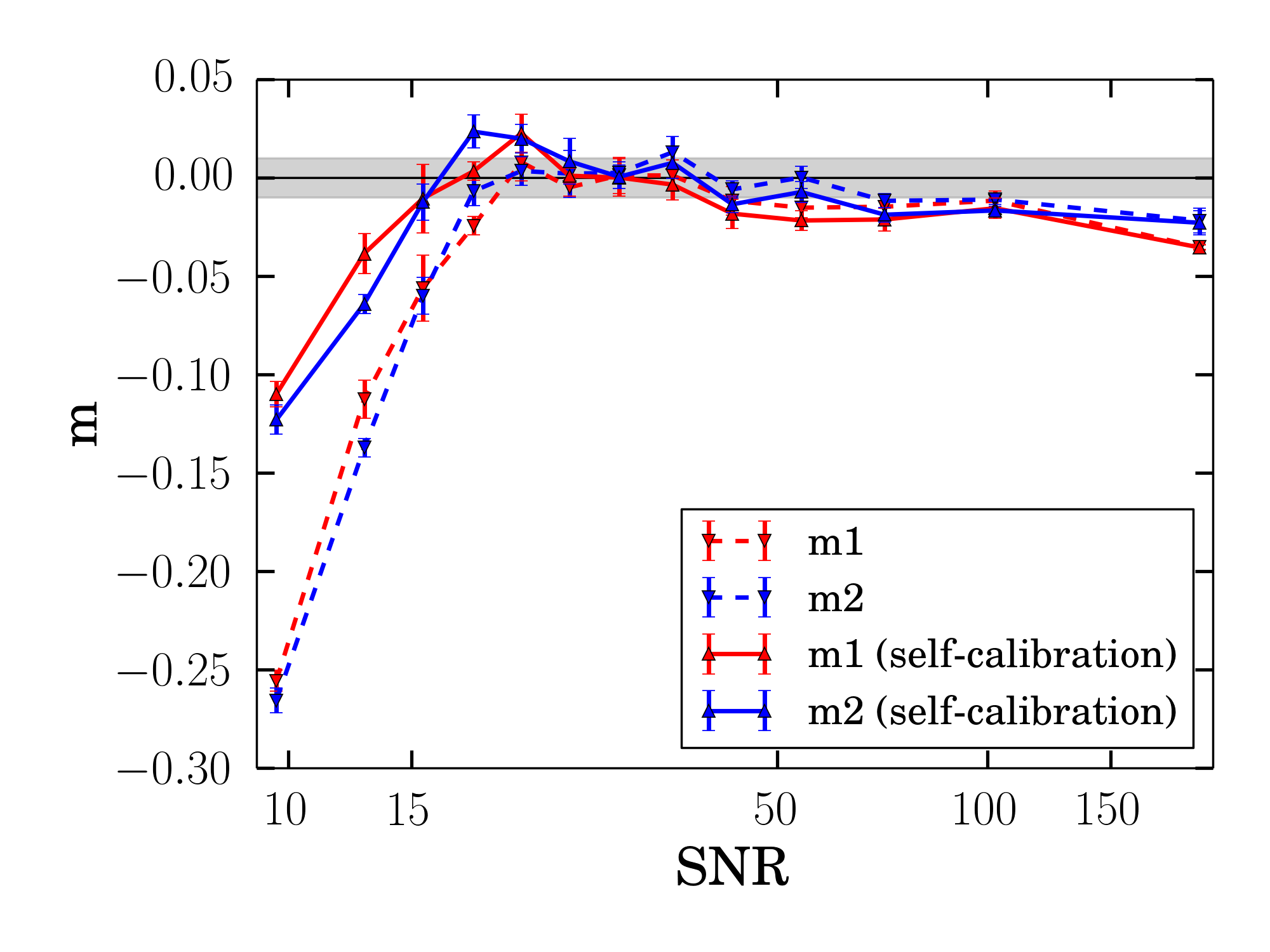} & 
  \includegraphics[width=.48\textwidth]{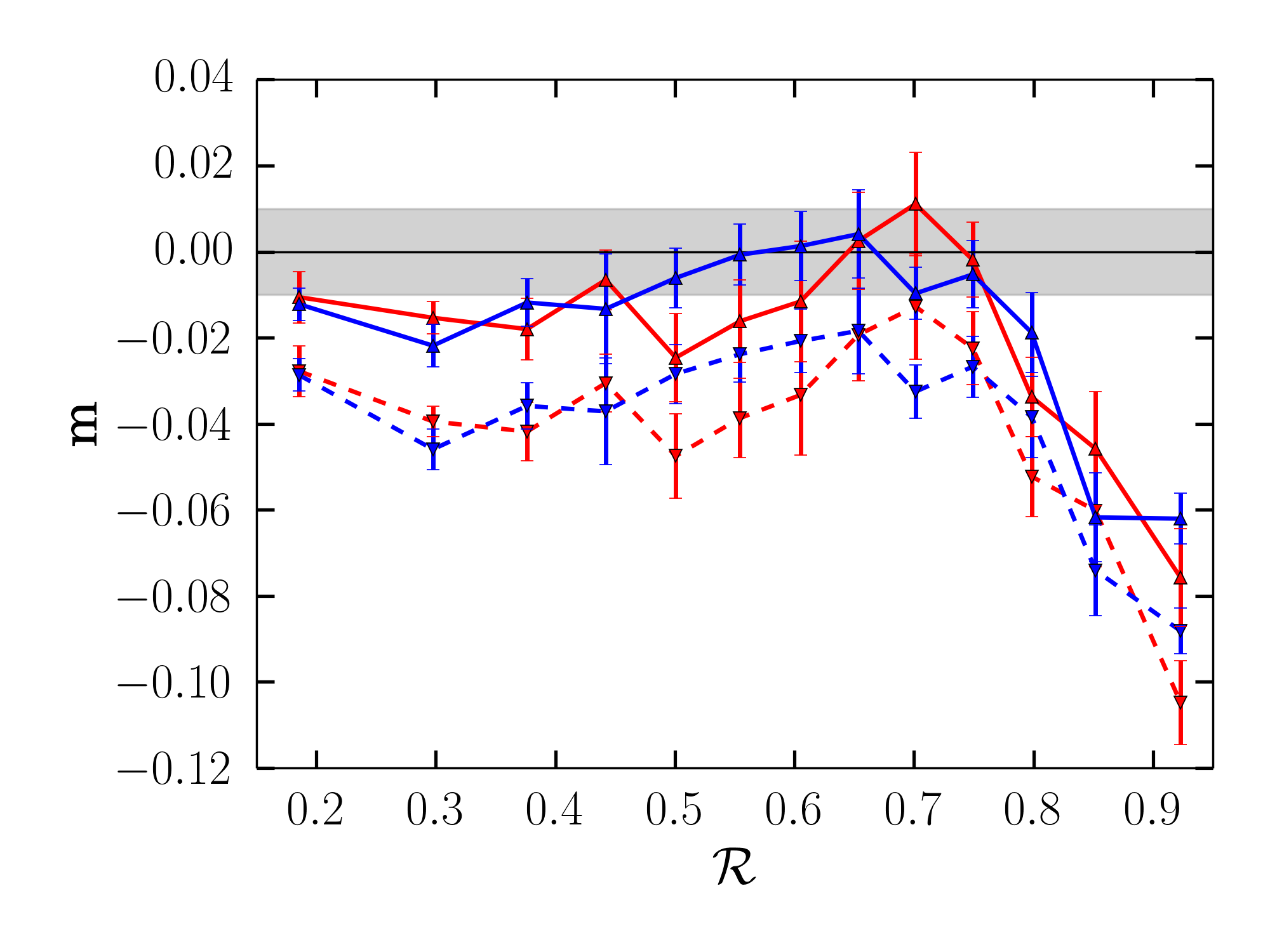} \\
\includegraphics[width=.48\textwidth]{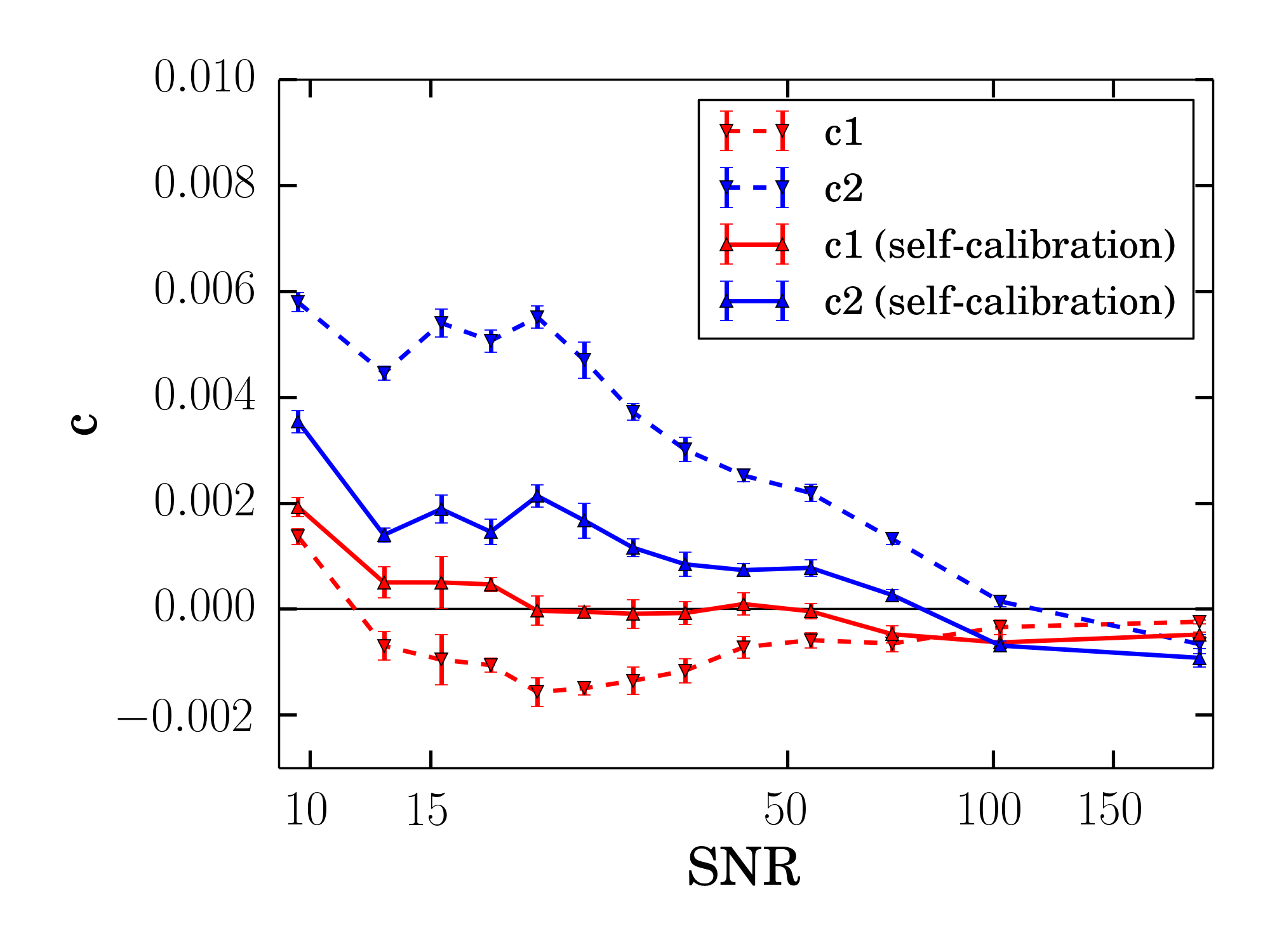} & 
\includegraphics[width=.48\textwidth]{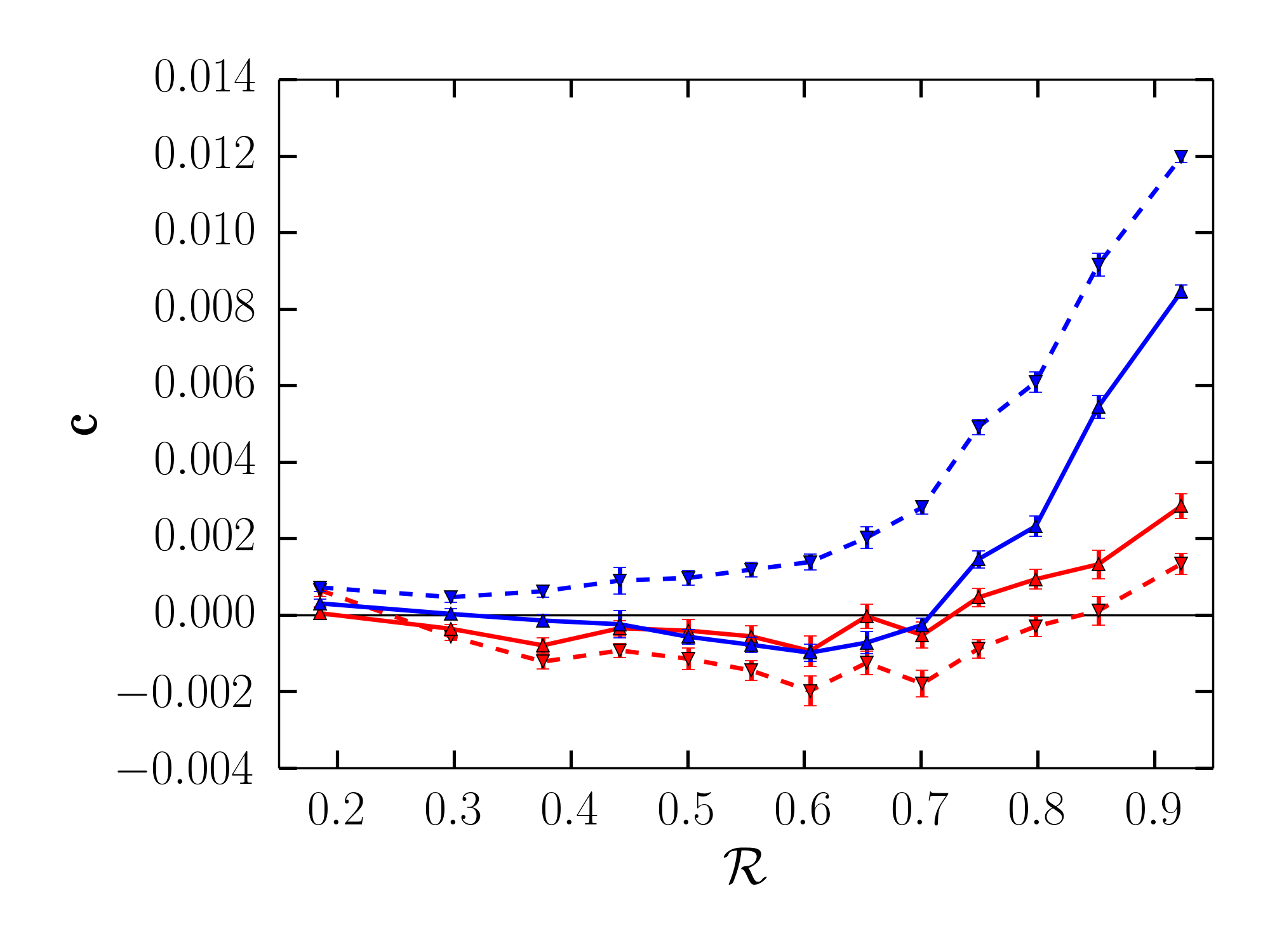}\\
 \end{tabular}
\caption{The multiplicative shear bias $m$ (top) and additive shear bias $c$ (bottom) as a function of measured galaxy properties. The left panels shows the bias with and without \lf{} self-calibration as a function of measured model SNR. The right panels show the same measurements as a function of $\cal{R}$. The grey band in the top panels indicates the requirement on the knowledge of the multiplicative bias set by \citet{KiDS450} in the context of a cosmic shear analysis. \label{CAL::LENSFIT::mult::obs}} 
\end{figure*}

However, if we attempt to deduce a shear calibration that depends on observed quantities, the
correlations between observed quantities may cause calibration relations themselves to be biased, and
may even mislead the investigator into believing that their shear measurement is biased when it is not.
In this section, we discuss biases in calibration relations that arise artificially as a result of correlations between size and ellipticity, and thus shear, when following a calibration approach such as that adopted for
CFHTLenS \citep{miller13a} or Dark Energy Survey \citep{jarvis15a}.  
We distinguish this `calibration selection bias' from the `galaxy selection bias' discussed above, in \S\ref{sec:selectionbias}.

First, we consider the choice of size parameter.
The definition of galaxy size measured by \lf{} is the scale length, $r$, along the galaxy's major axis: for disk galaxies, where the ellipticity arises from the inclination of the disk to the line-of-sight, this
choice of size measure is the most invariant with the galaxy's ellipticity. However, at low SNR, pixel noise leads to a strong statistical correlation of the major axis size with the ellipticity. 
The distribution of observed ellipticity directly affects the inferred shear in a population, and 
thus a calibration relation that depends on major axis size causes large, apparent size-dependent
biases that in fact arise from the choice of observable. 

This difficulty may be mitigated by adopting instead $r_{ab}$, the geometric mean of the major and minor axis scale lengths.  In noisy data $r_{ab}$ has  significantly lower correlation with the measured ellipticity, but a bias on calibration relations still exists. This selection bias is illustrated in Fig.\,\ref{fig:sizeplot}.  Here, we follow \S\ref{sec:selectionbias} and again calculate the apparent shear bias that is deduced from using the true, noise-free sheared galaxy ellipticity values. It is important to realise that the biases 
seen here do not arise from any process in the noisy measurement of shear, other than through the correlation between the size parameter and shear.  The blue and red lines show the bias on the input (true) galaxy size, for the $r_{ab}$ and major axis $r$ size definitions respectively: 
it is this bias that we wish to minimise in order to achieve cosmological results that are unbiased.  It may be seen that the $r_{ab}$ measure yields a somewhat lower apparent bias, compared with $r$, which is a reflection of how the small, unmeasurable galaxies enter each plotted bin.  As a comparison, the green curve shows the 
results for the $r_{ab}$ input size definition, but where now the sheared major and minor axis values have been used to calculate $r_{ab}$: a very large bias results.

However, any calibration relation that we adopt must instead be a function of the noisy, measured galaxy size, rather than the true size, which is unknown in real data.  In Fig.\,\ref{fig:sizeplot} (magenta line), 
we also show that the correlation with the noisy, measured $r$ parameter has a bias that vastly exceeds the input size bias, and which is strongly dependent on the size value.  The $r_{ab}$ size definition (black line in Fig.\,\ref{fig:sizeplot}) is better behaved in this regard, although the bias observed using output size still does not reflect the bias on the input size.  On the other hand, the $r$ size definition should be less correlated with ellipticity in the true, astrophysical joint distribution. Hence, we continue to parameterise the \lf{} models in terms of $r$, and marginalise over $r$ when estimating galaxy ellipticity as described in \S\ref{sec:selfcal}, 
but we adopt $r_{ab}$ as the size parameter in our calibration relation. We then test how well the bias as a function of input parameters is corrected.

An alternative strategy that would mitigate the selection effects shown in Fig.\,\ref{fig:sizeplot} is to
subtract the true, intrinsic ellipticity value from every galaxy, before forming any shear estimates:
this accurately compensates for the calibration selection bias. This was the procedure adopted for
the CFHTLenS shear calibration \citep{miller13a}, but it has the severe disadvantage
that it also removes both the primary selection bias described in \S\ref{sec:selectionbias} and the 
weight bias described in \S\ref{sec:weight_bias}. As these are percent-level effects, we must 
include them in our KiDS calibration, and accordingly do not use this strategy here.  We note in passing that
neglect of these biases in CFHTLenS may have resulted in larger amplitude shear values (and hence a larger
value of the $\sigma_8$ cosmological parameter), by a few percent, than reported by \citet{heymans13a} and
other related cosmology analysis papers.

Finally, we note that \citet{clampitt16a} found significant size-dependent shear bias in their null test of
{\em Dark Energy Survey} galaxy-galaxy lensing: this bias may have been the result of the selection-induced
size bias we have discussed here, and in general, tests of the dependence of shear on measured galaxy size should be
avoided as a null test.

In the following sections, we investigate the full bias introduced by the noisy measurement process: this bias
includes the object selection bias discussed in \S\ref{sec:selectionbias} and we should be mindful of the 
artificial biases of this section when investigating the size dependence and when deriving a calibration
relation: biases as a function of galaxy size measured in noisy simulations may have a significant
contribution from the calibration selection bias. Provided the simulated galaxy distributions match well the
data distributions, any derived calibration relation should correctly include such effects and should result in
correctly calibrated data, but it makes sense to minimise the effect of the choice of size definition by calibrating
using $r_{ab}$ rather than $r$, as this should minimise the sensitivity to any mismatch between data and
simulations.

\begin{table*}
 \caption{The total multiplicative and additive shear bias, both with (`self-cal') or without (`no-cal') the \lf{}\
self-calibration having been applied.
Biases are quoted for components measured either in the
co-ordinate system of the sky simulations (upper Table section), or where shear and ellipticity
components have been rotated to be aligned, $m_{\vert\vert},c_{\vert\vert}$, or cross-aligned, $m_{\times},c_{\times}$, with the PSF orientation
(lower Table section).}
 \label{CAL::TAB::total_bias1}
 \begin{tabular}{lcccccccc}
  \hline
  sky-frame analysis &$m_{1}$& $\Delta m_{1}$(regr)/(BS)&$m_{2}$& $\Delta m_{2}$&$c_{1}$& $\Delta c_{1}$&$c_{2}$& $\Delta c_{2}$\\
  &$[10^{-2}]$&$[10^{-2}]$&$[10^{-2}]$&$[10^{-2}]$&$[10^{-3}]$&$[10^{-3}]$&$[10^{-3}]$&$[10^{-3}]$\\
\\
  no-cal   & -4.09 & 0.33/0.25 & -3.84 & 0.21/0.22 &-0.73 &0.09/0.07 & 3.32 &0.06/0.05 \\
  self-cal & -1.90 & 0.33/0.25 & -1.68 & 0.19/0.22 & 0.12 &0.05/0.05 & 1.10 &0.05/0.05 \\
  self-cal, no stars & -1.40 & 0.30/0.29 & -1.22 & 0.18/0.19 &0.15 &0.09/0.08 &1.26 &0.05/0.05 \\
  self-cal, low density, no stars & -1.39 & 0.19/0.21 & -0.93 & 0.18/0.26 &0.09 &0.05/0.06 &0.80 &0.05/0.06 \\
\hline
  PSF-frame analysis &$m_{\vert\vert}$& $\Delta m_{\vert\vert}$(regr)/(BS)&$m_{\times}$& $\Delta m_{\times}$&$c_{\vert\vert}$& $\Delta c_{\vert\vert}$&$c_{\times}$& $\Delta c_{\times}$\\
  &$[10^{-2}]$&$[10^{-2}]$&$[10^{-2}]$&$[10^{-2}]$&$[10^{-3}]$&$[10^{-3}]$&$[10^{-3}]$&$[10^{-3}]$\\
\\
    no-cal & -3.96 & 0.22/0.43 & -3.97 & 0.20/0.42 &-2.51 &0.06/0.10 &-0.84 &0.06/0.09 \\
  self-cal & -1.78 & 0.18/0.21 & -1.79 & 0.18/0.27 &-0.55 &0.05/0.07 &-0.15 &0.05/0.09 \\
  \hline
 \end{tabular}
\end{table*}

\subsection{\lf{} results}
\label{sec:shearCalib:lf}

We start the analysis of the noisy measurement biases 
by quantifying the impact of the \lf{} self-calibration (see \S\ref{sec:selfcal_detail}) on the recovered shear biases. This is done by simply removing the self-calibration corrections (which are reported in the catalogue) from the measured galaxy ellipticities before computing the shear. Without the self-calibration we find that the average multiplicative bias  for the full galaxy sample is $\sim$-4\% in both components. This number reduces to $\sim$-2\%  in each component once we use the \lf{} self-calibration. We report the exact values, together with their errors, in Table\,\ref{CAL::TAB::total_bias1}. Even more dramatic is the reduction of the additive bias when we use the self-calibrated version of \lf{}: it reduces by a factor five in $c_{1}$ and by a factor of three in $c_{2}$. This is extremely encouraging, in particular for cosmic shear analysis, where a large additive bias hampers the ability to measure the cosmological signal at large angular separations \citep[e.g.][]{heymans13a, KiDS450}. 

We also explore the impact of misclassified stars on the average bias in the simulations. In fact, \lf{}  occasionally classifies true stars as galaxies and assigns them a non vanishing weight. As stars are not sheared, the net effect is a reduction of the measured shear and hence a multiplicative bias. By measuring the shear bias either including or excluding these misclassified stars, we quantify the effect of star misclassification on the multiplicative bias as approximately $5\times 10^{-3}$. In the following analysis we keep misclassified stars in the catalogue used to estimate the shear bias. We also ran a set of simulations where the density was lowered by 50 \% to explore the effect of galaxy number density on the recovered biases. We found the multiplicative bias to differ by only $2\times 10^{-3}$, suggesting that at the current level of accuracy, simulating the correct number density of galaxies is not crucial for shear calibration, which in turn also implies that galaxy clustering should not impact the shear bias at the KiDS-450 measurement accuracy.

Despite the significant improvements of the self-calibrating \lf{}, residual shear bias remains, arising from both selection bias and from residual uncorrected noise bias,  and we now investigate how the total bias budget is distributed over bins of key input and observed quantities.   As discussed above, we expect the shear bias to depend predominantly on the galaxy SNR and on the ratio of the PSF size and galaxy relative size ${\cal R}$, defined by equation\,\ref{CAL::EQU:resolution} \citep{massey13a}. This is confirmed by Fig.\,\ref{CAL::LENSFIT::mult::obs}, which shows the multiplicative and additive 
bias  from the simulated data as a function of \lf{} model SNR and ${\cal R}$  with, and without, self-calibration. We notice that at low SNR (and faint magnitude) the self-calibration reduces the multiplicative bias by more than a factor of 2; similar improvements are seen as a function of ${\cal R}$.
However, even with self-calibration, the residual multiplicative bias can still be substantially above the 5\% level for very faint (low SNR) and very small (large ${\cal R}$) objects. This emphasises the need for an additional, post measurement bias calibration based on the results of the image simulations.

When the self-calibration corrections are included,
the residual bias almost vanishes, within its errors, for $c_{1}$ but remains significant for $c_{2}$. 
Motivated by the difference in the two components and in order to explore whether the residual additive bias depends on PSF properties, we perform the same analysis in the PSF frame, by rotating all ellipticity and shear values into a frame where the two axes of the PSF align with the coordinate frame. Once we repeat the bias analysis in the PSF frame, we find that the additive bias is now consistent with zero in the cross-aligned component and that for the PSF-aligned component it has risen to the level we found for the second component in the sky frame. This indicates a dependence of the measured bias on PSF properties and motivates a more detailed investigation in \S\ref{sec:shearCalib:cbias}.

\begin{figure*}
 \begin{tabular}{cc}
  \includegraphics[width=.48\textwidth]{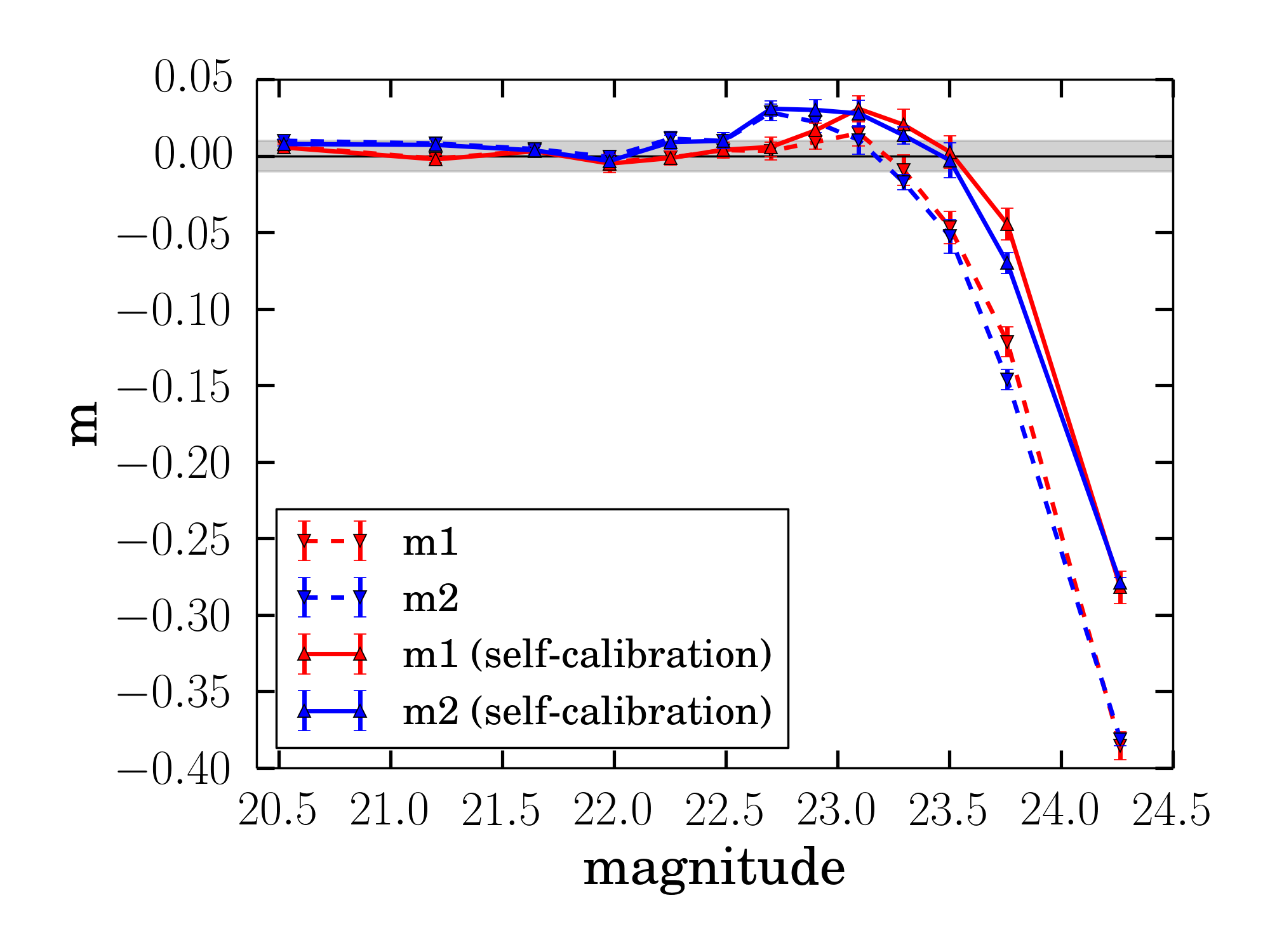} & \includegraphics[width=.48\textwidth]{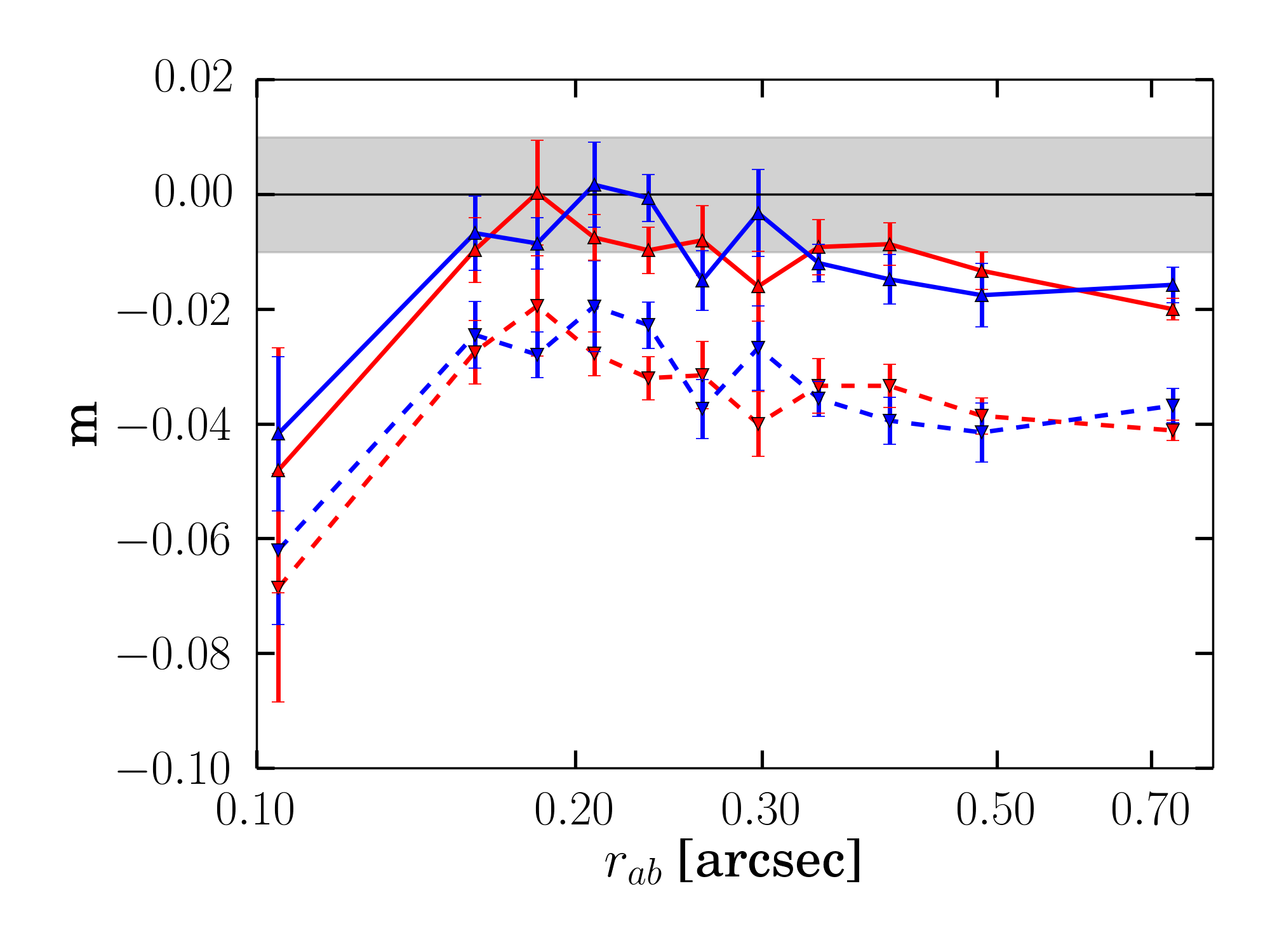} \\
 \includegraphics[width=.48\textwidth]{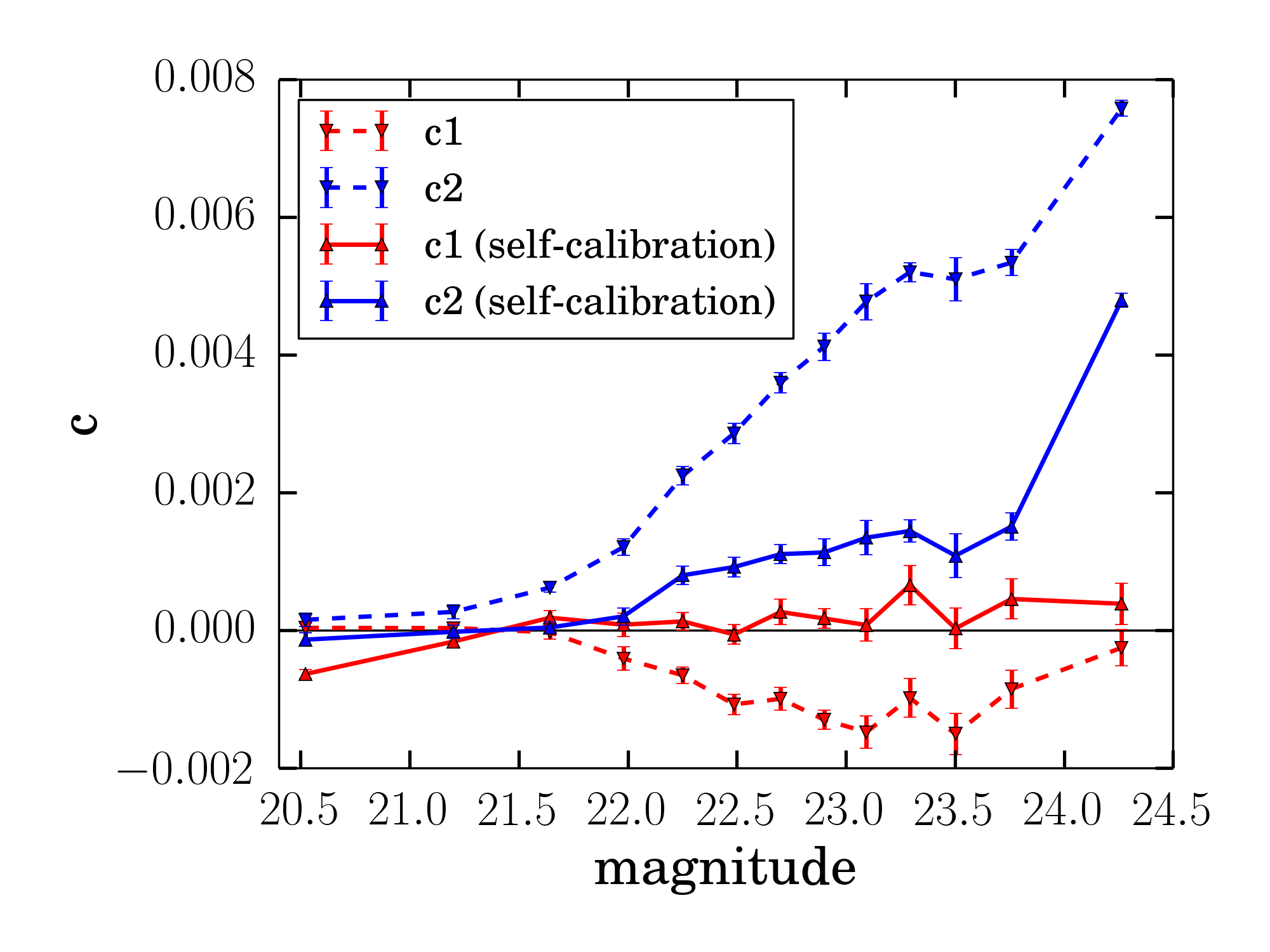} & \includegraphics[width=.48\textwidth]{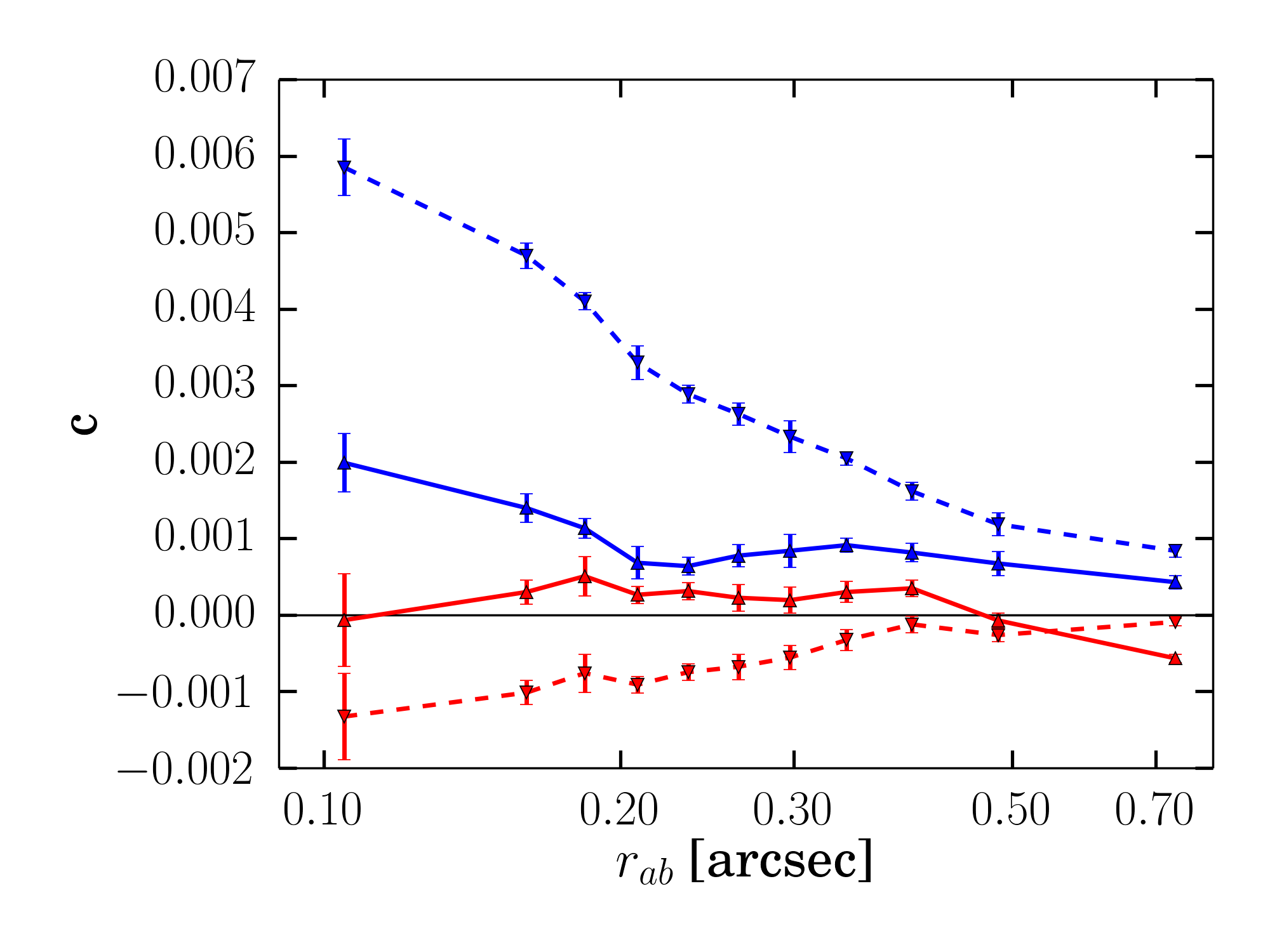}\\  
 \end{tabular}
\caption{The multiplicative bias $m$ (top) and additive bias $c$ (bottom) as a function of simulation input galaxy properties. The left panels shows the bias with and without \lf{} self-calibration as a function of input magnitude. The right panels shows the same measurements as a function of input size. The grey band in the top panels indicates the requirement on the knowledge of the multiplicative bias set by \citet{KiDS450} in the context of a cosmic shear analysis. \label{CAL::LENSFIT::mult::sim}}
\end{figure*}

To explore the dependencies on input parameters, Fig.\,\ref{CAL::LENSFIT::mult::sim} shows the bias in $m$ and $c$ as a function of input magnitude and size. Selection effects are clearly important for the multiplicative bias for faint galaxies,
although it should be noted that the most dramatic effects arise at magnitudes $m > 23$, where the galaxy detection
is incomplete (Fig.\,\ref{fig:distr_sim_data}) and where the weighted contribution to shear measurement is small.
In the case of the additive bias, in particular, 
the utility of self-calibration is evident, as the dependences on input parameters are significantly reduced.

\subsection{Multiplicative shear bias calibration}
\label{sec:shearCalib:mbias}
The self-calibrated \lf{} already delivers excellent results in terms of total residual shear bias, as shown in Table\,\ref{CAL::TAB::total_bias1}. However, emphasised by Fig.\,\ref{CAL::LENSFIT::mult::obs} and Fig.\,\ref{CAL::LENSFIT::mult::sim}, 
multiplicative biases significantly larger than 5\% are still possible, most prominently for faint and small galaxies, although we must be cautious in interpreting any size dependence, owing to the selection bias demonstrated in
\S\ref{sec:calibselectionbias}.
We aim here to derive a calibration for the residual multiplicative bias after self-calibration as a function of \lf{}-measured SNR and $\cal{R}$. 
While $\cal{R}$ is a good choice for characterising the size of a galaxy with respect to the PSF \citep{massey13a}, 
one could consider flux-related calibration quantities other than SNR, for example the observed magnitude, to use as
a calibration parameter. 
However, as discussed in \S\ref{sec:compsimdata}, the real KiDS imaging data has quite some variation of the pixel noise rms, mostly owing to varying observing conditions,  while in the simulations we used a fixed value.  As the shear bias depends on the noise level and not on the actual flux of the object, it is not possible to derive a robust calibration based on output magnitude. 

\begin{figure*}
\includegraphics[width=1.0\textwidth]{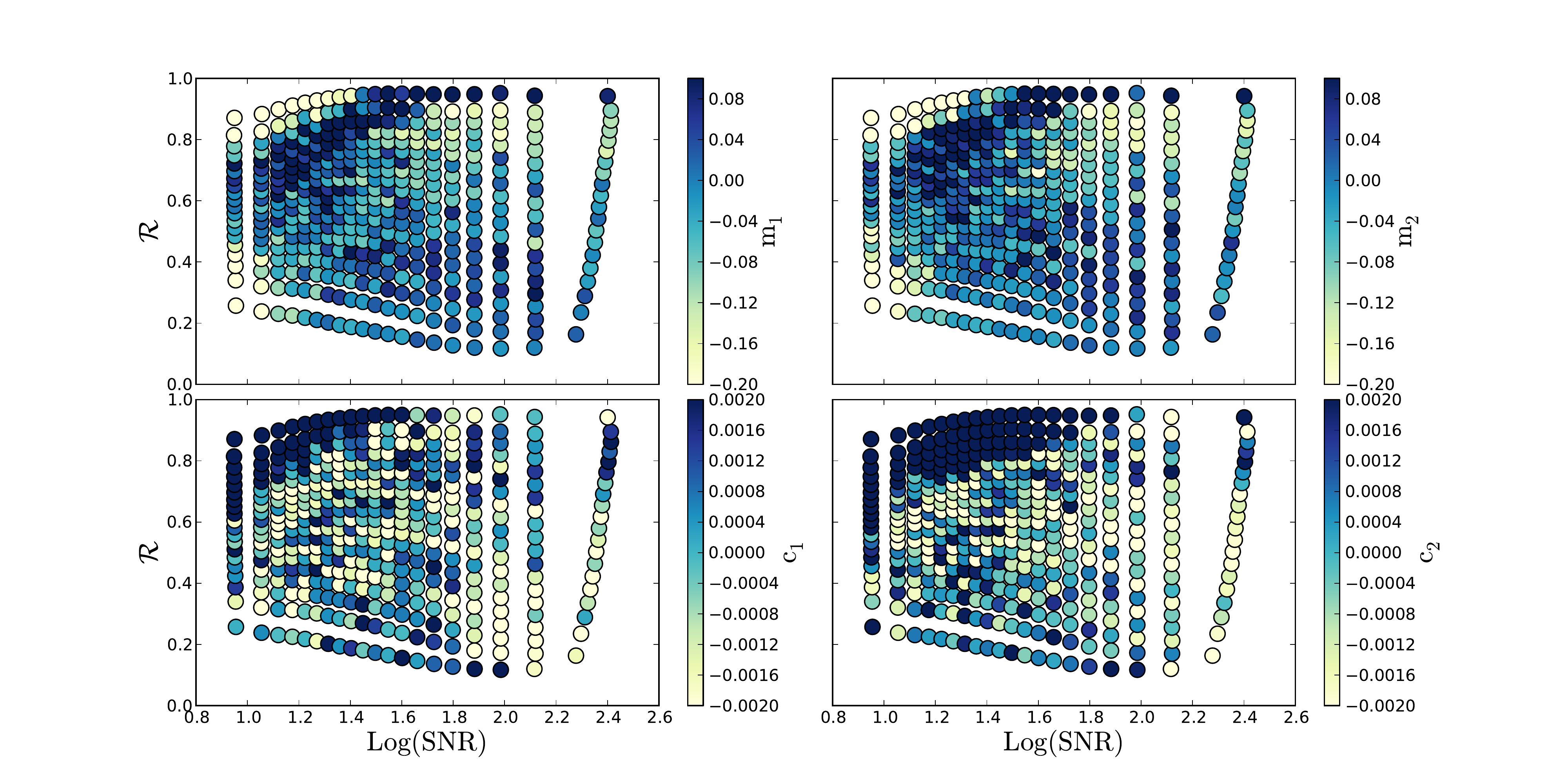}
\caption{The 2D bias surface as a function of model SNR and $\cal{R}$. The \textit{top panels} show the multiplicative bias surface, $m_{1}$ on the \textit{left} and $m_{2}$ on the \textit{right}. The \textit{bottom} panels show the additive bias components, $c_{1}$ on the \textit{left} and $c_{2}$ on the \textit{right}. Each point in the plot has equal \lf{} weight. \label{CAL::BIAS_SURFACE}}
\end{figure*}

We bin our simulated data according to the measured galaxy model SNR and $\cal{R}$, again requiring equal \lf{} weight in each bin and we use the self-calibrated \lf{} measurements as the default. The two dimensional multiplicative bias surface  as a function of SNR and $\cal{R}$ is shown
in Fig.\,\ref{CAL::BIAS_SURFACE}. A crucial parameter in such analyses is the total number of bins used to characterise the bias surface. On the one hand, we would like to have a fine enough grid to capture every real feature in the bias surface, but, on  the other hand, we have to ensure that there is enough statistical power in each bin so that measurements are not dominated by noise. We tried a variety of grids ranging from only two up to 40 bins on each axis.
A coarse $10 \times 10$  binning scheme results in an average m-bias error of 2\% in both components per bin and increases to an average 10\% per bin for the $40 \times 40$  scheme. This results in a vanishing signal-to-noise ratio for bins
with a small measured bias while using a very fine binning scheme. We found that a $20 \times 20$ bin grid provides the best compromise with an average signal-to-noise of 2.5 per bin over the full SNR-\cal{R} surface and enough resolution to capture the complicated structure of the bias surface in the low SNR, large \cal{R} regime.

Fig.\,\ref{CAL::BIAS_SURFACE} reveals that the multiplicative bias surface is complex. 
Our initial characterisation attempt is based on a fit of an analytic 2D function to the bias surface, as was done for example in \cite{miller13a, hoekstra15a,jarvis15a}. Unfortunately, even a complex 16-parameter functional form 
\begin{equation}
 m_{1/2} = f_{0} + f_{1}\mathcal{R}^{-1} + f_{2}\mathcal{R} + f_{3}\mathcal{R}^{2},
\end{equation}
where the pre-factors $f_{i}$ depend on the 16 parameters and the \lf{} SNR
\begin{equation}
 f_{i} = p_{4i+1} + p_{4i+2}\mathrm{SNR}^{-1} + p_{4i+3}\mathrm{SNR}^{-2} + p_{4i+4}\mathrm{SNR}^{-1/2},
\end{equation}
for $i\in(0,1,2,3)$ gave only a poor fit to the surface ($\chi^{2}$-values of 3.9 and 3.6 for $m_1$ and $m_2$ respectively). From now on we will refer to this form of characterisation of the bias surface as method \textit{A}.

Our second attempt to characterise the surface, method \textit{B},  is based on an interpolation of the bias surface. Simple spline interpolation fails
to robustly interpolate the bias due to its complicated structure in SNR and $\cal{R}$ space.   
We applied an interpolation scheme based on a Gaussian radial basis function with a spatially varying shape parameter \citep[see][and references therein]{2014arXiv1412.5186M}.
The interpolation was trained beforehand using the best-fit analytic functional form of method \textit{A}, 
to optimally adapt its shape-parameters to the spatial structure of the SNR-$\cal{R}$ grid and the general features of the bias surface. The resulting interpolation allowed us to query the multiplicative bias
in both components for any parameter pair, at least in the area covered by the given SNR and $\cal{R}$ range shown in Fig.\,\ref{CAL::BIAS_SURFACE}.   

Finally, we tried a simpler calibration strategy, method \textit{C}, 
which was to not fit or interpolate the bias surface, but rather
to assign the bias determined in each of the $20 \times 20$ bins to the galaxies that fall in each bin.

\begin{figure*}
 \centering
 \begin{tabular}{cc}
  \includegraphics[width=.48\textwidth]{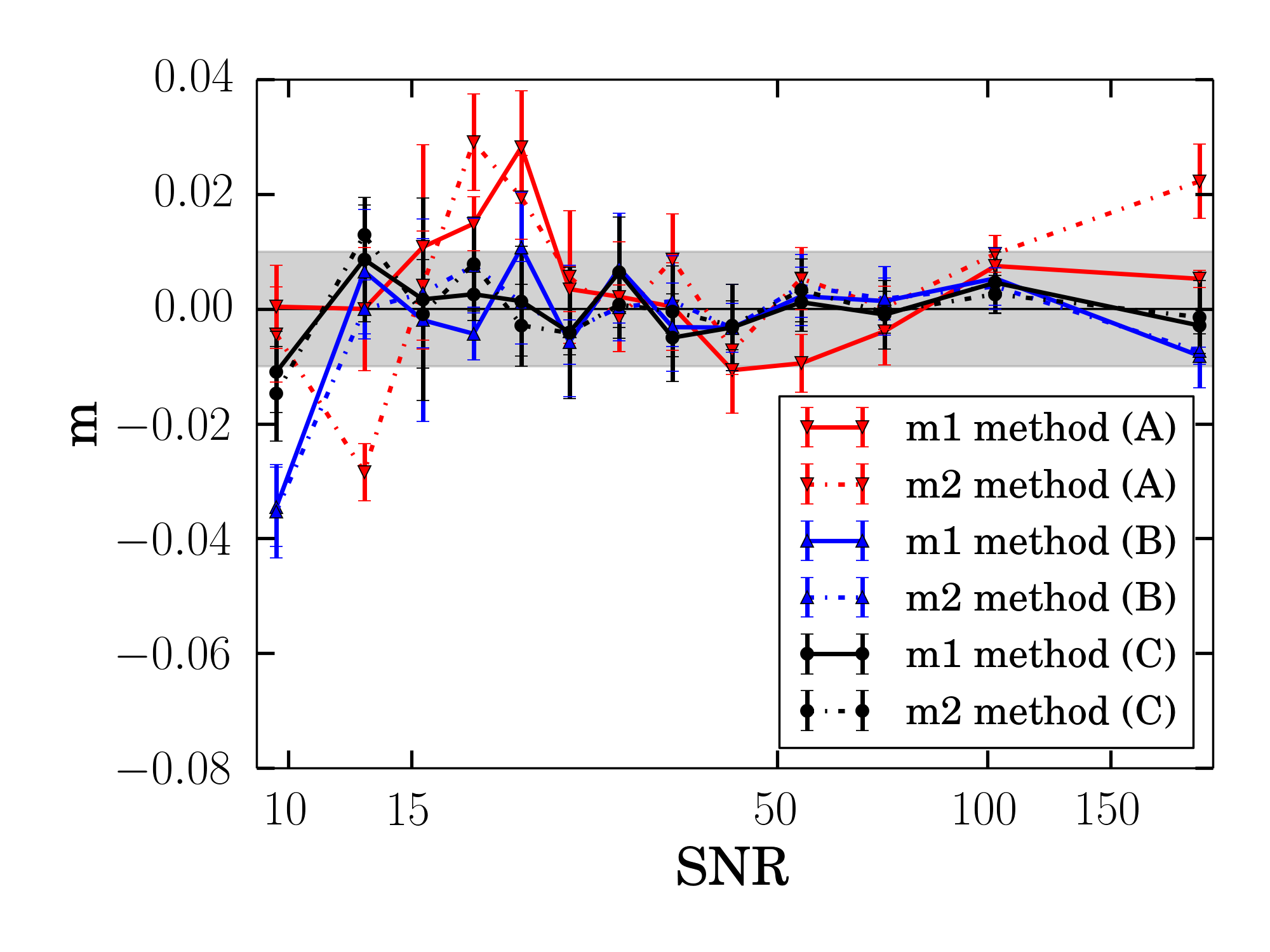} & \includegraphics[width=.48\textwidth]{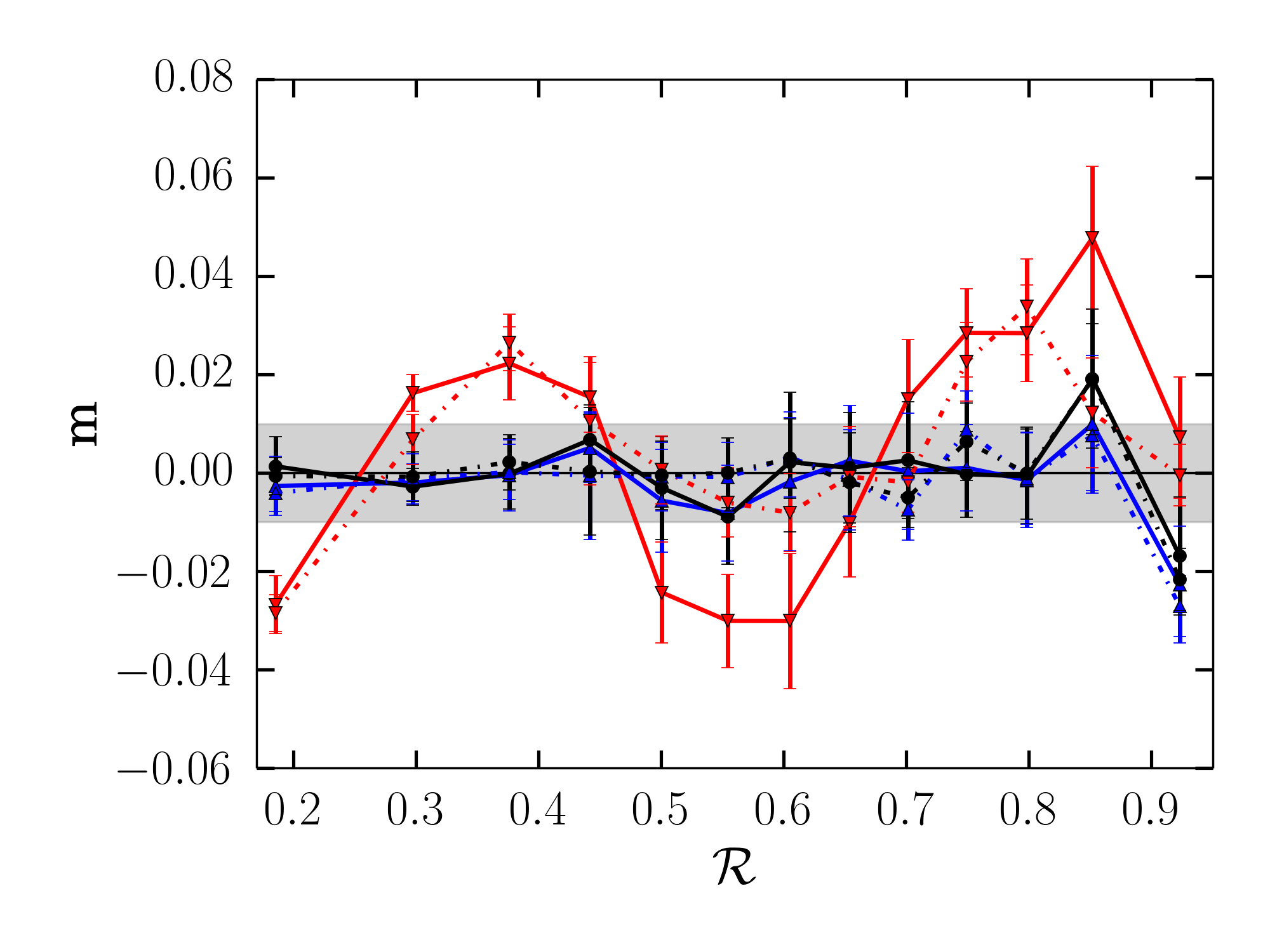}
 \end{tabular}
\caption{The multiplicative bias after empirical calibration using different methods. Method {\it A} is based on a function form fit to the bias surface, method {\it B} performs an interpolation of the bias surface and {\it C} assigns a constant bias correction in 2D bins. The \textit{left panel} shows the residual multiplicative bias after calibration as a function of model SNR and the \textit{right panel} as a function of $\cal{R}$.The grey band indicates the requirement on the knowledge of the multiplicative bias set by \citet{KiDS450} in the context of a cosmic shear analysis.
  \label{CAL::M1CAL::METHODS::obs}}
\end{figure*}

\begin{figure*}
 \centering
 \begin{tabular}{cc}
  \includegraphics[width=.48\textwidth]{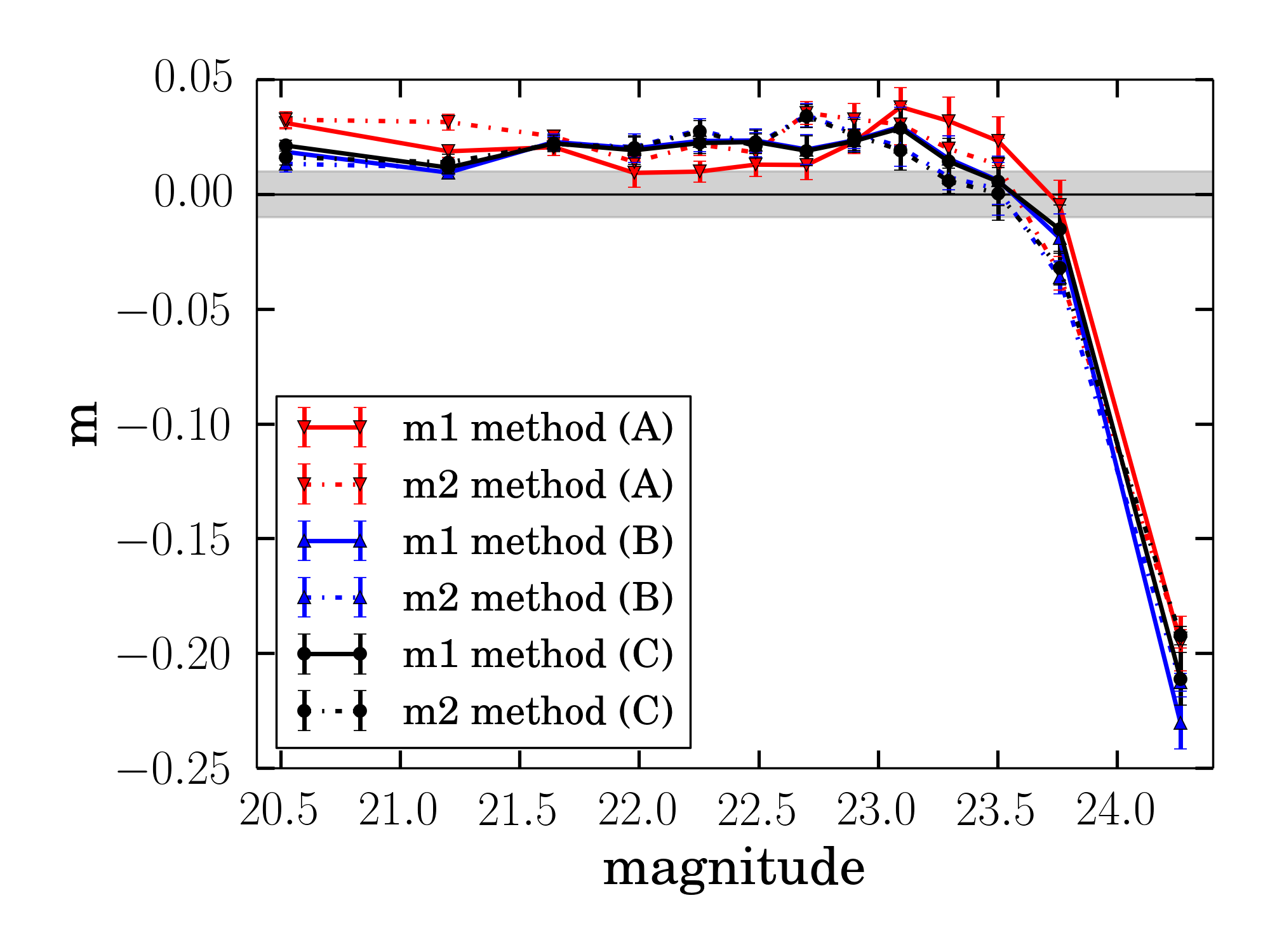} & \includegraphics[width=.48\textwidth]{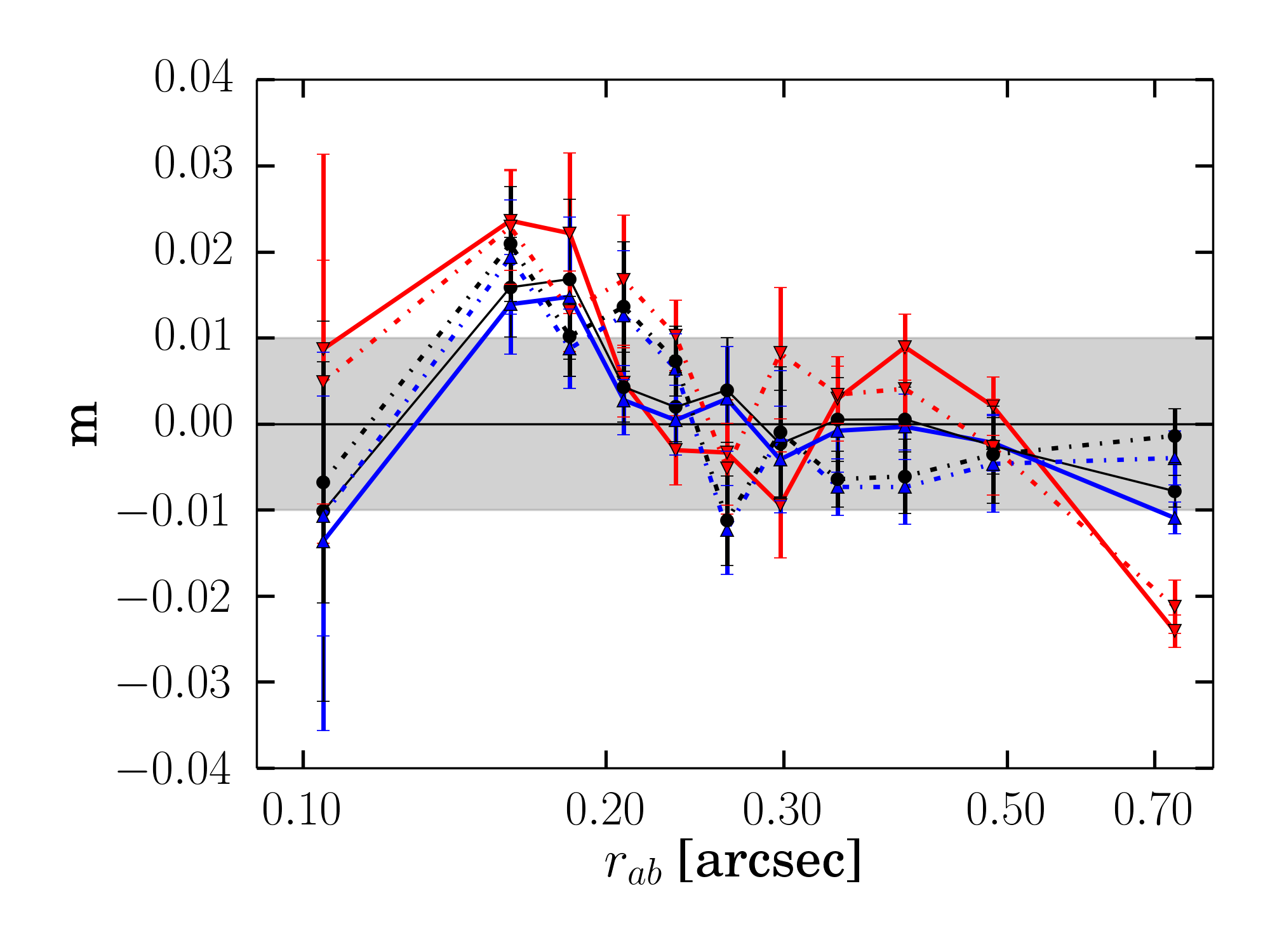}
 \end{tabular}
\caption{This plot is equivalent to Fig.\,\ref{CAL::M1CAL::METHODS::obs}, but shows the residual multiplicative bias as a function of input magnitude in the \textit{left panel} and as a function of input size in the \textit{right panel}. 
\label{CAL::M1CAL::METHODS::sim}}
\end{figure*}

We test the differing calibration strategies, by investigating the derived multiplicative bias as a function of SNR and $\cal{R}$ according to methods \textit{A}, \textit{B} or \textit{C}, for all galaxies with shape measurement in the simulation. In each bin of the analysis we calculate the \lf{}-weighted average multiplicative bias correction and apply it to the average measured ellipticity in the bin according to equation\,\ref{EQU::CAL::BIAS}. Afterwards, we recalculate the bias.  The results for each method are presented in  Table\,\ref{CAL::TAB::total_bias1} in terms of the total bias and in Figs.\,\ref{CAL::M1CAL::METHODS::obs} and \ref{CAL::M1CAL::METHODS::sim} as a function of the key output and input quantities. The total multiplicative bias after we apply the calibration is around or below the percent level in both shear components for all three methods. It vanishes completely, by construction, within its error bars for the bin-based calibration method \textit{C}. In terms of our 1\% target window, method \textit{A} fails to deliver a robust calibration over the full $\cal{R}$ range. Methods \textit{B} and \textit{C} do clearly better and robustly calibrate the residual bias over the full $\cal{R}$ range.   An exception are extremely small, high $\cal{R}$ objects, which represent only a small population in the image simulations.  The very last bin in $\cal{R}$, where methods \textit{B} and \textit{C} show a residual bias of 2\%, accounts for 7\% of the total \lf{} weight in the sample.

The picture is similar in terms of the calibration performance as a function of SNR. Method \textit{C} performs best and only marginally falls out of our target accuracy for objects with SNR $< 7$. The reason why this method shows a residual bias at all, is the fact that the binning scheme we used for this analysis differs in both the number of bins and its 1-dimensional nature from the $20 \times 20$ SNR-$\cal{R}$ binning scheme that we used to derive the calibration.  The first SNR bin in Fig.\,\ref{CAL::M1CAL::METHODS::obs}, where methods \textit{B} and  \textit{C} show residual multiplicative biases of -3.5\% and 1.5\%, respectively, contributes 7\% to the \lf{} weight in the full sample. In the extremely low SNR regime ($\sim10$), the interpolation based method \textit{B} performs much worse than \textit{C}, likely due to less robust interpolation result near the edges of the initial bias surface. In the final analysis and considering all mentioned effects, we find that method \textit{C} provides the most robust calibration of the multiplicative bias and it will be our default method. 

In order to  test the dependence of this calibration on the number of bins used to characterise the multiplicative bias surface, we investigated the measured bias as a function of the number of 2D bins used. We find that if the number of bins
is too small,
the calibration is not able to pick up all relevant features in the bias surface and hence existing residual bias remains uncalibrated. Using more than ten bins starts to remedy the problem and a 20 bin scheme is the first calibration that delivers a robust calibration within 1\% for the full range of SNR and $\cal{R}$, with the exception of very small objects with ${\cal{R}} > 0.9$, which contribute only a small fraction of the sample's total \lf{} weight.

We might hope that when the residual bias, after applying the calibration, is measured as a function of input magnitude
and size, it should be consistent with zero.  However, this is not the case, as shown in 
Fig.\,\ref{CAL::M1CAL::METHODS::sim}. All the calibration schemes show a small positive bias for objects with bright input magnitudes ($m \la 23$) and small galaxies ($r_{ab} \la 0.2''$), and a negative bias at faint magnitudes which becomes large
for galaxies below the selection completeness limit. The average weighted bias, however, for the entire simulation, 
is consistent with zero.  The cause of this effect is that the calibration on noisy output quantities relies on there being
a stationary correlation between the true quantities and their measured, noisy counterparts.  At magnitudes below the
completeness limit, the relationship between true size and measured size in the selected galaxies changes, which in turn
impacts the calibration relation.  In effect, there is a third axis of ``magnitude'' in our calibration space which has not
been included in the calibration relation.  In fact, it is not possible to reliably include this third axis, as the three
quantities are highly correlated, and also correlated with galaxy ellipticity, and correct calibration in this
space would require the joint distributions in the simulations and in the data to match precisely, which is difficult
to achieve and is not the case in our simulations.

As by construction, the net residual bias after calibration in the simulations is zero, 
if the data that we seek to calibrate has the same
distribution of true magnitude and size as the simulations, application of the calibration relation should also
result in zero residual bias in the calibrated data.  However, in reality the data and simulation distributions
differ, as shown in Fig.\,\ref{fig:distr_sim_data}, and in the cosmic shear analysis \citep{KiDS450}
the data are divided into
tomographic subsamples, with their own size and magnitude distributions.  We investigate the amount of residual
bias that might leak into the tomographic analysis presented in \citet{KiDS450} via this effect in \S\,\ref{sec:resampling}.

\begin{table*}
 \caption{The total multiplicative and additive bias after residual bias calibration.}
 \label{CAL::TAB::total_bias2}
 \begin{tabular}{ccccccccc}
  \hline
  method &$m_{1}$& $\Delta m_{1}$(regr)/(BS)&$m_{2}$& $\Delta m_{2}$&$c_{1}$& $\Delta c_{1}$&$c_{2}$& $\Delta c_{2}$\\
  &$[10^{-3}]$&$[10^{-3}]$&$[10^{-3}]$&$[10^{-3}]$&$[10^{-5}]$&$[10^{-5}]$&$[10^{-5}]$&$[10^{-5}]$\\
\hline
  \textit{A}       &  3.80 & 3.35/4.62 &  4.90 & 1.88/1.90 &--&--&--&-- \\
  \textit{B}       & -1.99 & 3.35/3.72 & -1.89 & 1.90/2.44 &--&--&--&-- \\
  \textit{C}       &  -0.008 & 3.37/3.89 &  -0.01 & 1.91/2.49 &--&--&--&-- \\
  \textit{C} (m+c) &  -0.008 & 3.36/4.22 &  -0.005& 1.90/2.72 &-0.007&9.51/9.38&0.014&5.37/6.66\\
  \hline
 \end{tabular}
\end{table*}

 \subsection{Additive shear bias calibration and PSF properties}
 \label{sec:shearCalib:cbias}
We have identified the $20 \times 20$ grid, bin-based method \textit{C} as the most robust to calibrate for the remaining residual multiplicative bias. 
Using exactly the same methodology and 
by again following equation\,\ref{EQU::CAL::BIAS} we also characterise the small remaining additive bias not accounted for by \lf{}'s self-calibration.
When calibrating for both, multiplicative and additive
bias, simultaneously, we find the residuals shown in the last line of Table\,\ref{CAL::TAB::total_bias2}, which is our best and final result. 

\begin{figure*}
 \centering
 \begin{tabular}{cc}
  \includegraphics[width=.48\textwidth]{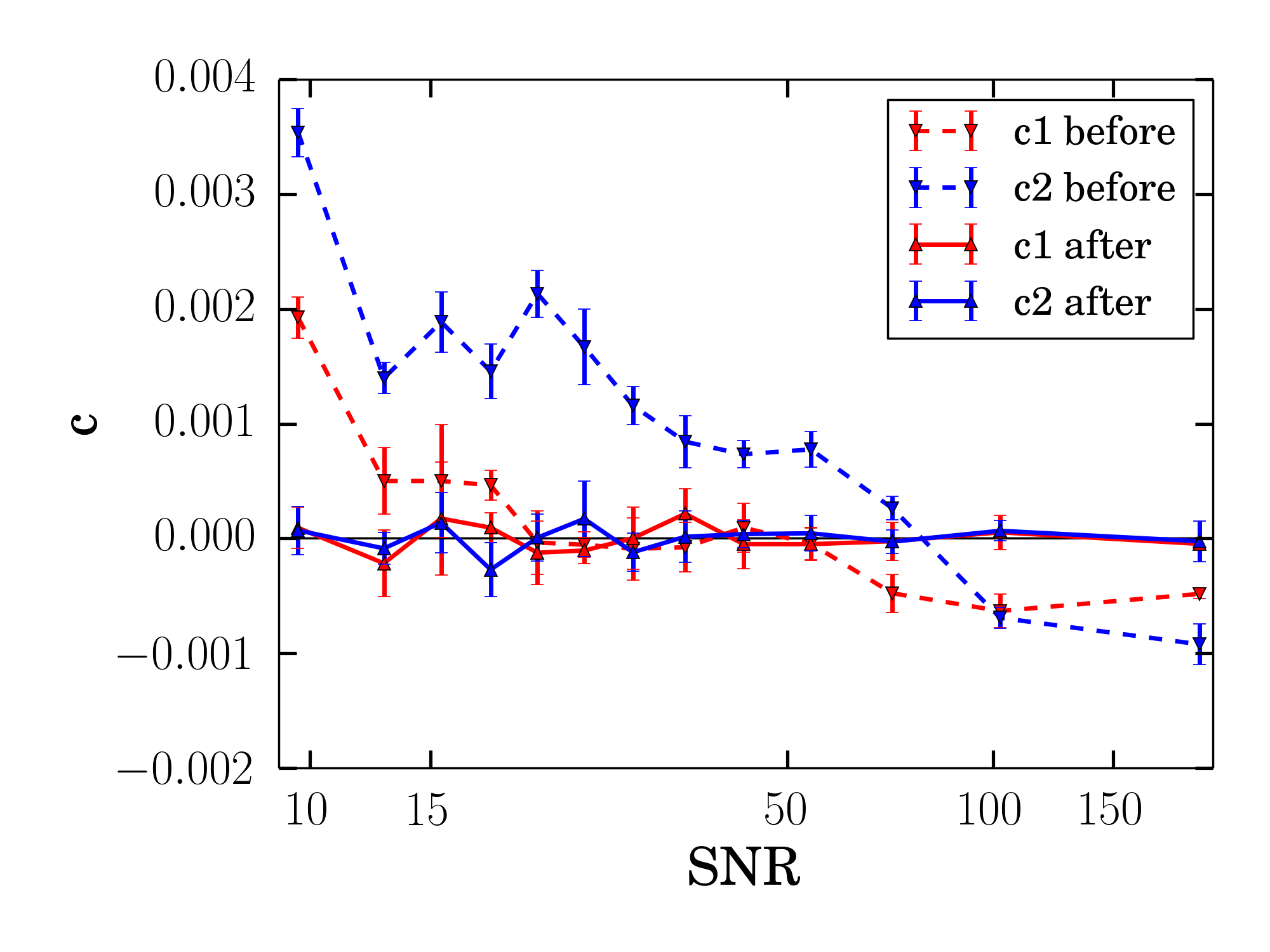} & \includegraphics[width=.48\textwidth]{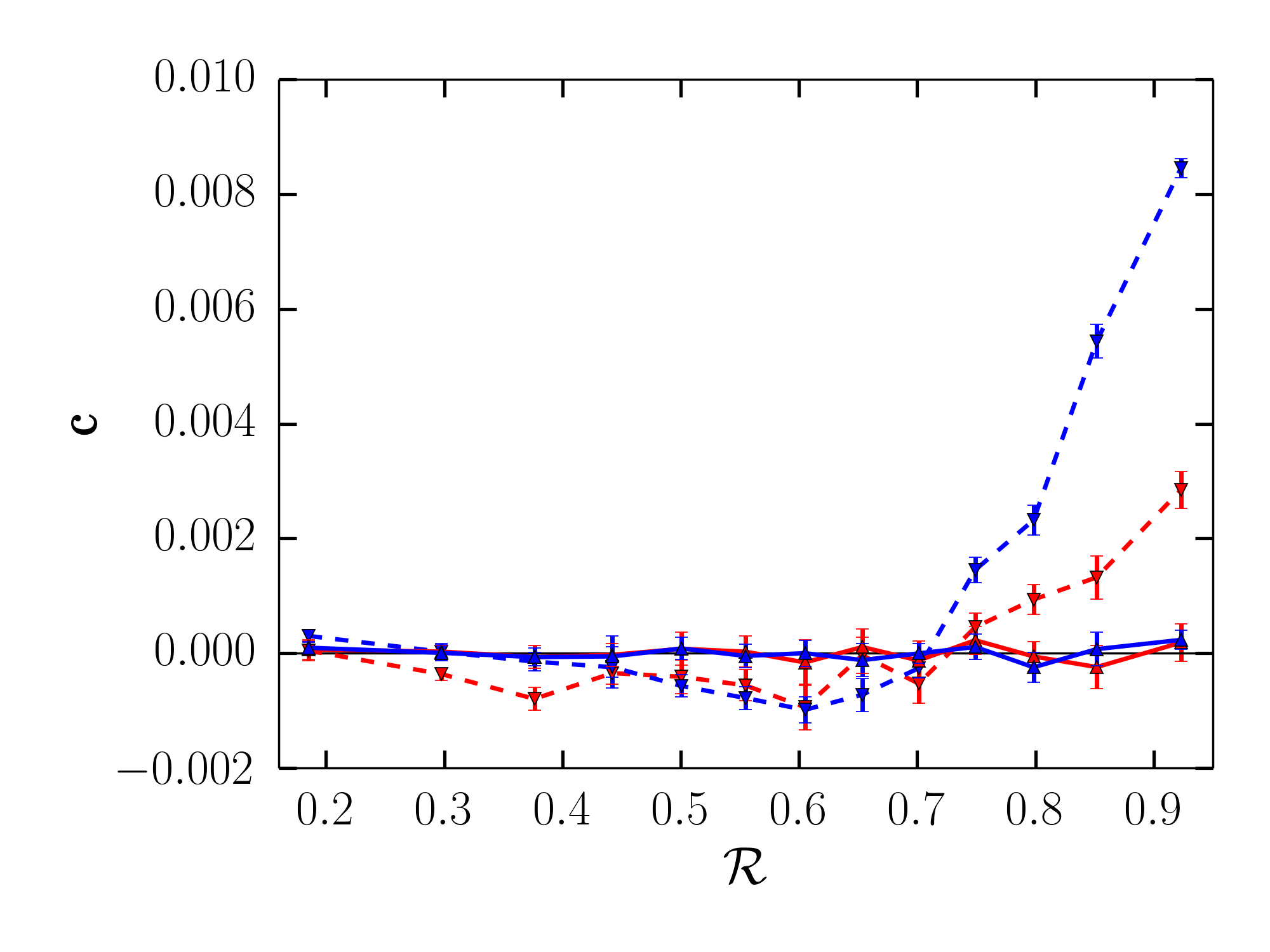}\\
 \end{tabular}
\caption{The residual additive shear bias before and after calibration using method {\it C}. The \textit{left panel} shows residual bias as a function of model SNR and the \textit{right panel} in bins of $\cal{R}$.  \label{CAL::CCAL}}
\end{figure*}

\begin{figure*}
 \centering
 \begin{tabular}{cc}
  \includegraphics[width=.48\textwidth]{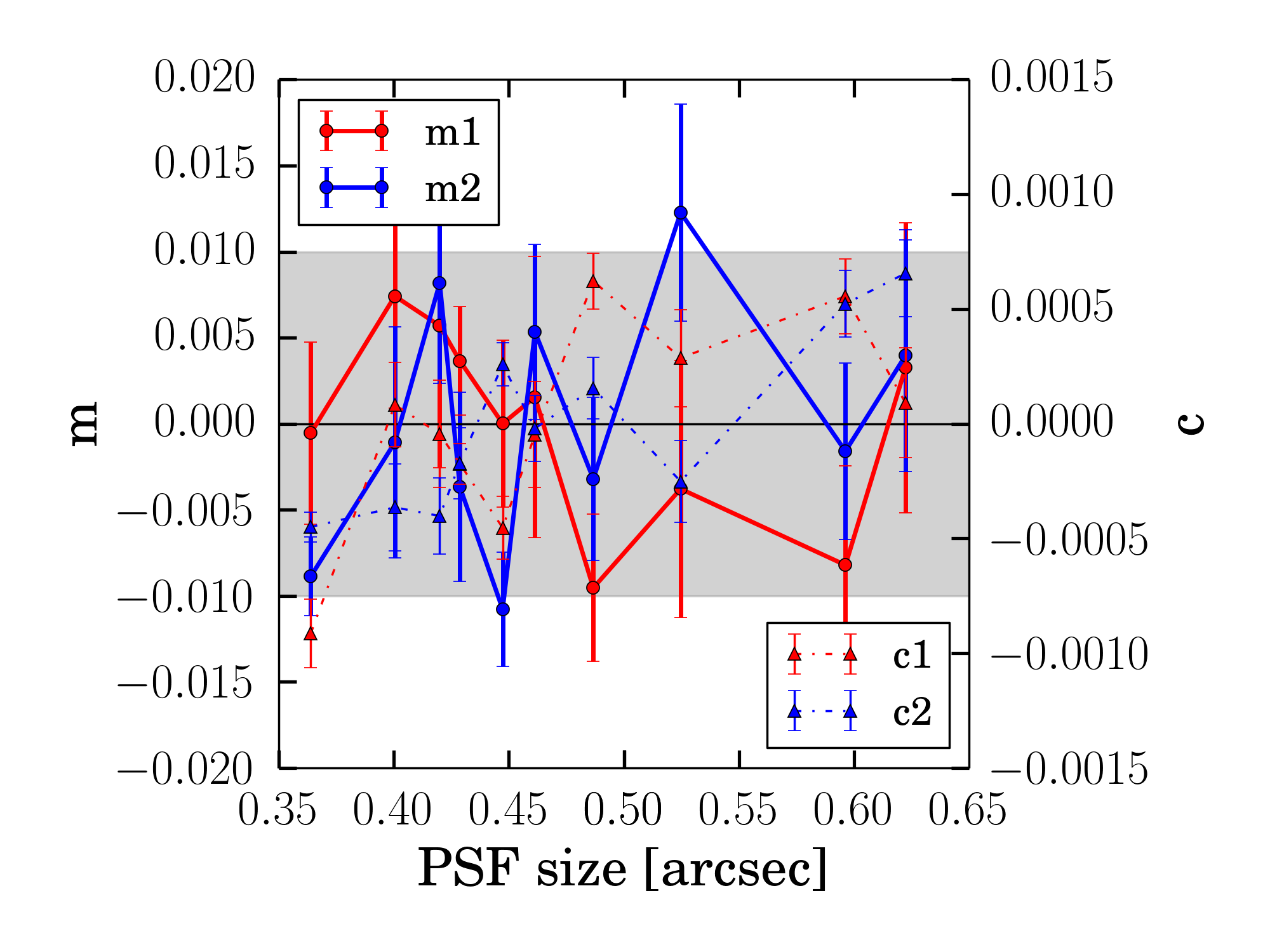} & \includegraphics[width=.48\textwidth]{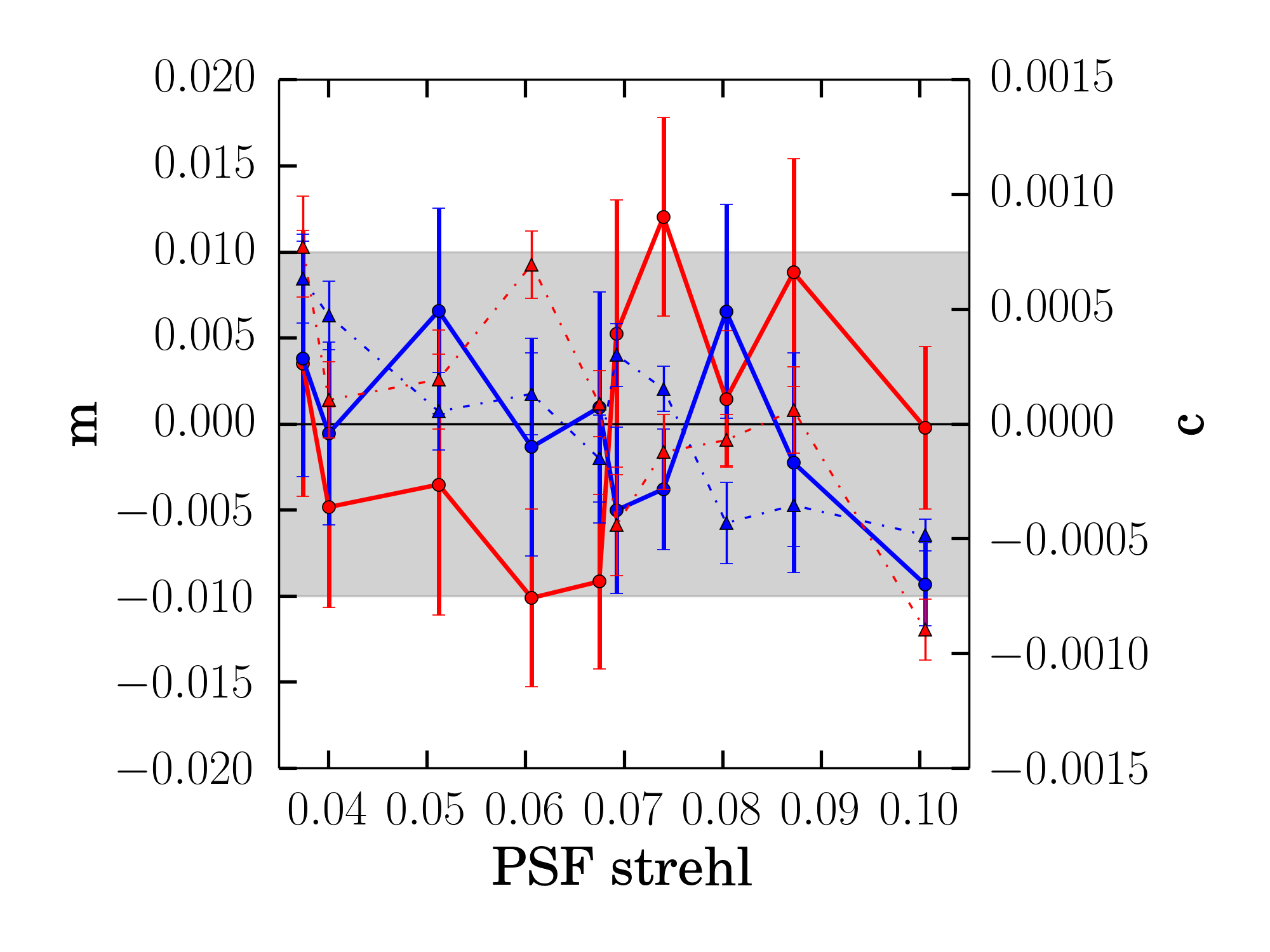}\\
  \includegraphics[width=.48\textwidth]{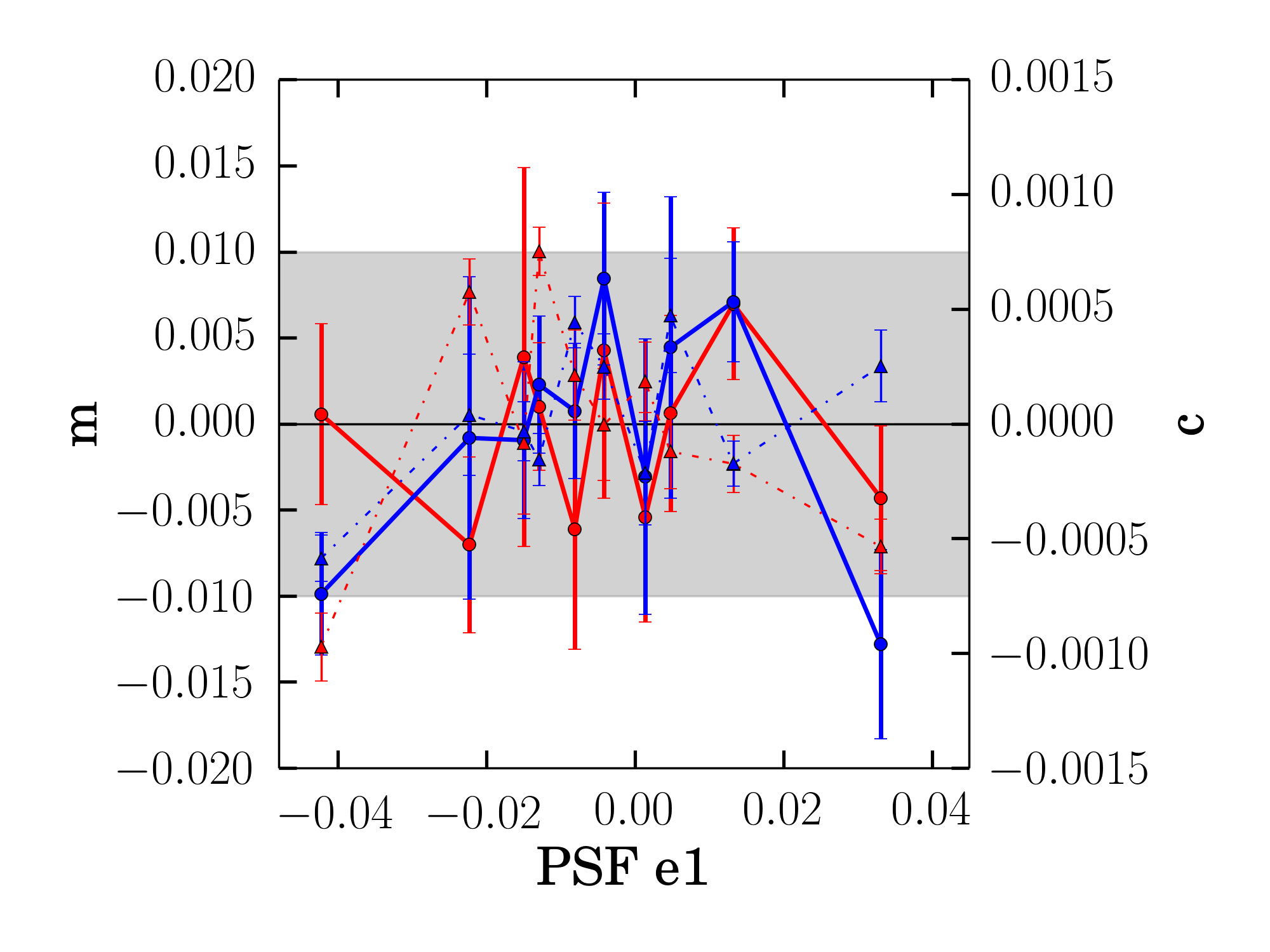} & \includegraphics[width=.48\textwidth]{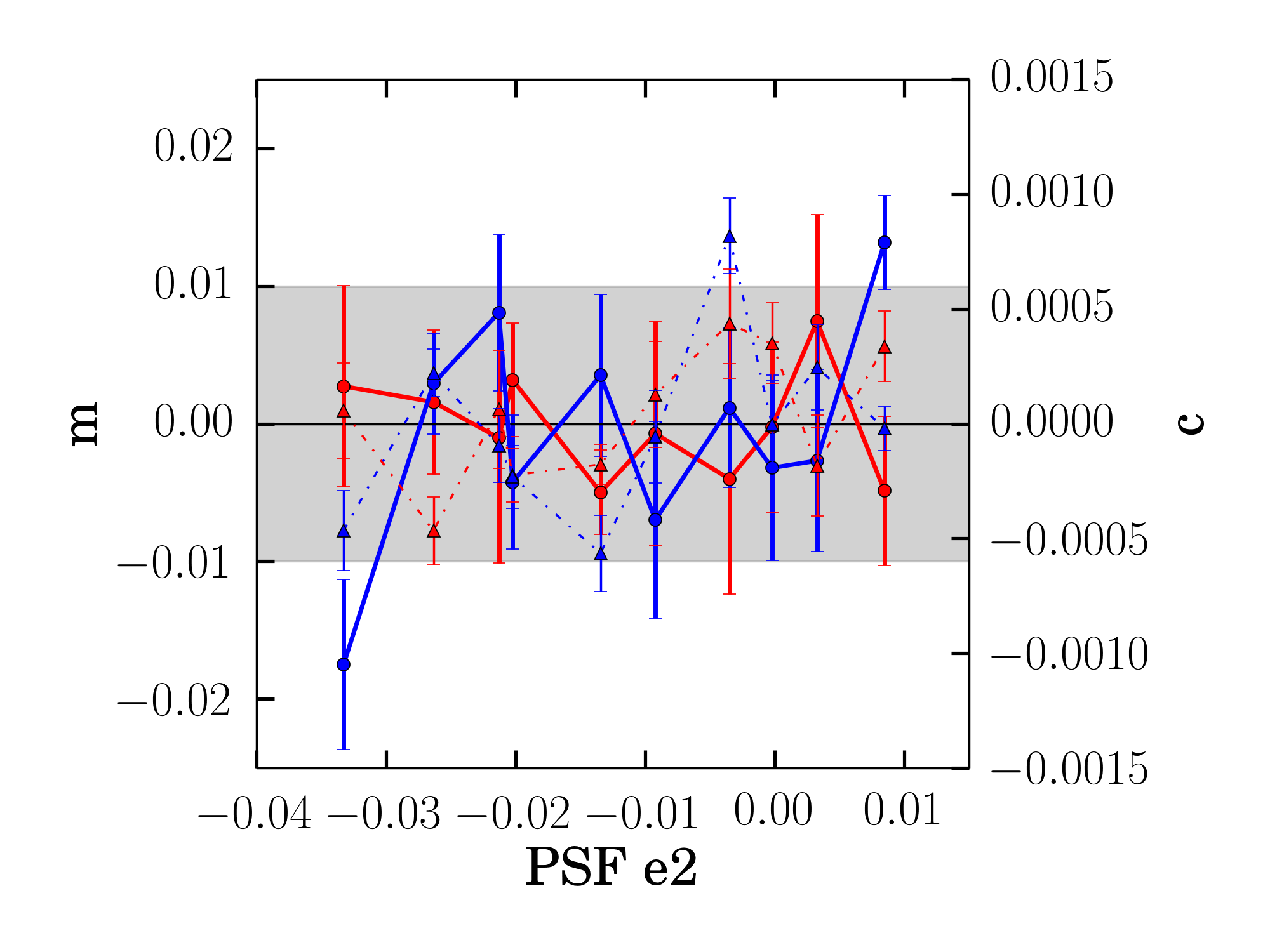}
 \end{tabular}
\caption{The residual bias as a function of PSF properties. The solid lines refer to the residual multiplicative bias with the scale given by the left y-axis. The dot-dashed lines refer to residual additive bias with the scale 
on the right y-axis in each plot, respectively. The four panels show the biases in clock-wise order starting on the top-right as a function of: measured PSF size, PSF pseudo-Strehl ratio, second PSF ellipticity component and first ellipticity component.
\label{CAL::CAL::PSF}}
\end{figure*}

Fig.\,\ref{CAL::CCAL} shows the residual additive bias as a function of SNR and $\cal{R}$ before and after calibration and Fig.\,\ref{CAL::CAL::PSF} shows the remaining multiplicative and additive bias as a function of PSF properties. 
This includes the two PSF ellipticity components, the PSF size and ``pseudo-Strehl ratio'' 
(defined as the fraction of light contained in the central pixel of the PSF). 
All the analyses show no systematic dependence of m -and c-bias on PSF properties and all reported residual biases fulfil, within their errors, our target of 1\% residual bias. 
However, as summarised earlier in Table\,\ref{CAL::TAB::total_bias2}, we do detect bias when performing the analysis in the PSF and not in the sky frame. This is expected from the additive selection bias 
of \S\ref{sec:calibselectionbias} and should also have a contribution arising from residual uncorrected noise bias
\citep{miller13a}.
In order to characterise this effect we
extend our bias description by including a PSF ellipticity dependent term $\alpha$, following \citet{jarvis15a}:
\begin{equation}
 g^{\textrm{meas}}_{j}=\left(1+m_{j}\right)g^{\textrm{true}}_{j}+\alpha_{j}\epsilon_{j}^{\textrm{PSF}}+c_{j}.
 \label{EQU::CAL::BIAS::ALPHA}
\end{equation}

\begin{figure*}
 \centering
 \begin{tabular}{cc}
  \includegraphics[width=.48\textwidth]{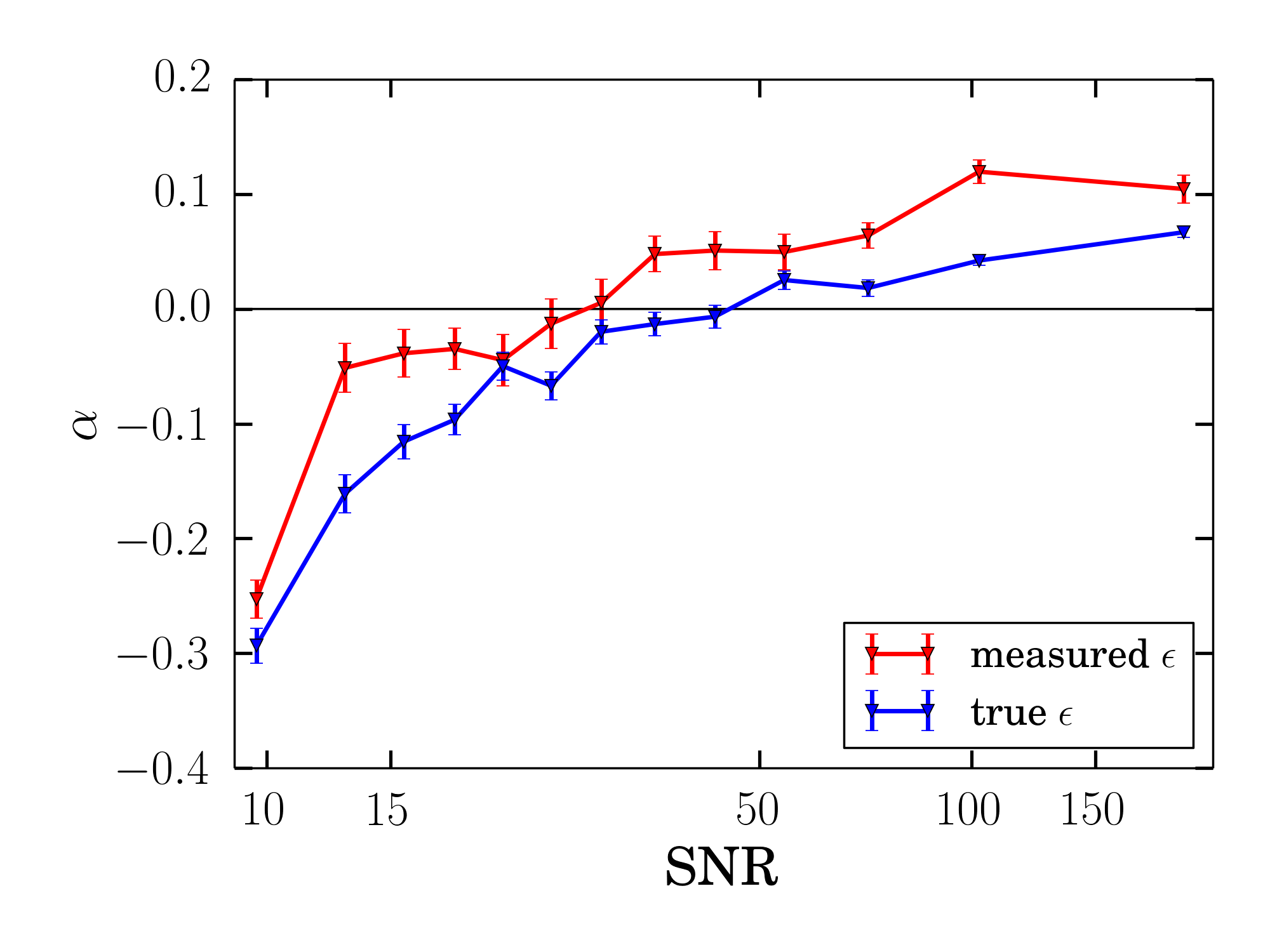} & \includegraphics[width=.48\textwidth]{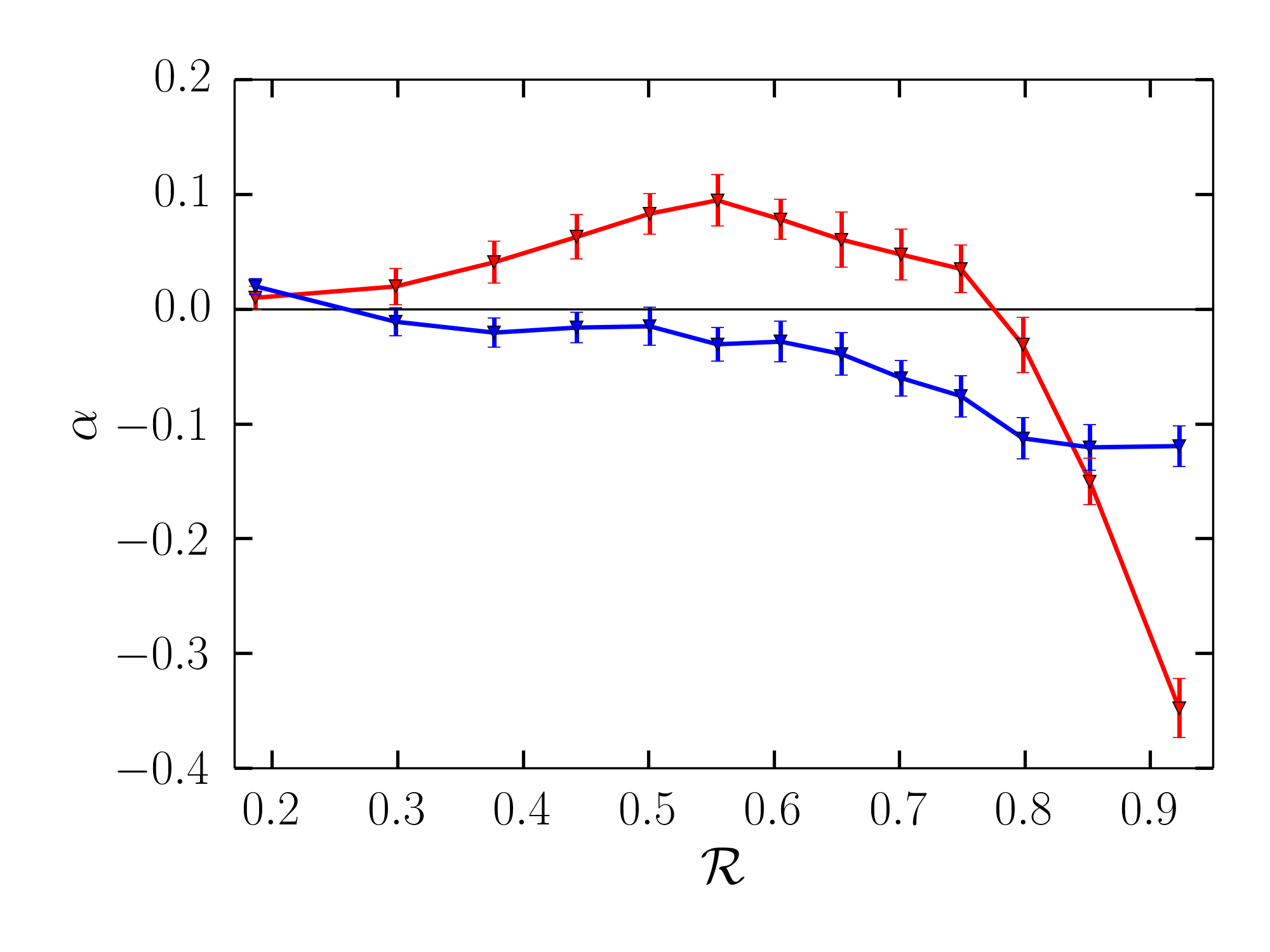}
 \end{tabular}
\caption{The average of the two PSF leakage components, $\alpha$, as a function of measured galaxy properties,
showing the leakage deduced from measured \lf{}~ellipticities (red curves and points) and 
from true, sheared input ellipticities (blue curves and points), as a test of selection bias. 
The \textit{left panel} shows $\alpha$ as a function of model SNR, the \textit{right panel} as a function of $\cal{R}$. 
\label{CAL::ALPHA::OBS}}
\end{figure*}

\begin{figure*}
 \centering
 \begin{tabular}{cc}
  \includegraphics[width=.48\textwidth]{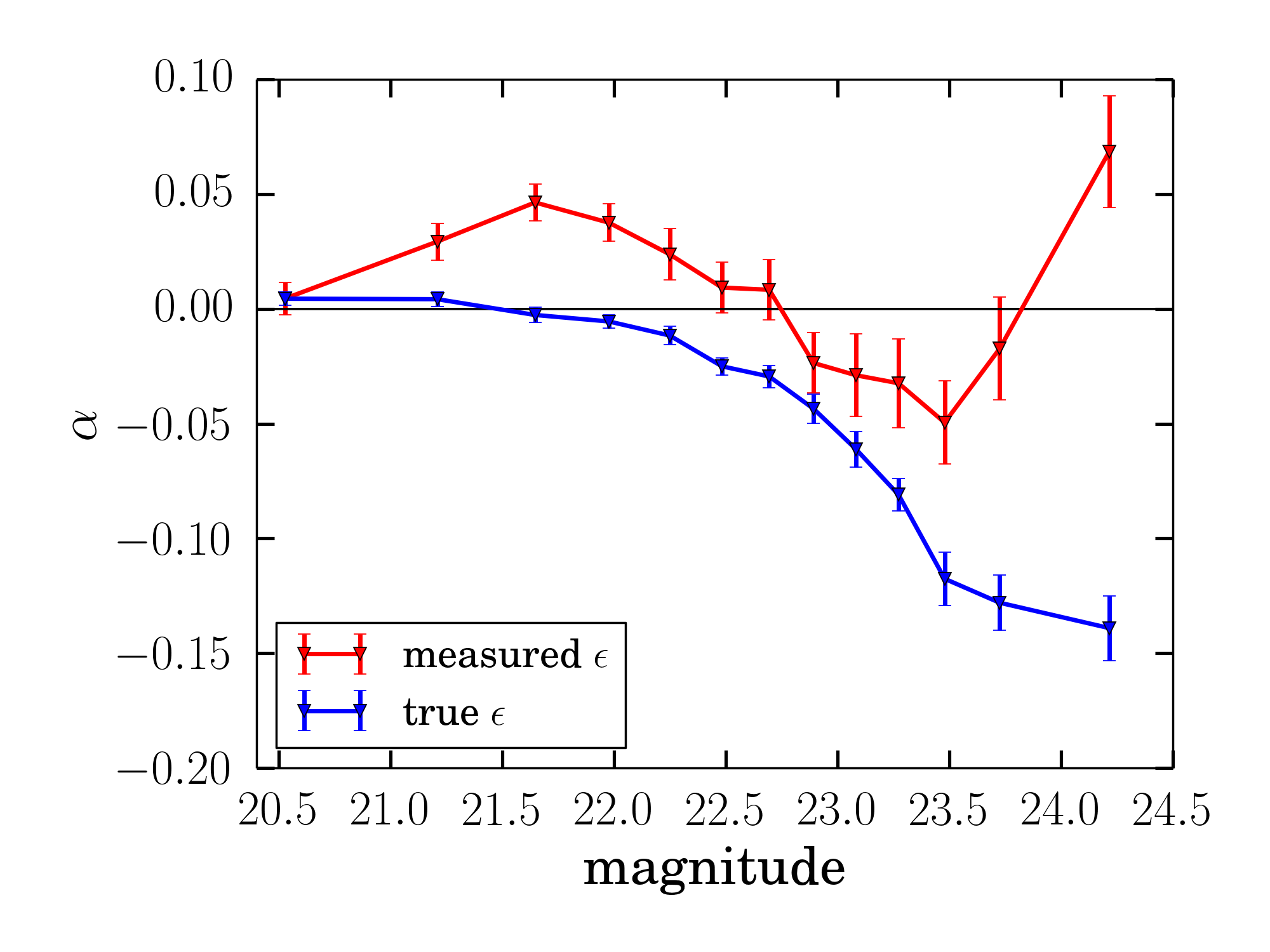} & \includegraphics[width=.48\textwidth]{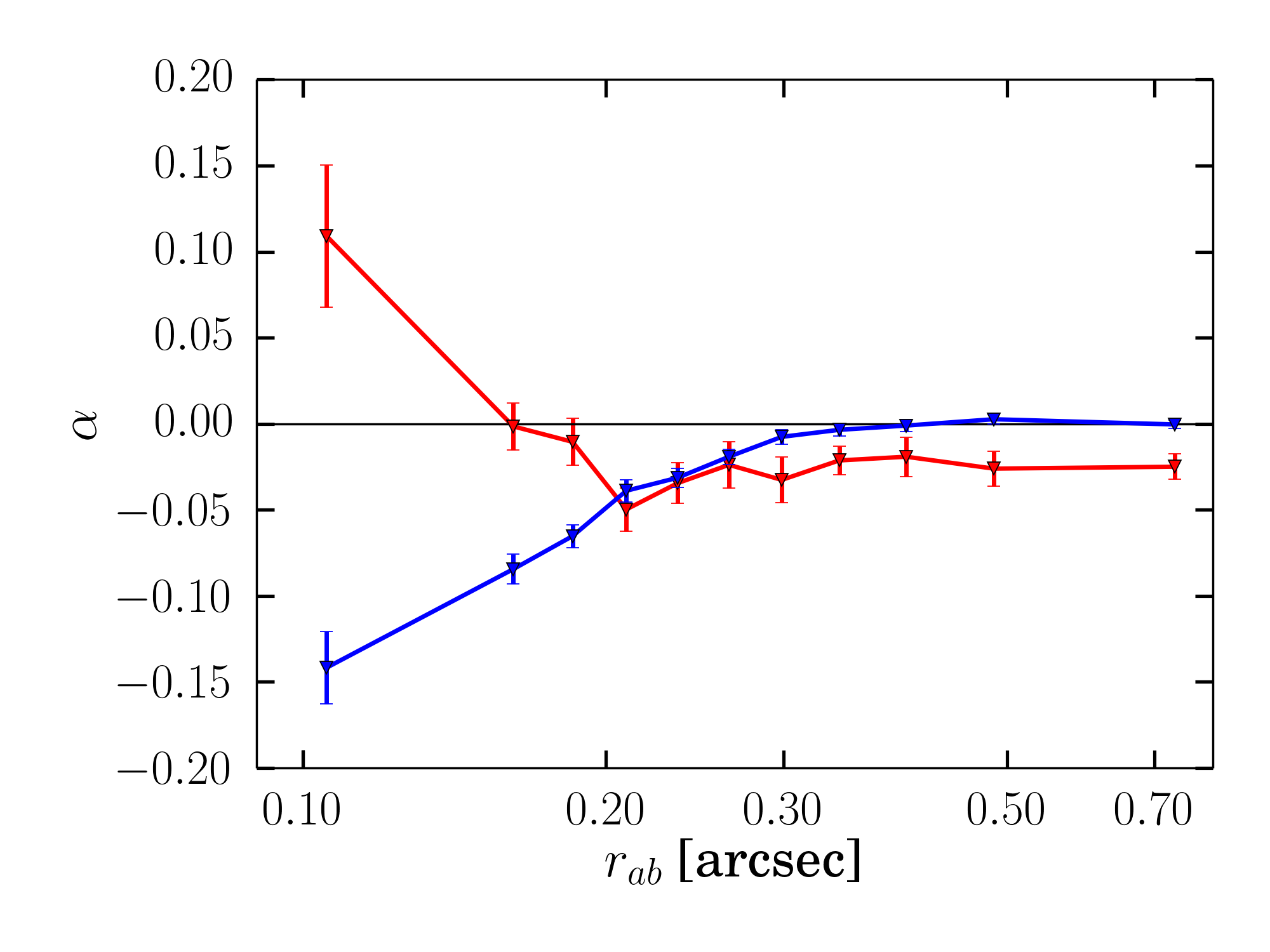}
 \end{tabular}
\caption{The PSF leakage for measured and true ellipticities as a function of simulation input quantities. Input magnitude in the \textit{left panel} and input size in the \textit{right panel}. \label{CAL::ALPHA::SIM}}
\end{figure*}

We measure the two $\alpha$ components by subdividing the galaxy sample into bins of the respective PSF ellipticity component.
For the full sample, without any further subdivision into bins of galaxy properties we determine $\alpha_{1}=-0.006 \pm 0.002$ and $\alpha_{2}=0.005  \pm 0.003$ for the 
self-calibrated \lf{} output. It is important to note that no additional residual bias calibration, as described in  \S\ref{sec:shearCalib:mbias} and \S\ref{sec:shearCalib:cbias} is applied here.  
Fig.\,\ref{CAL::ALPHA::OBS} shows the dependence of $\alpha$, which is sometimes also called PSF leakage, on measured galaxy properties and
Fig.\,\ref{CAL::ALPHA::SIM} shows it as a function of simulation input quantities. Clearly, the measurement is significant over the full property range, but
is most significant for the low SNR and the small size regime. 
Fig.\,\ref{CAL::ALPHA::OBS} also shows the bias obtained when true, sheared ellipticity
values are propagated through the analysis, as in \S\ref{sec:selectionbias}. We observe that the $\alpha$-dependence on SNR is well explained by the selection bias, but that there remains $\alpha$-dependence on the relative galaxy size that appears to have an additional contribution to the selection bias. 

In summary, referring to our preferred calibration scheme (method \textit{C}), all m, c -and $\alpha$ biases vanish for the galaxy sample in its entirety. When looking closer into the biases as a function of 
measured galaxy properties we find small, of the order 2\% residual multiplicative biases for extremely low SNR and extremely high $\cal{R}$ objects. All c-biases vanish after our calibration and while residual $\alpha$ terms are presented in the 
self-calibrated \lf{} output, they vanish after the additional residual bias calibration. 
We do expect the PSF-dependent additive biases to be sensitive to the PSF properties, and thus we recommend that the
additive bias measured from the simulations is not simply applied blindly to any science analysis.  
In \citet{KiDS450}, the additive bias is investigated empirically in the data, and the results compared with
those from the simulations, rather than relying on the simulations to be an exact representation of the data regarding
its PSF and noise properties.

\section{Calibration by resampling the simulated catalogue}\label{sec:resampling}

\subsection{A resampling approach to calibration}
Once the bias has been characterised in terms of relevant observed properties, it can be applied to virtually any selection of the real galaxies used to measure shear. For example, a tomographic cosmic shear analysis requires splitting the galaxy sample into redshift bins; a galaxy-galaxy lensing analysis requires selecting a source sample behind lenses at a given redshift.  However, as we saw in \S\ref{sec:shearCalib}, the bias surface may be complex and thus difficult to characterize, and may itself be biased (see \S\ref{sec:calibselectionbias}). This may be a concern, given the tight requirements from current and especially future lensing surveys.

The \lf{} measurements are, however, made for individual objects, and as an alternative to the approach presented in  \S\ref{sec:shearCalib}, we may instead resample the output from the image simulations, such that the measured galaxy parameter distributions match those of any (sub-)selection of galaxies. The multiplicative and additive biases may then be calculated from the resampled catalogues and applied to the galaxy sample of interest. Note, however, that this approach will only give reliable
results if the multi-dimensional parameter space of simulated galaxy properties covers the full parameter space of the real galaxies. Whilst this approach is less flexible than the one described in \S\ref{sec:shearCalib}, as the simulations need to be resampled for each galaxy sample used to measure shear, it avoids having to characterise the bias as a function of galaxy properties. 
 
Comparison of the biases determined using the different schemes provides an important check on the robustness of the calibration. As described in more detail below, we therefore implemented the resampling approach and applied it to the four tomographic bins used in the cosmic shear analysis presented in \cite{KiDS450}. 

\subsection{Application to the multiplicative bias in KiDS data}

For a given selection of real galaxies, the population of simulated galaxies may be resampled using a $k$-nearest neighbour search of an $N$-dimensional volume, defined by a combination of $N$ observed properties of the simulated galaxies. As the search is done by minimising the Euclidian distance between the simulated and real galaxies in that space, 
it is important to map the distributions of the chosen properties onto a unit length vector. Moreover, there are two important points to consider in order to successfully apply this technique:
\begin{itemize}
\item The galaxy properties that define the $N$-dimensional volume must be correlated with the shear bias;
\item The $N$-dimensional volume of the simulations has to be at least as large as the corresponding volume defined using the properties of the real galaxy sample. 
\end{itemize}

\begin{figure}
\includegraphics[width=9cm, angle=0]{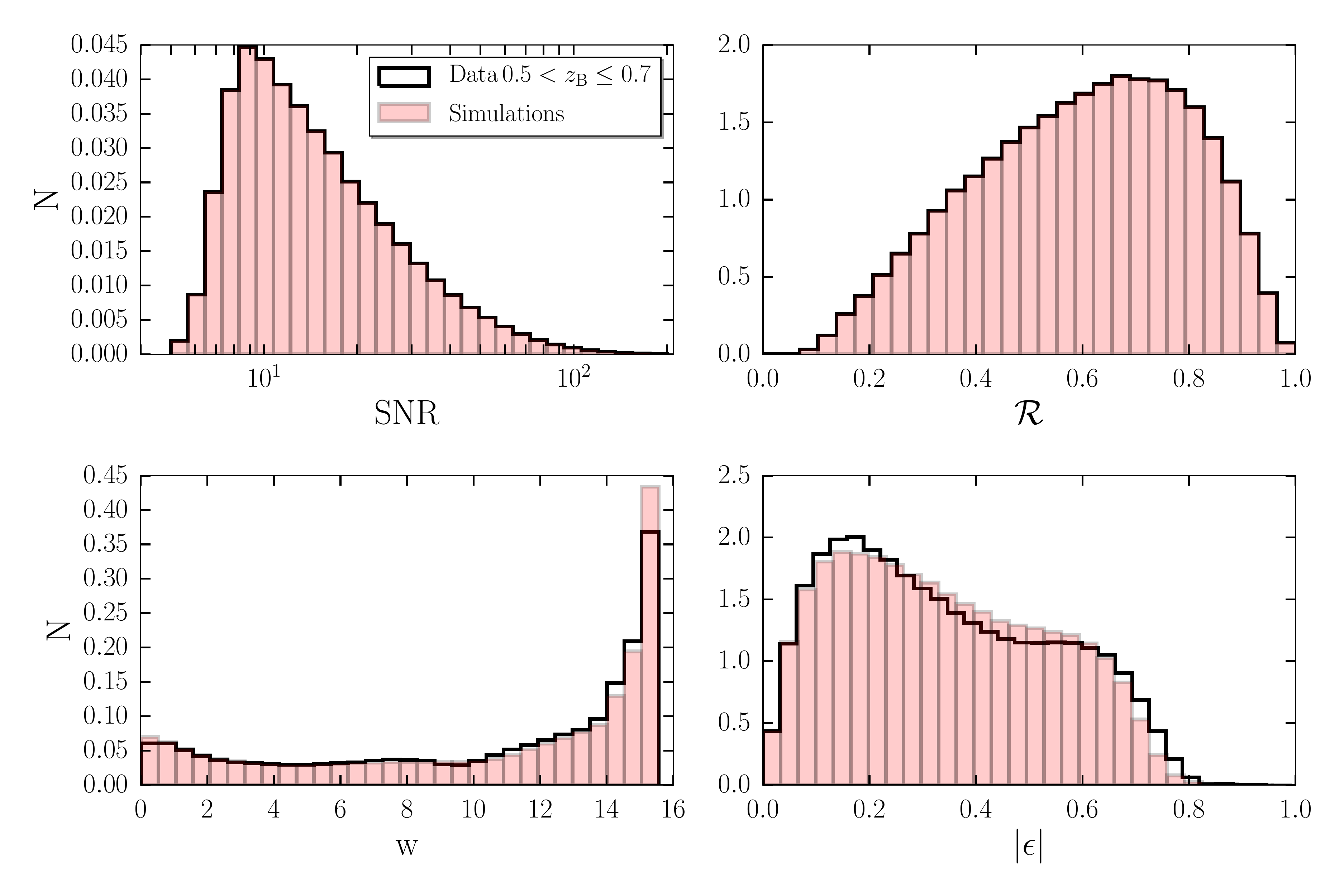}
\caption{\textit{Top panels:} SNR and ${\cal R}$ distributions measured from the KiDS-450 data (black line) and using the resampled simulations (red histogram). \textit{Bottom panels}: 
The distribution of \lf{}  weight (left) and  weighted ellipticity (right) measured from the KiDS-450 (black line) and using the resampled simulations (red histogram). All distributions are computed using galaxies in the redshift range $0.5 < z_{B} \le 0.7$, which corresponds to the third tomographic bin used in the cosmic shear analysis presented in \citet{KiDS450}}.\label{fig:Dis_res_data}
\end{figure}

Motivated by the results presented in \S\ref{sec:shearCalib}, we define the resampling volume based on the galaxy SNR  and the ratio of the PSF size and observed galaxy size (${\cal R}$), for which the simulations cover the same space as the data, as we have shown in \S\ref{sec:compsimdata}. We apply the resampling technique to the selection of galaxies defined by the four tomographic bins used for the cosmic shear analysis presented in \cite{KiDS450}. Our simulations do not contain any simulated redshift information: we implicitly assume that matching the size and SNR distributions of each tomographic bin is adequate, and that there is no redshift dependence of the bias beyond that conveyed by the bias as a function of SNR and size.

The tomographic bins are defined using the peak of the posterior photometric redshift distribution $z_{\mathrm{B}}$ as measured by BPZ \citep{benitez00a} using observations in four optical bands $ugri$ \citep{kuijken15a}. The KiDS-450 data are further divided in five contiguous regions on the sky (designated G9, G12, G15, G23 and GS). We resample the simulations using each region individually, in order to test the robustness of the method, although we note that the SNR and ${\cal R}$ distributions are very similar between the regions.   

The top panels in Fig.\,\ref{fig:Dis_res_data} show the SNR and ${\cal R}$ distributions measured from the KiDS-450 data (all regions combined) and those obtained from the resampled simulations for the third tomographic bin, $0.5<z_{\mathrm{B}}\leq 0.7$, used in \citet{KiDS450}. The excellent agreement between them validates the resampling technique and confirms that the simulations are representative of the data. 
In the bottom panels of Fig.\,\ref{fig:Dis_res_data} we show the
distributions of the \lf{} weight and the weighted distribution of the modulus of the ellipticity. 
As those two quantities were not used in the resampling, it is not surprising that the distributions differ slightly.
However, the amplitude of the noise bias depends on the galaxy ellipticity distribution \citep{viola14a}: we will assess 
the possible impact of this mismatch on the derived average biases in \S\ref{sec:sensitivity}.

\subsection{Robustness of the tomographic calibration}
\label{sec:robustness}
  
From the $k$-nearest neighbour search we can define a `resampling' weight $w^{\rm res}$, which
is the number of times that a simulated object was matched to an object in the data.  We use this new weight in combination with the \lf{} weight to measure the shear from the resampled simulations:
\begin{equation}
g^{\rm obs,res}_j\equiv\frac{\sum_i w_iw_i^{\rm res}\epsilon_{ij}}{\sum_i w_iw_i^{\rm res}},
\end{equation}
\noindent and compute the multiplicative and additive bias using equation\,\ref{EQU::CAL::BIAS}. We verified that the estimate for the bias is robust against the choice of the number of nearest neighbours. The errors on the biases are also unchanged for $k>4$. Unless explicitly stated, all the results quoted in this paper have been derived using $k=5$. 

\begin{figure}
\includegraphics[width=9cm, angle=0]{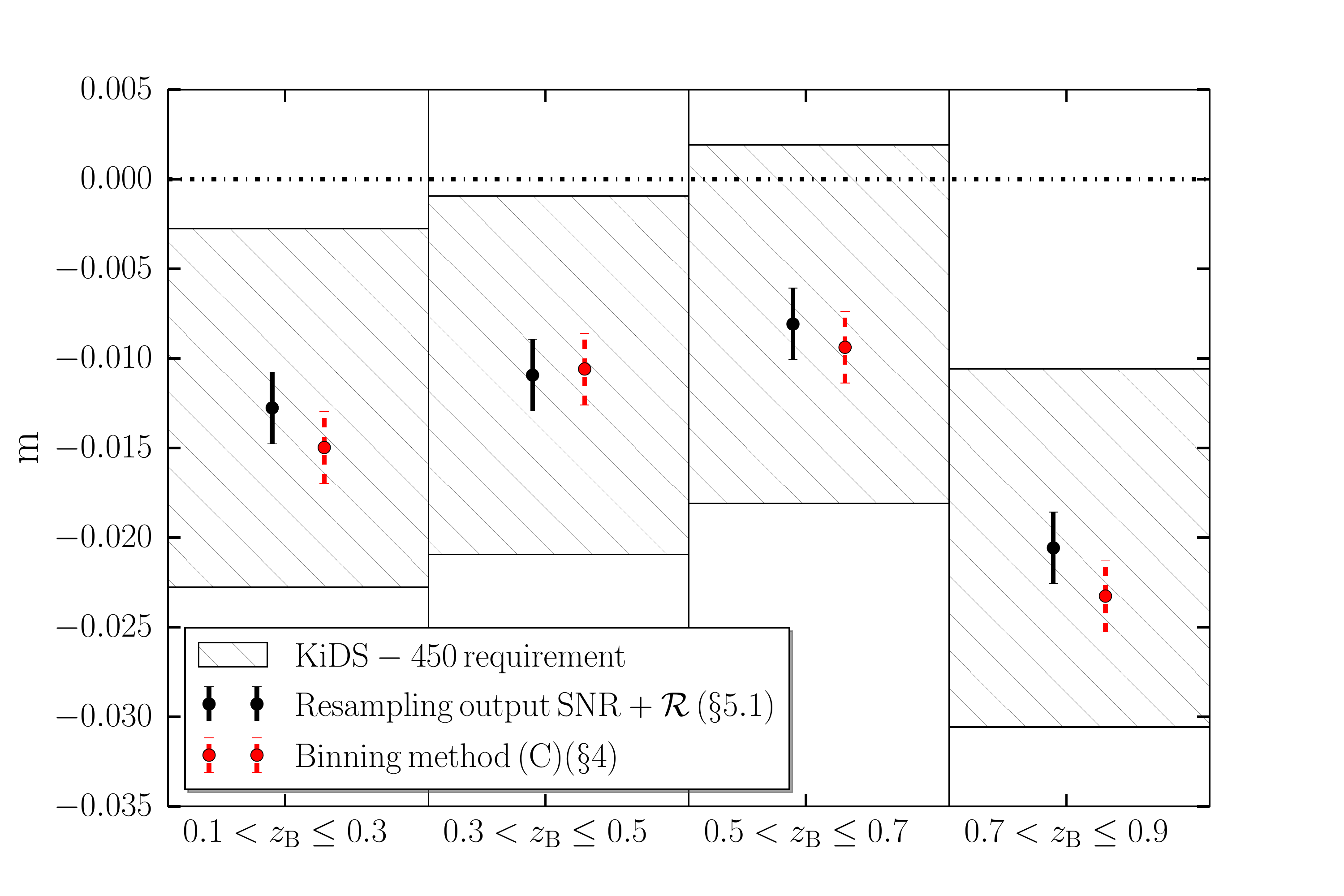}
\caption{Multiplicative bias calculated using the resampling technique  
and the bias calculated employing the calibration scheme described in \S\ref{sec:shearCalib} as a function of the tomographic bins used in the cosmic shear analysis described in \citet{KiDS450}. The hatched area indicates the requirement on the knowledge of the multiplicative bias for KiDS-450.\label{fig:Bias_comp}}
\end{figure}

The measured multiplicative bias does not depend on the PSF properties, in agreement with what we found in \S\ref{sec:shearCalib}. As an additional test we compared the average biases derived from resampling each individual PSF set individually with the results derived from resampling the whole simulation volume. Also in this case we found statistically equivalent results. Fig.\,\ref{fig:Bias_comp} shows the multiplicative bias derived using the resampling technique and the calibration method presented in  \S\ref{sec:shearCalib}. The hatched regions, centered on the bias measured using the resampling technique indicate the requirements in the knowledge of the multiplicative bias as derived by \cite{hildebrandt16a}. We compare the results from the two calibration schemes for the four tomographic bins 
used in \cite{KiDS450}. The average difference, combining all tomographic bins, is $\Delta m = -0.001 \pm 0.003$. 

\section{Calibration sensitivity analyses}
\subsection{Sensitivity to the magnitude distribution}
\label{sec:sensitivity_mag}

In \S\ref{sec:shearCalib:mbias} we noted that there might be a residual shear bias that arises from differences between the magnitude distributions of the simulations and of the selection of galaxies in the tomographic bins. 
We estimate this effect by first applying the method $C$ calibration scheme to the simulations.  Then, a new resampling weight is derived for each galaxy,  by comparing the \textit{lens}fit-weighted distributions of measured magnitudes in the simulations and in the KiDS-450 data in each tomographic bin, and reweighting the simulated galaxies so that those distributions match.

We measure the residual bias in these reweighted simulations, for each tomographic bin. First, we confirm that the residual bias is consistent with zero in the absence of any magnitude reweighting, as expected. Then, for each tomographic bin reweighting, we find residual bias levels of approximately $-0.001, 0.001, 0.0004, -0.012$ in each of the four bins.  The residual bias is consistent with zero in the first three bins, but shows a percent-level
residual in the highest-redshift bin.  We cannot know whether this effect is as large in the data as in the simulations, for two reasons: first, we have reweighted using noisy, measured magnitudes rather than true magnitudes, and second we know that the simulations become incomplete at a slightly brighter magnitude limit than the data, so the residual bias effect is expected to be larger in the simulations than in the data. However, this test does indicate the possible size of the residual bias, which is either much smaller than (tomographic bins $1-3$) or comparable to (tomographic bin 4) our nominal requirement on calibration accuracy.

To explore further the effect of the simulation magnitude limit on the measured shear bias we run another suite of simulations, which are identical to the reference simulations described in Section \ref{galsims}, except that we change the noise level, such that the magnitude limit increases by 0.3 magnitude. These simulations are 0.2 magnitude deeper than the KiDS-450 data. We apply the method $C$ to these new simulations and we compute the multiplicative shear bias in the four tomographic bins. Compared to the fiducial results we find a change in the bias of$-0.008,-0.003,-0.006,-0.014$ in each of the four bins. We can use this result to estimate the sensitivity of the bias to the magnitude limit from which we can calculate that the 0.1 magnitude limit different between the reference simulations and the KiDS-450 data should result in sub-percent residual biases of $-0.003,-0.001,-0.002, -0.005$ in the four bins.

\subsection{Sensitivity to the galaxy size distribution}
\label{sec:sensitivity_size}

The output galaxy size distribution also differs between the data and the simulations, as shown in Fig.\,\ref{fig:distr_sim_data}, which might arise from a difference between the input size distribution we used to create the simulations and the true size distribution of the KiDS-450 galaxies. 
To examine in more detail the impact of such a difference, we again reweight the galaxies such that the output
size distributions of data and simulations match.  However, in this case we cannot simply weight by the distribution
of output size, as that would not capture correctly the joint dependence of the correlated output size and
ellipticity measurements. Instead, we choose to reweight simulated galaxies as a function of their true, input
size.
We first define an alternative target input size distribution and calculate a `size weight' that may be applied to each galaxy, such that the fiducial input size distribution is transformed from the nominal distribution to the target distribution. The size weight is just the ratio of the values of the target and nominal distributions for each galaxy.  The target distribution was varied until a good match of output size distributions was found. The simplest target distribution that was tried had the same functional form as the input size distribution, but with a shift of the median relation by a constant factor to larger sizes, while preserving the magnitude dependence. The factor was varied to obtain the best match between the simulation and data size distributions (as measured by the Kolmogorov-Smirnov statistic), however differences in the distributions remained.  

Hence, we also tested a lognormal target distribution, where the median size was again scaled by some factor and where the standard deviation of the distribution of the logarithm was also varied to obtain the best match between data and simulations. This produced a better match, but with some magnitude dependence: a final sophistication then was to allow the slope of the $r_{\rm med}-m$ relation to vary.  The new relation was found to be $r_{\rm med} = \exp(-1.07 - 0.19(m-23))$ with standard deviation of the logarithm $\sigma = 0.48$. A good match was then found between the size distributions of the data and the reweighted simulations.  The size reweighting also causes some variation in the measured distributions of other quantities, but does not on its own remove the discrepancies between the data and simulations in the distributions of magnitude and SNR.

To test the possible effect on the deduced bias, we apply the size reweighting globally to the entire simulation, repeat the bias estimation using method C, and then deduce again the bias for each tomographic bin, as described above.
The reweighted bias values differ from the nominal values by
$-0.0011, -0.0014, -0.0013, 0.0085$ in each tomographic bin.
The differences in the first three bins are again negligible, with only a sub-percent level effect in the
final tomographic bin.  That effect has the opposite sign to that found in the magnitude reweighting, which suggests
that the joint effect of magnitude- and size-reweighting may be close to zero in all tomographic
bins.  
We conclude that the effect of the uncertainty in either the size or magnitude distributions does not
impact our tomographic bin calibration at the level of accuracy required here.

\subsection{Sensitivity to accuracy of the galaxy ellipticity distribution}
\label{sec:sensitivity}

A remaining concern is that the recovered ellipticity distribution in the simulations does not match precisely those from the KiDS-450 observations. This may indicate either that the intrinsic ellipticity distribution in the simulations is not the same as in the real Universe, or that some other observed property that is correlated with ellipticity is biasing the distribution. Such a discrepancy in the ellipticity distribution may result in a bias measured from the simulations which may not be applicable to the observations \citep{melchior12a,viola14a}. To quantify how our results change for different input ellipticity distributions, we perform a further resampling sensitivity analysis, similar to those done by \citet{bruderer15a} and \citet{hoekstra15a}, that investigates the effect of possible
variations in the ellipticity distribution on the resampling calibration, in tomographic bins 
(\S\ref{sec:resampling}).

\begin{figure*}
 \centering
 \hbox{
  \includegraphics[width=0.48\textwidth]{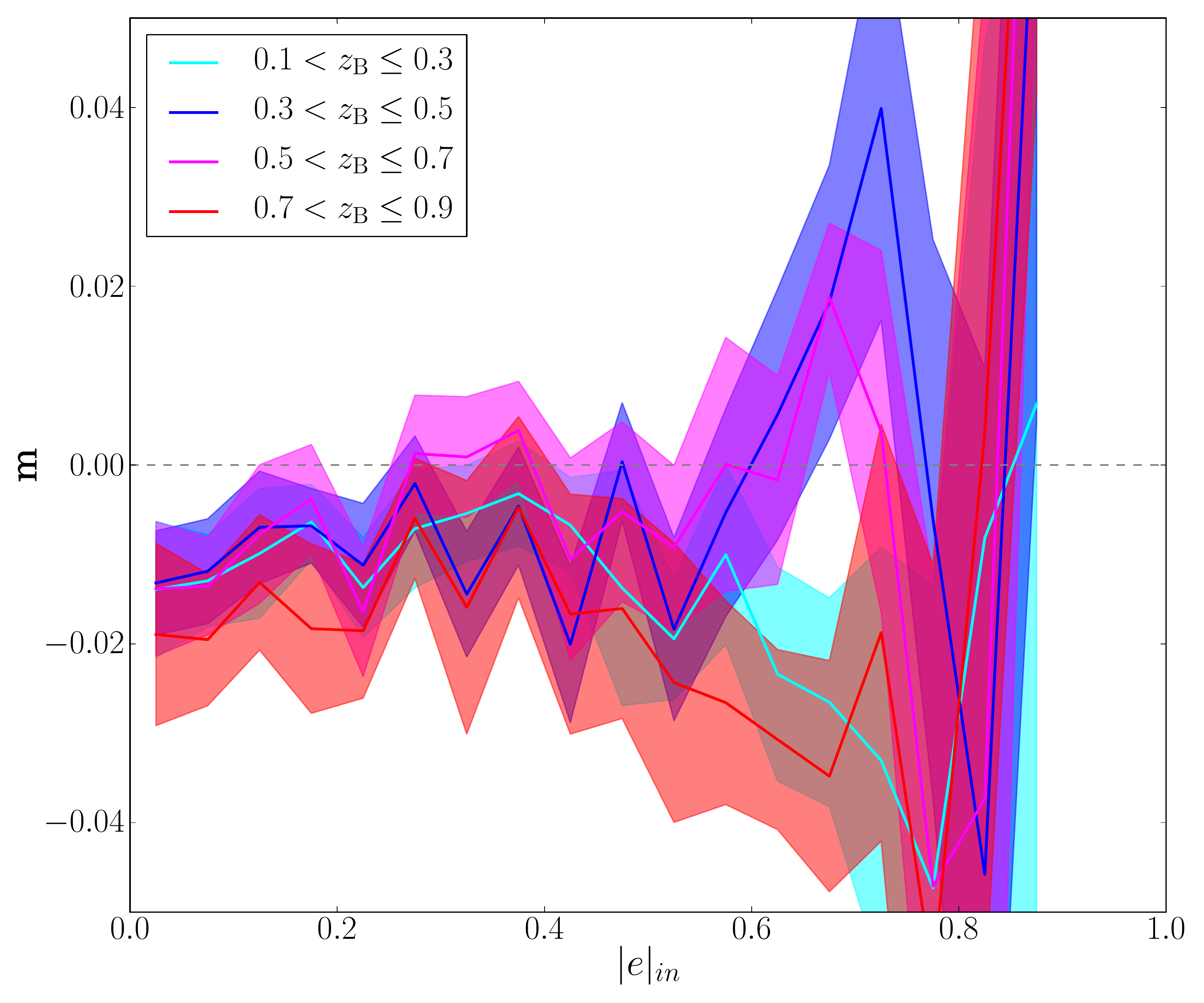}
   \includegraphics[width=0.48\textwidth]{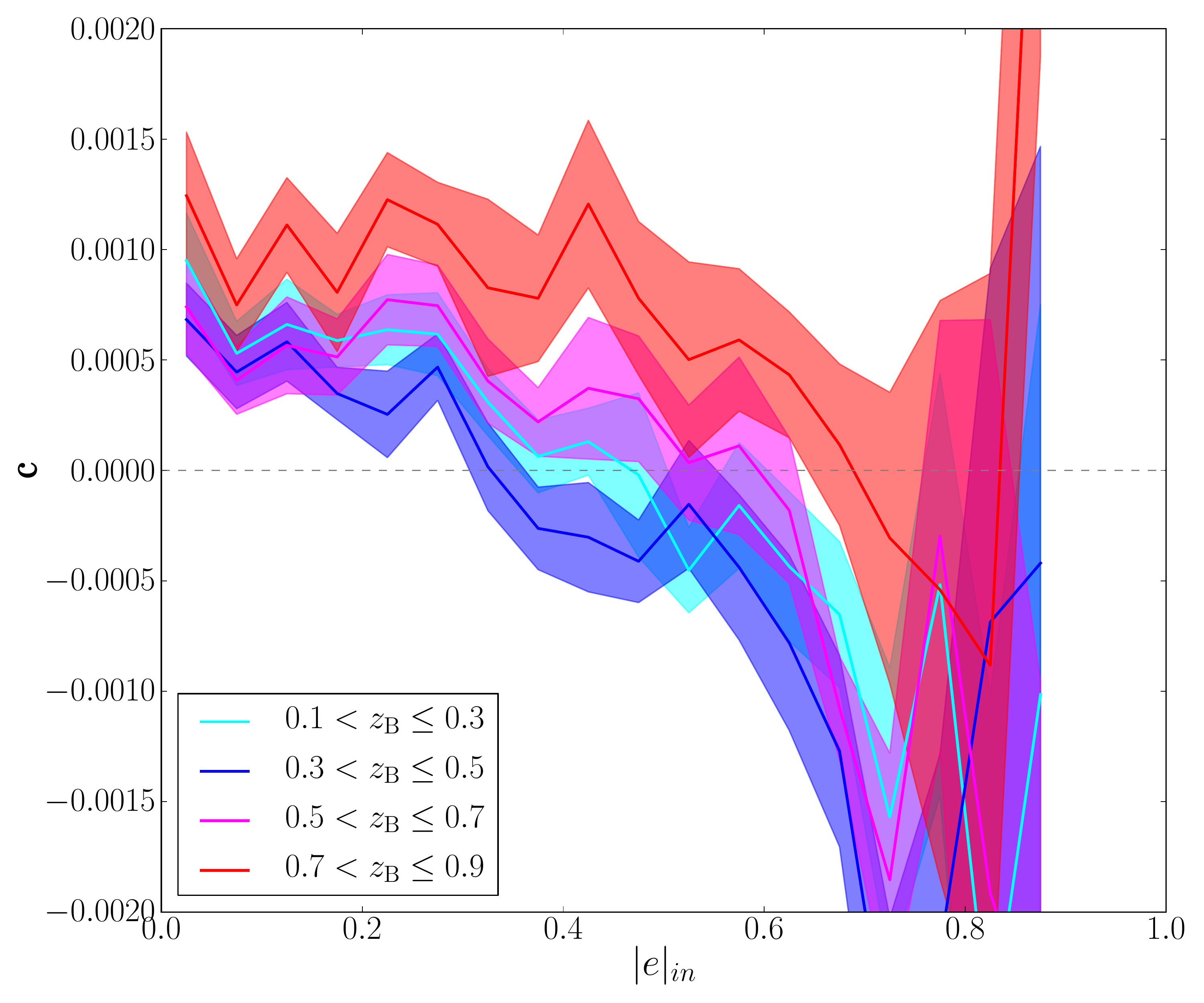}} 
   \caption{Multiplicative bias (left panel) and additive bias (right panel) for bins in input ellipticity for the four tomographic resampled catalogues with $1\sigma$ uncertainties. A redder colour indicates a higher redshift tomographic bin. 
 \label{fig:sensitivity}
}
\end{figure*}

We first quantify the sensitivity of the shear measurement to the input ellipticity distribution, by binning the simulated galaxies  according to their input ellipticity, $\epsilon^{s}$, and computing the multiplicative and additive bias in each ellipticity bin. The results are presented in Fig.\,\ref{fig:sensitivity} for the  resampled catalogues for the four tomographic bins (see \S\ref{sec:resampling}). Thanks to the resampling, these catalogues have the same observed SNR and resolution distributions as the KiDS-450 data in each tomographic bin. The multiplicative bias depends only weakly on the intrinsic ellipticity for objects with low ellipticities, although the biases differ between tomographic bins.
For the additive bias we observe a clear trend with $\epsilon^{s}$, but we note that the amplitude is low
and we do not, in any case, apply our simulated additive bias measurements directly to the data.
These findings are in line with the expectations  from \citet{viola14a} and show that modest changes to the input ellipticity distribution should result in at most a percent level effect on the overall multiplicative bias.

The results for the four tomographic bins shown in Fig.\,\ref{fig:sensitivity} indicate that the sensitivity 
of the multiplicative bias to the adopted intrinsic ellipticity distribution is small. Nonetheless, we aim to quantify this further by considering possible variations of the input ellipticity distributions in the simulations.
To do so, we follow a similar method to that in \S\ref{sec:sensitivity_size}, by
applying additional weights to the catalogue entries as a function of their input intrinsic ellipticity, 
and then computing the new, reweighted bias. The difficulty in this approach is that there may be many possible
variations of the true ellipticity distribution that result in the same, or similar, measured ellipticity
distributions.  
So, although the principle of resampling is analogous to that done in 
\S\ref{sec:sensitivity_size}, 
here we follow a Monte-Carlo approach to the reweighting, in which we test many possible
variations of the true ellipticity distribution, only selecting those that produce a match with the KiDS-450 data.
As the input ellipticity is uncorrelated to any other input galaxy property in the simulations, the new weight does not introduce any further bias due to selection effects in our measurements.
Here we focus on the ellipticity distribution, but note that this method could be used for other, or multiple, distributions, provided that the simulated volume is large enough. The steps for our sensitivity analysis procedure are as follows:
\begin{itemize}
 \item We bin the \lf{} weighted input ellipticity distribution in equally spaced bins 
 $P_{i}^{s}(|\epsilon|)$. 
 \item For each input ellipticity bin we determine the corresponding observed ellipticity distribution 
 $\widetilde{P}_i^{\mathrm{out}}(|\epsilon|)$.
 \item We assign a weight $\tilde{w_i}$ to each input ellipticity bin, resulting in a modification of both the input and output ellipticity distributions.
\end{itemize}

In this way we can mimic image simulations with differing input ellipticity distributions, without the need to create and analyse such simulations. For our analysis we have chosen to use 50 bins in input ellipticity. The weights $\tilde{w_i}$ are chosen such that the simulated output ellipticity distribution matches the observed ellipticity distribution in the KiDS-450 data. The intrinsic ellipticity distribution in the Universe varies due to cosmic variance, which limits the precision with which the bias can be determined from our sensitivity analysis. An estimate for cosmic variance can be obtained from the variation in the observed ellipticity distributions between the KiDS-450 patches. We found that these variations are very similar to the Poisson errors on the observed ellipticity distribution. When comparing the ellipticity distributions from simulations and data we therefore assign Poisson errors to the latter. 

Matching the observed and simulated ellipticity distributions can only be done reliably if the full range of  ellipticities found in the data is encompassed by the simulations. In the course of performing the analysis, we found that the KiDS-450 data contain a small fraction of galaxies with $\epsilon>0.8$, which are absent in the simulations
(see the inset in the lower left panel of Fig.\,\ref{fig:distr_sim_data}). In the simulations, such high ellipticities are caused either by measurement noise or by blending of galaxies with close neighbours. To check whether the objects in the data are also caused by noise or blending, we inspected {\it HST} images of the COSMOS field \citep{scoville07a} for which we also have VST $r$-band data. To ensure a fair comparison, we restricted the comparison to images in the $F606W$ filter, which is similar to the $r$-band.

Unfortunately, the $F606W$ imaging in the COSMOS field only covers 240 arcmin$^2$, resulting in a comparison sample of only about 100 galaxies. We found that 70\% of these objects were genuinely high-ellipticity, edge-on galaxies, while the rest were either spurious detections or blended objects. The likely cause is that there exists a distribution
of the ratio of galaxy disk scale-heights to their scale-lengths \citep[e.g.][]{unterborn08a}, with a tail of galaxies having very thin disks,
which are not represented by the nominal ellipticity prior that we assume.
Even though the comparison sample is small, this test suggests that the high-ellipticity tail of the \lf{} prior is not representative of the Universe in this regime. However, the sample is too small to allow us to derive an updated ellipticity prior. 
Instead, to compensate this incompleteness, we augment our catalogues with very elliptical objects. We created and analysed additional simulations with 2000 galaxies per exposure, adopting a flat input ellipticity distribution with 0.5$\le | \epsilon | \le$0.95. All other properties of the simulations remained unchanged from what has been described in \S\ref{galsims}. Note that the number density of these
very elliptical galaxies does not reflect reality, but rather was chosen to provide adequate information
for  the sensitivity analysis.

We use Monte Carlo Markov Chains (MCMCs) to sample the $\tilde{w_i}$ parameter space. We found that convergence was slow, and the resulting input ellipticity distribution very irregular and spiky if no priors on $\tilde{w_i}$ were imposed. This result is not physical, and does not agree with our limited knowledge of 
the ellipticity distribution based on high quality data, which indicates a much smoother distribution.
To speed up the MCMC runs in finding a more physical solution, we applied a prior to regularise the result. The form of the prior is 
\begin{equation}
\pi(K, |\epsilon^{s}|):=K \times \left |{ 1 - \frac{P_{i+1}(|\epsilon^{s}|)}
{P_i(|\epsilon^{s}|) }}\right| \frac{|\epsilon^{s}|_i}{|\epsilon^{s}|_{i+1}} \, ,
\end{equation}
\noindent which penalises a spiky distribution where subsequent bins have very different values. The extra factor of $|\epsilon^{s}|_{i} / |\epsilon^{s}|_{i+1}$ lessens the effect of the prior near $|\epsilon|=0$, where the distribution turns over. The strength of the prior $K$ should be chosen so that the prior does not dominate.  We explored several values of $K$ and found a good compromise for $K=500$; this
choice produced physical distributions in a reasonable amount of computing time.

\begin{figure*}
 \centering
  \includegraphics[width=1.0\textwidth]{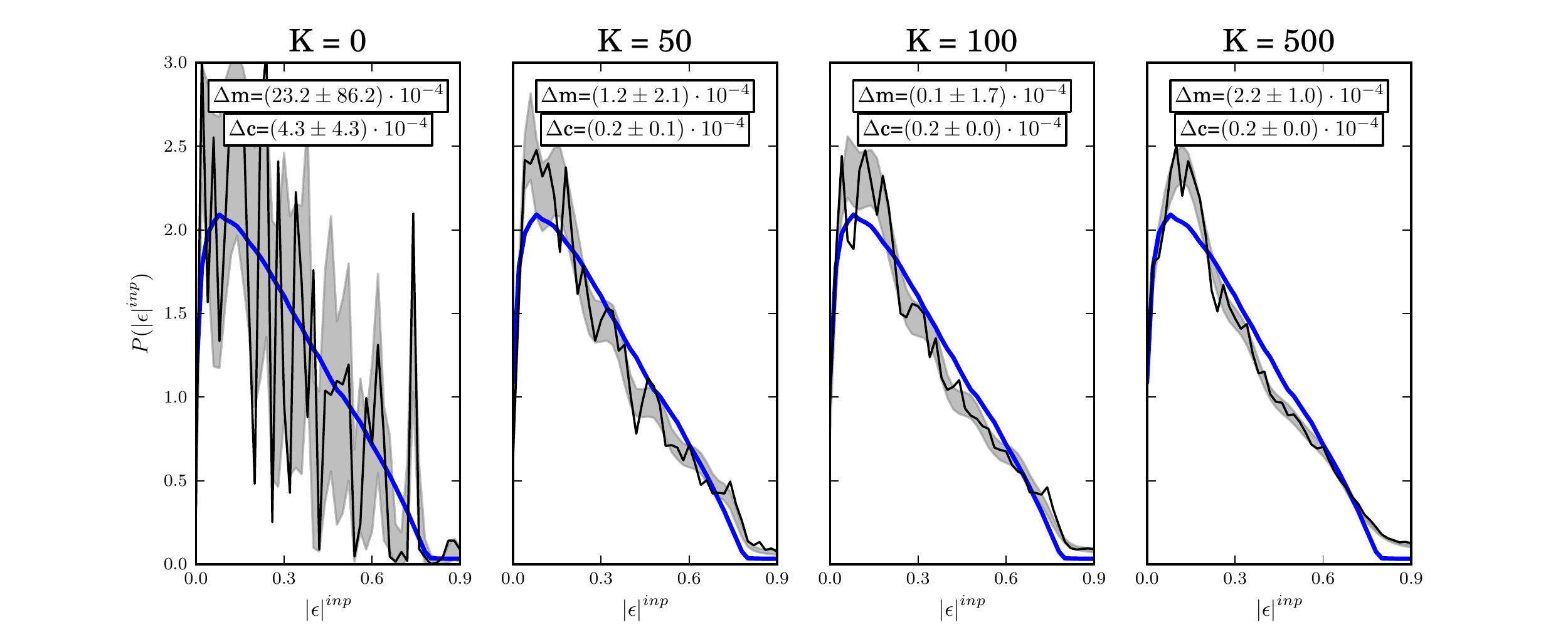}
   \caption{ Results from the sensitivity analysis based on $0.5 \leq Z_B < 0.7$ galaxies in the G15 patch of the KiDS DR3 data. The intrinsic ellipticity distribution in the resampled catalogue in blue and the distribution which best fits the measured KiDS data and the grey band shows the possible variations from the MCMC tests. To suppress the spiky nature of the best fit we demanded smoother intrinsic ellipticity distribution, finding a strength of the smoothness prior $K=500$ to be adequate, as indicated at the top of the plot. The bottom row shows how similar the observed ellipticity distribution is to the KiDS-450 data for the resampled catalogue in blue and the best fit in black. The textboxes show the difference in multiplicative (top box) and additive (bottom box) bias between the blue and black distribution. The biases change with $K$, but all biases are much smaller than the 1\% required for cosmic shear.}
 \label{fig:mcmc_smooth}
\end{figure*}

The third tomographic bin ($0.5<z_{\mathrm{B}}\leq 0.7$) shows the largest discrepancy between the observed ellipticity distribution in the simulations and KiDS DR3 data and thus serves as a worst case scenario for the sensitivity analysis. We use the ellipticity distribution from patch G15 in the sensitivity analysis and use the 1$\sigma$ variation between the patches as the error on the distribution. The results of our sensitivity analysis and the effect of the smoothing prior are shown in Fig.\,\ref{fig:mcmc_smooth}, which shows the input ellipticity distribution of the \imsim \ simulations $P(|\epsilon^{\mathrm{s}}|)$ in blue and the best fit model $\sum_i \tilde{w_i} P(|\epsilon^{\mathrm{s}}|)_i$ from the MCMC results in black. The MCMC chains converged for every run, so that the observed ellipticity distribution was identical to the KiDS ellipticity distribution within the errorbars.

The MCMC framework was able to match the simulations to the data.
For the family of modified ellipticity distributions from the MCMC, we compute the standard deviation in input ellipticity for each bin and show this as the grey band. From left to right the strength of the smoothness prior increases, resulting in smoother distributions. Importantly, the unphysical spike around $|\epsilon^{s}|=0.75$ is no longer present in this case. For 1\% of the $\sim2 \times 10^7$ MCMC solutions we computed the shear bias from the corresponding (observed) ellipticity distributions. The difference between the average bias and that measured from the resampled catalogue is shown in the boxes and the error is the $1\sigma$ spread of all the computed biases. The difference in ellipticity distribution thus results in only a small change in bias. The biases also change very little as a function of the applied smoothing; the change in multiplicative and additive bias never exceeds 0.3\% and 0.01\%. These tests show that the shear measurement is quite insensitive to changes in the intrinsic ellipticity distribution and any reasonable variations are within the 1\% errors. The discrepancy between the observed ellipticity distribution in the simulations and the data is therefore not a concern for the cosmic shear analysis.

\section{Conclusions}

The large areas covered by ongoing and future imaging surveys dramatically reduce the statistical uncertainties in the measurement of the alignments of galaxies caused by lensing by intervening large-scale structure. This increase in precision needs to be matched by a corresponding improvement in the accuracy with which 
weak lensing shear can be measured. This can only be achieved by evaluating the performance
of shear measurement algorithms on realistic mock data \citep[e.g.][]{miller13a, hoekstra15a}.
In this paper we use extensive image simulations created using \textsc{GalSim} \citep{rowe15a}, to test and calibrate  the \lf{} algorithm used by \cite{KiDS450} to analyse 450\,deg$^2$ (360.3\,deg$^2$ after accounting for masking) of KiDS-450 data. This large survey area implies that the multiplicative bias needs to be determined to better than about $1$ percent.

We have shown that the average multiplicative bias over the simulation volume using the self-calibrating \lf{} algorithm is $\sim$ 2\%, and the average additive bias is $\sim 5\times 10^{-4}$. Although this is close to the required level of accuracy, a final correction is nonetheless required. We have investigated the behaviour of the bias as a function of observed properties of galaxies, such as SNR and size. 
The measured bias as a function of galaxy properties is a combination of measurement bias, caused by noise, and selection bias, caused by the inability to measure small galaxies and by the weighting of galaxies in the
shear measurement process. While it is possible to disentangle those effects in the simulations, it is not possible to do the same in the data. In our analysis, we find that selection bias is at least as important as measurement bias, which implies that even shear measurement methods that are free from, or that perfectly correct for, noise bias may still show shear biases that are present at the percent level or larger. 

We have successfully derived a calibration relation that corrects for the dependence of bias on galaxy properties, but we have also shown that this calibration itself may be biased by its use of noisy, measured galaxy properties rather than their unobservable true properties, and these `calibration bias' effects need to be assessed when deriving any new shear calibration. We have tested the accuracy of the application of the calibration relation, including the effect of calibration bias, by a number of resampling tests that were designed to test the accuracy 
in the four tomographic bins used in the cosmic shear analysis presented by \cite{KiDS450}. 
Although there are sub-percent uncertainties in the calibrations arising from the differences between the data and the simulations, and from the effects of calibration bias, the accuracy of the calibration appears to satisfy the specification required for cosmic shear analysis of the KiDS-450 data set, at 1 percent accuracy of multiplicative bias. In deriving cosmological constraints it is therefore necessary to marginalise over the uncertainty in the shear bias employing a gaussian prior with $\sigma_m=0.01$. As the SNR and $\cal{R}$ distributions in the four tomographic bins are very broad, the shear biases derived from the simulations described in this paper are strongly correlated among tomographic bins. For this reason we conservatively recommend to assume a correlation coefficient of r=0.99 between all bins.

\section*{Acknowledgements}

We would like to thank Joe Zuntz for his detailed referee report which helped improving the paper substantially.
We would like to thank Ami Choi, Thomas Erben, Catherine Heymans, Hendrik Hildebrandt and Reiko Nakajima for pipeline testing during the development of self-calibrating \lf{} using images from the KiDS survey, in addition to the KiDS Collaboration for providing the meta-data upon which the image simulations were based. We would also like to thank Dr Alessio Magro, from the Institute of Space Sciences and
Astronomy for many hours of help and support whilst running the simulations pipeline on the cluster.
This research has been carried out using computational facilities procured 
through the European Regional Development Fund, Project ERDF-080 ``A supercomputing laboratory for the University of Malta''.
JM has received funding from the People Programme
(Marie Curie Actions) of the European Unions Seventh
Framework Programme (FP7/2007-2013) under REA
grant agreement number 627288. HH, RH, MV acknowledge support  from
the European Research Council FP7 grant number 279396. This work is supported by the Netherlands Organisation for Scientific Research (NWO) through grants 614.001.103 (MV). LM is supported by STFC grant ST/N000919/1.

{\it Author contributions:} All authors contributed to the development and writing of this paper. The authorship list is given in alphabetical order (IFC, RH, HHo, JM, LM, MV). MV leads the image simulation working group in the KiDS collaboration.




\bibliographystyle{mnras}
\bibliography{lensing} 

\appendix

\section{Model bias}
\label{sec:modelbias_sims}

The measurements used for KiDS-450 may suffer from ``model bias'', if the assumed model surface brightness
distributions are mismatched to the true distributions of galaxies \citep[e.g.][]{kacprzak14a, zuntz13a}.
Results from the {\sc great3} challenge suggest that the amplitude of such bias is sub-percent and hence is subdominant compared to the $\sim 1$\,percent systematic uncertainties on the shear calibration arising from other effects that we estimate in this work. To verify this, 
here we describe a differential measurement between the shear recovered from a population of synthetic galaxies generated
by \textsc{GalSim} \citep{rowe15a} using {\em Hubble Space Telescope} (HST) images of faint galaxies and the shear
recovered from a population of galaxies made with synthetic bulge-plus-disk models whose distributions of
sizes and shapes match the HST galaxies.

First, a simulation was created using postage stamps of high resolution HST galaxies, with i-band magnitude between 20 and 24.5, which are available in \textsc{GalSim}. Each galaxy was sheared and convolved with the median KiDS PSF (FWHM=$0.64''$, Moffat $\beta$=3.14, $\epsilon_1$=0.08, $\epsilon_2$=-0.05) and rendered to a pixel scale of $0.214''$. The flux is the same for each object and set high enough with respect to the noise level, so that noise bias in the measurements is small. The simulated images consist of a grid of approximately 50\,000 isolated galaxies, so that blended galaxy isophotes do not influence the shape measurement. As was done for the fiducial simulations (see \S\ref{galsims}), four rotations of each galaxy were used to reduce shape noise and the same 8 shear values were tested. 
Given the high SNR of the galaxies and the use of four rotations, the simulated volume is large enough to achieve per mille precision in the shear bias determination.

\textsc{SExtractor} was run on the simulated images with the same configuration used in the analysis of the KiDS-450 data. About 1\% of the HST galaxies were incorrectly segmented and flagged by \lf \ in the subsequent analysis as blended. We visually inspected several postage stamps and indeed confirmed that these HST images showed unphysical features, such as a large number of negative pixels, creating problems for \textsc{SExtractor}. Furthermore another $\sim$ 1\% of objects were flagged by \lf \ and assigned a weight of zero. In order to retain the rotational symmetry we used in the subsequent analysis only galaxies for which all the 32 renditions (4 rotations time 8 shears) have a weight larger than zero and are unflagged, as would be the case in a survey of real galaxies.

We then reran the same simulation without applying the shear to the galaxies. This was necessary to determine the distributions of intrinsic galaxy properties for the input for the synthetic galaxy simulation. 
The modulus of the intrinsic ellipticity of each HST galaxy was obtained by averaging the modulus of the measured \lf \ ellipticity of the four rotations. As before, only if all four rotations were properly detected and had non-zero weight, were they included in the average.
Similarly we obtained the intrinsic scale lengths and bulge fractions. 

The comparison set of simulations were created using synthetic galaxies, adopting a bulge plus disk model. The modulus of the intrinsic ellipticity, the size and the bulge fraction were drawn from the measured distribution in the real galaxy simulation. The intrinsic position angle of galaxies was randomly assigned from a uniform distribution. 
This procedure ensures that the distributions between the first and the second set of simulations are the same and it also removes any bias in the \lf \ measurements correlated with the shear. 
These galaxies were sheared, in the same way as it was done for the HST galaxy simulations, and convolved with the same PSF.

Finally, the same analysis was run as described in Section \S\ref{sec:shearCalib} on the two catalogues and we compared the average biases. The HST galaxies showed an average multiplicative bias $m=-0.002 \pm 0.002$, while the bulge-plus-disk galaxy simulations the average bias was $m=-0.001 \pm 0.002$. We conclude that there is no evidence of a \lf \ multiplicative bias larger than couple of permille. This is in line with the previous results achieved on the GREAT3 benchmark simulations.

\bsp	
\label{lastpage}
\end{document}